\documentclass[a4paper,12pt]{article}

\usepackage{ucs}              
\usepackage[utf8x]{inputenc} 	

\usepackage[T1]{fontenc}     
\usepackage{lmodern}         

\usepackage[text={160mm,250mm},centering]{geometry}

\usepackage{url}
\DeclareUrlCommand\email{\urlstyle{tt}}
\usepackage[breaklinks,colorlinks=true,plainpages=false,linkcolor=blue]{hyperref}

\usepackage[british]{babel} 
\usepackage{authblk}

\usepackage{latexsym}
\usepackage{amsmath,amssymb,amscd,amsbsy}
\usepackage{upgreek}

\usepackage{tensor}

\usepackage{makeidx}         

\usepackage{tikz-cd}

\usepackage{graphicx}        

\usepackage{tabularx,paralist}

\usepackage{rcs} \RCS $Revision: 1.12 $ \RCS $Date: 2025/04/01 14:11:13 $ \RCS $Author: hgm $ \RCS $RCSfile: p-4th_tensors.tex,v $ \RCS $Id: p-4th_tensors.tex,v 1.12 2025/04/01 14:11:13 hgm Exp $

\usepackage{float}
\usepackage[font=normalsize]{caption}
\usepackage{subcaption} 
\usepackage{booktabs}
\usepackage{tikz}
\usepackage{pgfplots}

\usepackage{refdef}
\usepackage{ifontdef}

\usepackage{amsmath}

\newcommand{\ignore}[1]{}

\newcommand{\Ho}{\mathring{\mrm{H}}}
\newcommand{\Lp}{\mrm{L}}

\newcommand{\Hp}{\mrm{H}}

\newcommand{\vepsilon}{\varepsilon}
\newcommand{\veps}{\varepsilon}

\newcommand{\vtheta}{\vartheta}

\newcommand{\vek}[1]{\mathchoice{\displaystyle\boldsymbol{#1}}
{\textstyle\boldsymbol{#1}}{\scriptstyle\boldsymbol{#1}}
{\scriptscriptstyle\boldsymbol{#1}}}

\newcommand{\mat}[1]{\mathchoice{\displaystyle\mathbf{#1}}
{\textstyle\mathbf{#1}}{\scriptstyle\mathbf{#1}}
{\scriptscriptstyle\mathbf{#1}}}

\newcommand{\ops}[1]{\mathchoice{\displaystyle\mathsf{#1}}
{\textstyle\mathsf{#1}}{\scriptstyle\mathsf{#1}}
{\scriptscriptstyle\mathsf{#1}}}

\newcommand{\tnb}[1]{\mathchoice{\displaystyle\mathboldsans{#1}}
{\textstyle\mathboldsans{#1}}{\scriptstyle\mathboldsans{#1}}
{\scriptscriptstyle\mathboldsans{#1}}}

\newcommand{\tns}[1]{\mathchoice{\displaystyle\mathsans{#1}}
{\textstyle\mathsans{#1}}{\scriptstyle\mathsans{#1}}
{\scriptscriptstyle\mathsans{#1}}}

\newcommand{\vbar}[1]{\vek{\bar{#1}}}
\newcommand{\vtil}[1]{\vek{\tilde{#1}}}
\newcommand{\vhat}[1]{\vek{\hat{#1}}}

\newcommand{\Mtil}[1]{\mat{\tilde{#1}}}
\newcommand{\Mhat}[1]{\mat{\hat{#1}}}

\newcommand{\Tbar}[1]{\tnb{\bar{#1}}}
\newcommand{\Ttil}[1]{\tnb{\tilde{#1}}}
\newcommand{\That}[1]{\tnb{\hat{#1}}}

\DeclareMathOperator{\diag}{diag}

\DeclareMathOperator{\im}{im}

\DeclareMathOperator{\tr}{tr}

\DeclareMathOperator{\spn}{span}

\newcommand{\di}{\mathop{}\!\mathrm{d}}

\newcommand{\ee}{\mathchoice{\displaystyle\mathrm e}
{\textstyle\mathrm e}{\scriptstyle\mathrm e}
{\scriptscriptstyle\mathrm e}}

\newcommand{\EXP}[1]{\mathbb{E}\left(#1\right)}

\newcommand{\ip}[2]{\left\langle #1 , #2 \right\rangle}

\newcommand{\nd}[1]{\left\Vert #1 \right\Vert}

\newcommand{\trpos}{{\ops{T}}}

\newcommand{\BIGOP}[1]{\mathop{\mathchoice%
{\raise-0.22em\hbox{\huge $#1$}} {\raise-0.05em\hbox{\Large $#1$}}
{\hbox{\large $#1$}}{#1}}}
\newcommand{\BIGboxplus}{\mathop{\mathchoice%
{\raise-0.35em\hbox{\huge $\boxplus$}}%
{\raise-0.15em\hbox{\Large $\boxplus$}}{\hbox{\large
$\boxplus$}}{\boxplus}}}

\newcommand{\frkt}[2]{{{\raise0.6ex\hbox{{\leavevmode$\textstyle #1$}}}{\raise0.25ex\hbox{\kern-0.35ex\hbox{/}}}{\raise-0.3ex\hbox{\kern-0.4ex\hbox{{\leavevmode$\textstyle #2$}}}}}}
\newcommand{\frks}[2]{{{\raise0.7ex\hbox{{\leavevmode$\scriptstyle #1$}}}{\raise0.2ex\hbox{\kern-0.4ex\hbox{\footnotesize /}}}{\raise-0.2ex\hbox{\kern-0.4ex\hbox{{\leavevmode$\scriptstyle #2$}}}}}}
\newcommand{\frkx}[2]{{{\raise0.75ex\hbox{{\leavevmode$\scriptscriptstyle #1$}}}{\raise0.17ex\hbox{\kern-0.45ex\hbox{\scriptsize /}}}{\raise-0.15ex\hbox{\kern-0.4ex\hbox{{\leavevmode$\scriptscriptstyle #2$}}}}}}
\newcommand{\frkz}[2]{{{\raise0.75ex\hbox{{\leavevmode$\scriptscriptstyle #1$}}}{\raise0.17ex\hbox{\kern-0.45ex\hbox{\tiny /}}}{\raise-0.15ex\hbox{\kern-0.4ex\hbox{{\leavevmode$\scriptscriptstyle #2$}}}}}}

\newcommand{\frk}[2]{{\mathchoice{{\frkt{#1}{#2}}}{{\frks{#1}{#2}}}{{\frkx{#1}{#2}}}{{\frkz{#1}{#2}}}}}

\newcommand{\citep}[1]{\cite{#1}}

\newcommand{\thebib}{./misc}

\newcommand{\olsi}[1]{\,\overline{\!{#1}}} 

\definecolor{mygreen}{rgb}{0.1, 0.8, 0.4}
\newcommand{\rvsi}[1]{\textcolor{black}{#1}}
\definecolor{myblue}{rgb}{0.2, 0.4, 0.9}
\newcommand{\rvsn}[1]{\textcolor{black}{#1}}

\ignore{     

\usepackage[sfdefault=cmbr,OMLmathsans]{isomath}  

\newcommand{\ignore}[1]{}

\newcommand{\mat}[1]{\mathchoice{\displaystyle\mathbf#1}
{\textstyle\mathbf#1}{\scriptstyle\mathbf#1}
{\scriptscriptstyle\mathbf#1}}
\newcommand{\vek}[1]{\mathchoice{\displaystyle\boldsymbol#1}
{\textstyle\boldsymbol#1}{\scriptstyle\boldsymbol#1}
{\scriptscriptstyle\boldsymbol#1}}

\newcommand{\F}{\mathfrak} 
\newcommand{\C}{\mathcal}  
\newcommand{\mrm}{\mathrm}     

\newcommand{\di}{\mathop{}\!\mathrm{d}}

\newcommand{\ee}{\mathchoice{\displaystyle\mathrm e}
{\textstyle\mathrm e}{\scriptstyle\mathrm e}
{\scriptscriptstyle\mathrm e}}

\newcommand{\ip}[2]{\langle #1 , #2 \rangle}

\newcommand{\nd}[1]{\| #1 \|}

\DeclareMathOperator{\tr}{tr}
\DeclareMathOperator{\diag}{diag}

}    

\newcommand{\feq}[1]{Eq.(\ref{#1})} 
\newcommand{\feqs}[2]{Eqs.(\ref{#1}) and (\ref{#2})} 
\newcommand{\feeqs}[2]{Eqs.(\ref{#1})--(\ref{#2})} 
\newcommand{\feqss}[3]{Eqs.(\ref{#1}),(\ref{#2}) and (\ref{#3})} 
\newcommand{\fsec}[1]{Section~\ref{#1}}
\newcommand{\fsecs}[2]{Sections~\ref{#1} and \ref{#2}} 

\newcommand{\ffig}[1]{Fig.\ref{#1}} 
\newcommand{\ffigs}[2]{Figs.\ref{#1} and \ref{#2}} 
\newcommand{\ftbl}[1]{Table~\ref{#1}}

\newcommand{\meansymb}[1]{\olsi{#1}}
\newcommand{\whsymb}[1]{\widehat{#1}}	
	
\newcommand{\rsymb}[1]{\mathring{#1}}

\newcommand{\sqz}{\sqrt{2}}
\newcommand{\sptldom}{\C{G}}

\newcommand{\Rd}{\D{R}^d}
\newcommand{\Rn}{\D{R}^n}
\newcommand{\Rk}{\D{R}^k}
\newcommand{\Physpc}{\C{V}}
\newcommand{\Strspc}{\C{T}}
\newcommand{\SymPh}{\C{S}}
\newcommand{\SymTen}{\C{E}}

\newcommand{\smplspc}{\Omega}
\newcommand{\evt}{\omega}
\newcommand{\sigalg}{\F{F}}
\newcommand{\prob}{\D{P}}

\newcommand{\SO}{\mrm{SO}}
\newcommand{\so}{\F{so}}
\newcommand{\ST}{\mrm{ST}(\Strspc)}
\newcommand{\st}{\F{st}(\Strspc)}
\newcommand{\Dgp}{\mrm{Diag^+}}
\newcommand{\dg}{\F{diag}}
\newcommand{\Syp}{\mrm{Sym}^+}
\newcommand{\sy}{\F{sym}}
\newcommand{\sS}{\ops{S}_p}
\newcommand{\sSl}{\F{s}_p}
\newcommand{\rpal}{\F{r}}
\newcommand{\sygrp}{\F{S}}
\newcommand{\sycls}{\sygrp_{\etns}}
\newcommand{\syclsm}{\sygrp_{\etns,m}}
\newcommand{\syclS}{\sygrp_{\eTns}}
\newcommand{\syclSm}{\sygrp_{\eTns,m}}

\newcommand{\vrep}{\ops{vrep}}  
\newcommand{\Vrep}{\ops{Vrep}}  
\newcommand{\skwr}{\ops{skw}}  
\newcommand{\Trep}{\ops{Trep}}
\newcommand{\trep}{\ops{trep}}

\newcommand{\etns}{\mat{c}}
\newcommand{\eTns}{\tnb{C}}
\newcommand{\strnt}{\vek{\varepsilon}}
\newcommand{\strst}{\vek{\sigma}}
\newcommand{\strnv}{\tnb{e}}
\newcommand{\strsv}{\tnb{s}}

\newcommand{\vI}{\vek{I}}

\newcommand{\vP}{\vek{P}}
\newcommand{\vQ}{\vek{Q}}
\newcommand{\vR}{\vek{R}}

\newcommand{\vV}{\vek{V}}
\newcommand{\vW}{\vek{W}}
\newcommand{\vLbd}{\vek{\Lambda}}
\newcommand{\tD}{\tnb{D}}
\newcommand{\tH}{\tnb{H}}
\newcommand{\tI}{\tnb{I}}
\newcommand{\tM}{\tnb{M}}
\newcommand{\tP}{\tnb{P}}
\newcommand{\tQ}{\tnb{Q}}
\newcommand{\tR}{\tnb{R}}
\newcommand{\tS}{\tnb{S}}
\newcommand{\tU}{\tnb{U}}
\newcommand{\tV}{\tnb{V}}
\newcommand{\tW}{\tnb{W}}
\newcommand{\tX}{\tnb{X}}

\newcommand{\te}{\tnb{e}}

\newcommand{\tv}{\tnb{v}}

\newcommand{\tn}{\tnb{n}}

\newcommand{\tx}{\tnb{x}}
\newcommand{\ty}{\tnb{y}}
\newcommand{\tz}{\tnb{z}}
\newcommand{\tLbd}{\tnb{\Lambda}}
\newcommand{\ve}{\vek{e}}
\newcommand{\vg}{\vek{g}}
\newcommand{\vh}{\vek{h}}
\newcommand{\vk}{\vek{k}}
\newcommand{\vp}{\vek{p}}
\newcommand{\vq}{\vek{q}}
\newcommand{\vr}{\vek{r}}
\newcommand{\vu}{\vek{u}}
\newcommand{\vx}{\vek{x}}
\newcommand{\vy}{\vek{y}}
\newcommand{\vz}{\vek{z}}
\newcommand{\ry}{\tns{y}}

\newcommand{\aTns}{\tnb{A}}
\newcommand{\bTns}{\tnb{B}}

\ignore{     

\newcommand{\conducttensor}{\vek{\kappa}}

\newcommand{\samplespace}{\varOmega}
\newcommand{\event}{\omega}

\newcommand{\DFcond}{\conducttensor(x)}

\newcommand{\Expvalue}{\mathbb{E}}

}    

\ignore{     

\newcommand{\genSecTensor}{\vek{C}}

\newcommand{\symdiag}{\vek{Y}}

\newcommand{\Real}{\mathbb{R}}

\newcommand{\eigval}{\vek{\Lambda}}

}    

\newcommand{\KL}{Karhunen-Lo\`eve }

\newcommand{\cov}{\text{cov}}

\newenvironment{keyword}{%
\noindent \emph{Keywords:}}{\newline}

\begin{document}

\newcommand{\authorsks}{Sharana Kumar Shivanand}
\newcommand{\authorbr}{Bojana Rosi\'{c}}
\newcommand{\authorhgm}{Hermann~G. Matthies}
\newcommand{\theauthor}{\authorsks \hspace{2.0em} \authorbr \hspace{2.0em} \authorhgm}
\newcommand{\affiltur}{{\small Scientific Computing Center, Karlsruhe Institute of Technology, Germany}}
\newcommand{\affilwire}{{\small Institute of Scientific Computing, 
   Technische Universit\"at Braunschweig, Germany}}
\newcommand{\affiltwen}{{\small Applied Mechanics and Data Analysis, University of Twente, The Netherlands}}

\author[a]{\authorsks\thanks{Corresponding author. {E-mail:} \texttt{\textcolor{black}{sharana.shivanand@kit.edu}}}}
\author[b]{\authorbr}
\author[c]{\authorhgm}
\affil[a]{\affiltur}
\affil[b]{\affiltwen}
\affil[c]{\affilwire}



\newcommand{\thetitle}{\textcolor{black}{Stochastic Modelling of Elasticity Tensor Fields}}

\newcommand{\thesubject}{35B30, 37M99, 41A05, 41A45, 41A63, 60G20, 60G60, 65J99, 93A30}
\newcommand{\thekeywords}{stochastic material modelling, tensor-valued random variable, anisotropy, 
	elasticity symmetry classes of ensemble and mean, elasticity distance, Fr\'{e}chet mean, 
	Lie algebra representation, directional and scaling uncertainty, uncertainty quantification}

\newcommand{\textdate}{\today}
\date{\textdate}

\newcommand{\thereport}{2023-XX}


\title{\thetitle}



\maketitle

\begin{abstract}
    \textcolor{black}{We present a novel framework for the probabilistic modelling of random fourth order material tensor fields, with a focus on tensors that are physically symmetric and positive definite (SPD), of which the elasticity tensor is a prime example.  
    Given the critical role that spatial symmetries and invariances play in determining material behaviour,
    it is essential to incorporate these aspects into the probabilistic description and modelling of material properties.
    In particular, we focus on spatial point symmetries or invariances under rotations, a classical subject in elasticity.
    Following this, we formulate a stochastic modelling framework using a Lie algebra representation via a memoryless transformation that respects the requirements of positive definiteness and invariance. With this, it is shown how to generate a random ensemble of elasticity tensors that allows an independent control of strength, eigenstrain, and orientation. The procedure also accommodates the requirement to prescribe specific spatial symmetries and invariances for each member of the whole ensemble, while ensuring that the mean or expected value of the ensemble conforms to a potentially `higher' class of spatial invariance. 
    Furthermore, it is important to highlight that the set of SPD tensors forms a differentiable manifold, which geometrically corresponds to an open cone within the ambient space of symmetric tensors. Thus, we explore the mathematical structure of the underlying sample space of such tensors, and introduce a new distance measure or metric, called the \textit{`elasticity metric'}, between the tensors. Finally, we model and visualize a one-dimensional spatial field of orthotropic Kelvin matrices using interpolation based on the elasticity metric.}

\end{abstract}


\begin{keyword}
	\textcolor{black}{stochastic material modelling, random elasticity, tensor-valued random field, 
	spatial symmetries of ensemble and mean, Fr\'{e}chet mean, Lie algebra representation,
	directional and scaling uncertainty, uncertainty quantification}
\end{keyword}

%
%
%
%
%
%
%
%
%


%
\section{Introduction} \label{S:intro}

%

%
%
%
%
%
%
%

The computational modelling of heterogeneous materials often has to resort to a 
statistical or probabilistic description, as the precise details are unknown or uncertain.
The material properties are usually collected in tensor quantities, 
and these tensors are often of even order, and due to Onsager's reciprocity relations 
\citep{deGrootMazur1984} physically symmetric --- in elasticity this is termed the \emph{major}
symmetry --- and additionally they are frequently positive definite as they are associated
with e.g.\ stored energy or entropy production \citep{ting1996}.  This class of even-order physically
symmetric and positive definite (SPD) tensors --- and here especially the important class of 
4th-order tensors and their associated symmetry classes --- is the main focus in this contribution.
The stochastic modelling procedure we shall describe is applicable to any such even-order physically
symmetric and possibly additionally positive definite tensor, but we shall spell
things out in detail for 4th order SPD tensors, a prime example of which is
the elasticity tensor, which will be taken here to stand for all such tensors.

Well known examples of such SPD tensors are e.g.\ thermal conductivity ---
a 2nd order tensor mapping temperature gradients to thermal fluxes --- or, as
already mentioned above, the 
elasticity tensor of a linearly elastic material as
a fourth order tensor, mapping strains to stresses.  
A 6th order more involved example is the complete piezo-elasticity tensor,
mapping the combination of strains and electric displacement to stress and electric field.
Such a tensor of a coupled problem would be best tackled using the tensor-block
matrices of \citep{Nikabadze2016}, which would be a more or less straightforward extension
of the concepts treated here.

The representation and stochastic modelling of 2nd order tensors, such as those
describing diffusion phenomena, was shown in a precursor paper \citep{shivaEtal2021}.
Here we want to follow along the general lines of this approach based on the spectral decomposition
and symmetry-group reduction, but for higher-order tensors a novel situation arises, namely
the situation that the symmetry-group reduction \citep{Richards1963, faessStiefel1992, Jones1998} 
can not diagonalise the tensor.  We outline the general 
idea for SPD tensors of any even order but concentrate in detail on 4th-order tensors.

As the interest here is in the probabilistic modelling of such tensors, a simple way
to realise this is to think of the tensor entries as random variables.
But an important aspect of a material law described by such
a tensor is its invariance under spatial symmetries  \citep{newn2005, slaw2007, tinder2008,
Richards1963, faessStiefel1992, Jones1998}.
Although the numerical values of the tensor entries may be uncertain resp.\ random variables, 
it is not uncommon that the class of spatial symmetries is known exactly, both for 
each individual realisation as well as for the mean.
These invariances under spatial symmetries are a classical subject \citep{malyarenko_tensor-valued_2019},
and the different symmetry groups allow one to assign such material laws,
resp.\ the tensors describing them into certain classes, e.g.\ like the well-known
isotropic class, or the orthotropic class of materials.   
2nd order tensors in 3D show, next to these just mentioned invariance resp.\ symmetry 
classes of iso- or ortho-tropy, additionally only plan-isotropic behaviour 
(e.g.\ \citep{malyarenko_tensor-valued_2019, shivaEtal2021}).  
4th order spatial tensors, however,  have a richer set of symmetry classes than 2nd order tensors 
(e.g.\ \citep{cowin_identification_1987, bona_coordinate-free_2007, malyarenko_tensor-valued_2019}), 
namely eight symmetry or invariance classes in 3D \citep{forteVianel1996}, and four in 
2D \citep{blin1996}, cf.\ also the general references \citep{newn2005, slaw2007, tinder2008}.

Any even-order tensor may be viewed as a linear map when acting on the space of tensors of half 
the order by contracting over half the indices, e.g.\ \citep{Moakher2008, Nikabadze2009, Nikabadze2017}.  
In the case of the 4th order elasticity tensor, this space of tensors half the order 
is the space of strains, i.e.\ 2nd order symmetric tensors, as the elasticity
tensor maps a strain tensor into the corresponding stress tensor.
By choosing appropriate bases in the vector space of tensors of half the order \citep{Moakher2008}, 
this linear map can be represented by a matrix; so the class of tensors we are interested in can
thus be represented by symmetric positive definite (SPD) matrices.  We shall switch to
this matrix representation whenever it is convenient to do so.
The spatial invariance resp.\ symmetry groups are then realised as linear transformation
groups on the relevant space of half-order tensors, or, equivalently, on the space underlying 
the representing symmetric matrices just mentioned.  These transformation groups then define 
invariant subspaces of the SPD matrix representing the tensor 
\citep{Richards1963, faessStiefel1992, Jones1998}.
By transforming into a basis adapted to these invariant subspaces, the matrix takes on a
block-diagonal form.  This transformation to block-diagonal form is termed the 
symmetry resp.\ invariance group reduction or decomposition.  
Here we shall prefer the term `invariance' of the tensor resp.\ matrix under the spatial
group operations
over the term `symmetry', in order to avoid confusion with the physical symmetry due to 
Onsager's reciprocity relations, which results in symmetric matrices. 

Any SPD matrix or linear map can be spectrally decomposed 
with positive eigenvalues, and has orthogonal eigenvectors.  
This spectral eigenvalue / -vector (EV) decomposition separates the size, strength, 
or stiffness information contained in the eigenvalues
from the information about the orientation of the eigenvectors.  For the sake of brevity
and simplicity, in the following we shall speak about the eigenvalues in terms
of stiffness or elastic moduli as they apply to the elasticity or Hooke tensor,
but obviously this can be easily translated to any other physical property.
The group reduction together with the spectral 
decomposition are the main building blocks
of our approach.   In the representation proposed, and in this we see one of the
novelties of our approach, we want to keep the modelling of strength / stiffness,
eigenstrain distribution, and spatial orientation
separate and thus be able to control it independently.
So, finally, one ends up considering 
random SPD matrices, where certain invariances are known for the whole population,
as well as for the mean.
There is already extensive work on random matrices, drawing on several sources,
which is mentioned only very briefly here
(e.g.\ \citep{gupta_matrix_1999, schwartzman_random_nodate, soize_nonparametric_2000, GuilleminotSoizeGhanemR.2011}).
 
The separation of stiffness, eigenstrains, and spatial orientation information is
actually an old subject, as the 
characterisation of elasticity tensors through their eigenvalues resp.\ moduli
was already started by Stokes (cf.\ \citep{Love1944, TodhunterPearson1960, Nikabadze2016}) and 
Lord Kelvin (W.K.~Thomson) \citep{kelvin1856, kelvin1878}.
It was little appreciated at the time and then apparently almost forgotten, to be
picked up again about a century later by \citep{Fed1968}, and subsequently, among others, by
\citep{rych1984, Ostrosablin1984, Ostrosablin1986a, Ostrosablin1986,
mehrCow1990, sutcl1992, cowMehr1992, cowMehr1995, cowMehr1995a, ting1996,
rych2000, rych2001, bonaEtal2004, annin2008, KowalczykOstrowska2009, Nikabadze2016, 
Nikabadze2017a, Nikabadze2017, NordmannEtAl2018}.  
Kelvin originally used this to classify material symmetries; a historical account may be 
found in \citep{Helbig1996, helbig2013}.  It is an early example of a tensor representation
through an orthogonal tensor
decomposition, and many other such representations and decompositions have been considered;
a compilation and comparison of different orthogonal 
decompositions may be found in \citep{cigdem2010, DinckalAkgoez2010, Dinckal2012}.  
Furthermore, topics of tensor representation are also treated in 
\citep{oliveAuffray2013, auffrayEtal2014, malyarenko_random_2017, malyarenko_tensor-valued_2019},
and more information on tensor decompositions may also be found in e.g.\ 
\citep{Backus1970, Walpole1984, Baerheim1993, BaerheimHelbig1993, Baerheim1998a, 
ForteVianello1998, ZhengZou2000, Moakher2008, 
Desmorat2009, Moakher2009, ItinHehl2013, BrowaeysChevrot2004, OliveEtal2017, 
OliveKolevEtAl2018, DesmoratAuffraytEtAl2022}.   
\rvsn{The representation preferred by us, based on
the spectral decomposition referred to above, offers the possibility  of separately
modelling stiffness, orientation, and possibly the eigenstrain, as well as
separately controlling their individual probability distributions.  
Relative to existing models this offers a direct and fine
control of those properties, while being free of any limitations.  
It additionally offers direct control of the symmetry resp.\ invariance properties
to be discussed below, both of the ensemble mean and of each individual realisation.
This feature distinguishes our approach from existing models.  This and further distinctions are
discussed in more detail in \citep{shivaEtal2021}, and will not be repeated here}.

It is well known that there are eight symmetry or invariance classes 
for symmetric 4th order tensors in 3D, they may be found in the 
references just quoted, 
as well as in the  reference works \citep{newn2005, slaw2007, tinder2008}.
To sketch the connection of the spectral or EV-decomposition with the idea of invariance under
symmetry groups, it was already mentioned that one considers the representation of the 
elasticity tensor as a linear map through a proper choice of a basis in the space of strains.  
\rvsn{In the case of the elasticity tensor this is known as the Kelvin notation.}
\rvsn{The eigenvalues of the Kelvin matrix are traditionally
called \emph{Kelvin moduli} and the corresponding eigenvectors \emph{Kelvin modes}.}

\rvsn{
The invariance class of the material or elasticity tensor
is identified by rotating the coordinate system of the physical space
in such a way as to introduce 
as many zeros as possible in the tensor.  
This pattern then turns out to be stable 
under a whole subgroup of rotations, which is the point invariance resp.\ symmetry group of the
material resp.\ the tensor or the Kelvin matrix. 
This process of \emph{group reduction} is an algorithm to find orthogonal invariant subspaces
which reduce the matrix to block-diagonal form (e.g.\ \citep{Richards1963, faessStiefel1992}),
thereby also identifying the invariance resp.\ symmetry group.
For 4th order tensors resp.\ Kelvin matrices this group is represented
as a subgroup of the  group of orthogonal transformations on strain-space,
cf.\ e.g.\
\citep{silvestrov_spectral_2016, malyarenko_random_2017, altenbach_tensor_2018, 
malyarenko_tensor-valued_2019}.  It also determines the eigenstrain distribution 
in the relevant elasticity class
\citep{nye_physical_1984, cowin_identification_1987, cowMehr1995, 
malgrange_symmetry_2014, bona_coordinate-free_2007}.
}

\rvsn{
The group-reduced block-diagonal form alluded to above is the representation 
usually chosen in the literature when the different elasticity classes are discussed, e.g.\
\citep{cowin_identification_1987, bona_coordinate-free_2007, 
malyarenko_tensor-valued_2019, newn2005, slaw2007, tinder2008}.
The block-diagonal form of the matrix can be  further reduced to diagonal
form in each block with a orthogonal transformation to a 
direct sum of specific eigenstrain subspaces.
This last orthogonal transformation in strain-space does not result from a
spatial rotation, but is purely in strain-space.
This  consideration of iso-spectral eigenstrain distributions is
a crucial extension to higher-order tensors of the analogous development 
in \citep{shivaEtal2021} for 2nd order tensors.
}

The 4th order elasticity tensor may thus be represented in a two-stage procedure
by a combined group reduction and spectral decomposition, 
i.e.\ one may represent the elasticity tensor in form of the
Kelvin matrix by a positive diagonal matrix of eigenvalues resp.\ Kelvin moduli, an
`eigenstrain distributor' rotation within an invariant subspace of the
block-diagonal form, and the spatially induced
rotation for the symmetry group reduction.  The task of representing random SPD tensor
fields thus is reduced to representing these three components.
It is worthwhile, and in our context essential, to observe that these three ingredients 
--- spatially induced orthogonal transformations for the group reduction, general
orthogonal transformation on strain-space to diagonalise each block, and diagonal SPD
matrices of Kelvin moduli --- are each members of matrix Lie groups.  

In the modelling of random tensor fields, the most comfortable and desirable way is 
to find an unconstrained representation on a linear vector space.
For the purpose of modelling or representing a random SPD tensor, a geometric point of 
view might be helpful.  In the vector space of physically symmetric tensors of a given 
even order, the positive definite ones (SPD tensors) form an open convex cone,
i.e.\ sums of SPD tensors with positive coefficients are again SPD tensors.
They thus form a differentiable manifold, but not a vector space.
So the desire is to represent this cone on a vector space without any constraints.
One such possibility is the matrix exponential and logarithm (e.g.\  
\citep{JungSchwartzmanGroisser2015, GroisserJungSchwartzman2017,
groissJungSchwman2017, shivaEtal2021}),
and this can be used in different guises, and will also be used here.  
The Lie groups just alluded to  --- diagonal 
SPD matrices and subgroups of orthogonal matrices --- which we want to use in the
representation, as the they allow one to independently model different aspects
of the SPD tensor, are geometrically speaking more complicated than the simple vector space
of physically symmetric tensors.  But the exponential and logarithm may be used in this
situation as well --- here coinciding with the Lie group exponential and logarithm --- and
furnish a representation on the associated Lie algebra of the corresponding group, which 
is geometrically the tangent space at the group identity; effectively giving a representation 
on a direct sum of linear spaces.

In this context it is worthwhile to note that the nature of the set of SPD tensors --- a differentiable
manifold which is geometrically an open cone in the ambient space of symmetric tensors --- suggests
that the calculation of averages, means, or expected values inherited from this ambient space
based on a Euclidean metric is not the natural one to use.  It has actually been observed that 
using the mean based on the Euclidean metric has some undesirable properties 
\citep{PennecFillardAyache2004, AndoLiMathias2004, Moakher2005, ArsignyFillardPennecEtAl2006, 
ArsignyFillardPennecEtAl2007, drydenEtal2009, DrydenEtal2010, WangSalehianEtAl2014, FujiiSeo2015,
FeragenFuster2017}, in particular 
the so-called swelling, fattening, and shrinking effects 
\citep{schwartzman_random_nodate, JungSchwartzmanGroisser2015, 
schwartzman_lognormal_2016, GroisserJungSchwartzman2017, groissJungSchwman2017, FeragenFuster2017}.  

To wit, when finding the midpoint (average) or interpolating between two points on 
earth's surface, one will usually not mean the midpoint of a straight connecting line
(the average of the ambient Euclidean space), but the intrinsic midpoint half way on a great circle.
This intrinsic kind of average or mean using a non-Euclidean metric is called a \emph{Fr\'echet} mean,
and will be shortly recapped in Appendix~\ref{S:appx-metric}, recalling the connection between
metric and averaging.  From a metric, one may define variationally the Fréchet or Karcher 
mean --- a generalisation of the arithmetic resp.\ Euclidean distance based mean, 
see Appendix~\ref{S:appx-metric} --- to more general metric spaces \citep{NielsenBhatia2013}.
This is an area of active research \citep{FeragenFuster2017, ThanwerdasPennec2019,
ThanwerdasPennec2019-2, ThanwerdasPennec2021, JungRooksEtAl2023, groissJungSchwman2023}, 
and the `best' metric and mean seem to be application dependent.

The effect of using a mean which is not adapted to the manifold can have various consequences.
When modelling a non-isotropic material, this may mean a loss of anisotropy for the mean,
which may be undesirable in some situations.  In the above literature --- particularly that
coming from the medical field of diffusion tensor imaging, where these undesirable
effects are not acceptable, but also in the estimation of covariance matrices --- there 
is discussion on what metric to use \citep{AndoLiMathias2004, Moakher2005, 
ArsignyFillardPennecEtAl2006, Pennec2006, Moakher2009, DrydenEtal2010, Hauberg2015, FeragenFuster2017, 
GroisserJungSchwartzman2017, SommerFletcherPennec2020, PennecSommerFletcher2020, GuiguiEtal2023,
ThanwerdasPennec2023, JungRooksEtAl2023, groissJungSchwman2023}.
For elasticity tensors, this topic of a metric  has also attracted some 
attention \citep{cowin_identification_1987, Moakher2006, MoakherNorris2006, 
BucataruSlawinski2008, FujiiSeo2015, WeberGluegeBertram2019, MorinEtal2020, 
StahnMuellerBertram2020, shivaEtal2021}, 
e.g.\ in finding the closest elasticity tensor of some invariance class to measurement data.

We formulate some desiderata for a mean resp.\ a metric for the case of elasticity.
This will be connected to the proposed representation on a direct sum of Lie algebras, which
allows one to choose inner products on the Lie algebras to define a metric, which is transported
via the exponential map to the Lie group and hence gives rise to a Riemannian structure 
on the group, which can then be used to define
a metric --- which we suggest to call the \emph{`elasticity metric'} --- on the cone of
positive definite tensors which satisfies the proposed desiderata; 
the details of this are to be found in the Appendix~\ref{S:appx-metric}.
Based on this distance one may then formulate the corresponding Fréchet mean,
which may be called the  \emph{`elasticity mean'}.

To generate random SPD matrices, a reduced non-parametric approach taking account 
of positive definiteness was presented in \citep{soize_nonparametric_2000, soize_2006}.  
Here the algebraic property that positive elements in the algebra are squares of other 
elements in the algebra \citep{segalKunze78, yosida-fa-1980} is used to ascertain positive 
definiteness.  This then forms the basis for the generation of
random elasticity tensors with specified invariance in the mean and fully anisotropic invariance
in each realisation, resp.\ controlled elasticity class invariance with constant spatial 
orientation of the symmetry axes in each realisation, which is shown in a series of papers
\citep{GuilleminotSoize2011, GuilleminotSoize2012,  GuilleminotSoize2012a, GuilleminotSoize2013, 
GuilleminotSoize2013a, GuilleminotSoize2014, NouySoize2014, StaberGuilleminot2017, Soize2021a}.  

In the area of modelling of random tensor fields with given invariance properties,
the culmination of a long series of works \citep{ostoja-starzewski_microstructural_2007,
malyarenko_statistically_2014, silvestrov_spectral_2016, malyarenko_random_2017, 
altenbach_tensor_2018} is reported in \citep{malyarenko_tensor-valued_2019},
specifically looking at stochastically homogeneous and isotropic random tensor fields 
of any order in a three dimensional domain.  The spatial modelling uses the spectral theorem for
the covariance function, to generate a spectral or \emph{Karhunen-Loève} like representation 
(e.g.\ \citep{hgm07}) of the random tensor field with the desired invariance properties.
No particular attention is paid to positive definiteness, as the methods used
are purely vector space like, such as linear combinations or series and integrals.
The danger is here that
although the limiting expression may well represent a positive definite tensor,
in a practical computational situations these series and integrals have to be
truncated resp.\ numerically approximated, and this step may lead to an unacceptable 
numerical loss of positive definiteness \rvsi{everywhere}. 
\rvsi{In some associated numerical examples of anti-plane shear \citep{zhang-malyarenko_fractal_2022},
positive definiteness everywhere is obtained by using a deterministic SPD base tensor and adding
a random rank-one tensor --- a dyadic or tensor product of identical vectors and thus
automatically positive --- and controlling the random coefficient so that the sum stays SPD.}
Here we shall also use a \KL like 
field model to represent the parameters in the Lie algebras mentioned above, 
but by modelling the logarithm of the tensor one can ensure positive 
definiteness \rvsi{everywhere} via the exponential map for each realisation at each point in space.

\rvsi{Ensuring positive definiteness everywhere through the use of the exponential function, 
resp.\ the logarithm in the opposite direction, is an often used approach; 
it typically based on the spectral or EV-decomposition.  As the spectral decomposition
may be written as a sum of dyadic or tensor products of eigenvectors, this furnishes
a connection to the above mentioned approach in \citep{zhang-malyarenko_fractal_2022}.
The exponential function} is employed in e.g.\ \citep{GuilleminotSoize2013a, NouySoize2014, 
JungSchwartzmanGroisser2015, schwartzman_lognormal_2016,
GroisserJungSchwartzman2017, groissJungSchwman2017,  shivaEtal2021, JungRooksEtAl2023}, 
and it may be combined with the squaring
approach \citep{GuilleminotSoize2013a, NouySoize2014, Grigoriu2016} to ensure compliance with
a particular invariance class.  This log-modelling is quite common
when measuring diffusion and conductivity tensors, i.e.\ focusing the description
on the logarithm of the tensor.  This takes advantage of the fact that the
invariance classes are the same for a tensor and its exponential resp.\ logarithm. 
Also in this work the exponential function is used in an essential way
to ensure positive definiteness numerically in each realisation.
The exponential function may be even used to model the rotations, essentially parametrising
the angles \citep{Grigoriu2016}.  Here we follow \citep{shivaEtal2021} and consequently
model the eigenvalues and the rotations via the exponential function, mapping from
a Lie algebra to the corresponding Lie group

\textcolor{black}{The structure of the paper is outlined as follows: \fsec{S:prelim} collects 
essential preliminaries, 
including the notation for even-order tensors in \fsec{SS:C-gen-notation} and Kelvin's matrix representation of Hooke's law in \fsec{SS:hooke-matrix}, with further details provided in Appendix~\ref{S:appx-Notation}. 
\fsec{SS:group-inv} discusses the transformation of the Kelvin matrix under rotations and briefly addresses how spatial rotations are modelled in the sequel, 
and how they influence the strain-space. Additional background on rotations can be found in 
Appendix~\ref{S:appx-Rotations}. 
This section also introduces the reduction of the Kelvin matrix to block diagonal form, its spectral decomposition, and the formulation of group invariance in elasticity. In order to make this work more easily accessible, the group theoretic material
will be kept to a minimum.
The representation and modelling of fourth-order SPD tensor fields, based on fields of parameters, is detailed in \fsec{SS:rand-fields}.}

\textcolor{black}{The main part of the paper is \fsec{S:modelling}, where \fsec{SS:Lie-Kelvin-represent} covers the Lie representation of the Kelvin matrix, and \fsec{SS:Frechet-means} introduces the concept of a Fr\'echet mean for SPD tensors. This uses the proposed `elasticity metric'
which is described in Appendix~\ref{S:appx-metric}.
In \fsec{SS:rep-requirem}, we formulate our requirements for a representation
of SPD tensors, as well as some general thoughts and desiderata on how to average
SPD tensors based on Fr\'echet means.
Next, in \fsec{SS:random-Kelvin}, we introduce the framework for generating random ensembles of Kelvin matrices at a spatial point, followed by a description of spatial variation. The focus here is on modelling randomness through the separation of stiffness, eigenstrains, and spatial orientation.
\fsec{S:sepc_decomp}
explores Kelvin’s spectral decomposition of Hooke’s law in detail, with the relationship to 2D and 3D elasticity classes covered in
\fsec{SS:el_cl_2D} and \fsec{SS:el_cl_3D}, respectively. Finally, the paper concludes in \fsec{S:concl}}

\ignore{  

}   

%
%
%
%
%
%
%
%
%
%
%
%
%
%


%
\section{Preliminaries} \label{S:prelim}

%

%
%
%
%

%

We consider physical space as given by
$\Physpc \equiv \Rd$, the physically interesting instances being the 2D situation
with $d=2$, and the full 3D case with $d=3$. For the sake of simplicity we shall 
consider only Cartesian coordinate systems and orthonormal bases,
so we do not distinguish co- and contra-variant indices.

\subsection{Even order tensors as linear maps}  \label{SS:C-gen-notation}
Vectors in $\Physpc$ like the position vector are denoted as $\vx = (x_i) \in \Physpc$, 
whereas 2nd order tensors (matrices) are denoted as $\vQ = (Q_{ik})\in\Physpc\otimes\Physpc = 
\Physpc^{\otimes 2}$, which will be identified with linear maps $\E{L}(\Physpc)$ on $\Physpc$
via the matrix-vector product by $\vy = (y_i)=  \vQ\vx = (Q_{ik} x_k)$
(Einstein's summation convention used).

Tensors of higher order than two will be denoted as $\mat{w} = \mrm{w}_{i_1 i_2 i_3 \dots i_q} = 
\mrm{w}_{\F{i}}\in \Physpc^{\otimes q}$ for a tensor of order $q$.  Considering a tensor 
of even order, say $2q$: $\mat{a} = (\mrm{a}_{i_1 i_2 i_3 \dots i_q j_1 j_2 j_3 \dots j_q}) = 
(\mrm{a}_{\F{i} \F{j}}) \in \Physpc^{\otimes 2q}$ --- where for ease of notation we have
used a multi-index $\F{i} = (i_1 i_2 i_3 \dots i_q) \in \D{N}^q$ ---
it can be naturally be viewed as a linear map $\mat{A}\in \E{L}(\Physpc^{\otimes q})$ 
on the space of tensors of half the order, by contracting the tensor product
of $\mat{a}$ and $\mat{w}$ over the last $q$ indices, thereby implicitly using the canonical
Euclidean inner product to be introduced below:
\begin{equation}  \label{eq:tns-as-linmap}  \mat{A}:\;
 \Physpc^{\otimes q} \ni \mat{w} \mapsto 
      (\sum_{j_1 \dots j_q} \mrm{a}_{i_1 \dots i_q j_1 \dots j_q}
     \mrm{w}_{j_1 \dots j_q}) = \sum_{\F{j}} (\mrm{a}_{\F{i} \F{j}} \mrm{w}_{\F{j}}) 
 = \mat{A} \mat{w} \in  \Physpc^{\otimes q},
\end{equation}
where the last sum is particularly suggestive, as it is a matrix-vector multiplication with
multi-indices.  By choosing bases in $\Physpc^{\otimes 2q}$, or, equivalently,
by ordering the multi-indices in some way,
one may easily construct a matrix representation for the map $\mat{A}$.  For the elasticity
tensor this will be done in \fsec{SS:hooke-matrix} by using the so-called \emph{Kelvin notation}. 

\paragraph{Inner product and symmetry:}
The canonical Euclidean inner product for $\vx, \vy \in \Physpc \equiv \Rd$ is 
denoted by $\vx \cdot \vy = \vx^\trpos \vy = \ip{\vx}{\vy} := x_i y_i$.
This induces an inner product for general tensors $\mat{v}, \mat{w} \in \Physpc^{\otimes q}$,
namely $\ip{\mat{v}}{\mat{w}} := (\mrm{v}_{j_1 \dots j_q}\mrm{w}_{j_1 \dots j_q})$.  
If the map $\mat{A}\in \E{L}(\Physpc^{\otimes q})$ in \feq{eq:tns-as-linmap}
induced by a tensor of twice the order is symmetric w.r.t.\ this inner product, i.e.\ if 
\begin{equation}  \label{eq:map-sym}
  \forall\; \mat{v}, \mat{w}: \quad \ip{\mat{A}\mat{v}}{\mat{w}} = \ip{\mat{v}}{\mat{A}\mat{w}},
\end{equation}
then the tensor $\mat{a}$ is \emph{physically symmetric}
(here $(\mrm{a}_{\F{i}\F{j}}) = (\mrm{a}_{\F{j}\F{i}})$).  And if in addition
\begin{equation}  \label{eq:map-SPD}
  \forall\; \mat{w} \neq \mat{0}: \quad \ip{\mat{A}\mat{w}}{\mat{w}} > 0,
\end{equation}
the map $\mat{A}$ and the tensor $\mat{a}$ are \emph{symmetric positive definite} (SPD).

It is not difficult to see that SPD maps resp.\ matrices form an open convex cone, denoted by 
$\Syp(\Physpc^{\otimes q})$, in the subspace $\E{L}_s(\Physpc^{\otimes q})$  of symmetric 
linear maps resp.\ matrices.
This is actually a differentiable manifold (e.g.\ \citep{Lang1995, Moakher2005, Moakher2006}),
and its tangent space at the identity is the ambient space of symmetric maps,
in this context sometimes denoted by
$\sy(\Physpc^{\otimes q}) := \E{L}_s(\Physpc^{\otimes q}) \subset \E{L}(\Physpc^{\otimes q})$.
If the underlying vector space is just some copy of $\Rn$, one often writes
$\Syp(n)$ and $\sy(n)$ instead of $\Syp(\Rn)$ and $\sy(\Rn)$.
The class of even-order SPD tensors is the one of interest here, and the elasticity tensor 
to be introduced below is in this sense symmetric, and also usually positive definite, i.e.\ SPD,
see also \citep{Moakher2008}.

\paragraph{Behaviour under rotation and invariance:}
A rotation $\vQ\in \mrm{O}(\Physpc)$ ($\mrm{O}$ denotes the orthogonal group, 
see Appendix~\ref{S:appx-Rotations}) (of the coordinate 
system) in $\Physpc$, $\vhat{x}=\vQ\vx$ induces a corresponding
transformation of a higher-order tensor $\mat{w} \in  \Physpc^{\otimes q}$,
this is sometimes termed the Rayleigh product \citep{NordmannEtAl2018}:
\begin{equation}  \label{eq:tens-Q-transform-q}
  \Mhat{w} = (\hat{\mrm{w}}_{\ell_1\dots\ell_q}) = \vQ \star \mat{w} := 
  (Q_{\ell_1 i_1}\dots Q_{\ell_q i_q} \mrm{w}_{i_1 \dots i_q})  \in \Physpc^{\otimes q} .
\end{equation}

In case an even-order tensor $\mat{a}$ does not change under a rotation, i.e.\ in case
\begin{equation}  \label{eq:tens-inv-rot}
  \Mhat{a} = \vQ \star \mat{a} = \mat{a},
\end{equation}
one says that it is \emph{invariant} under the rotation $\vQ$, or in a more fancy way, invariant
under the subgroup $\sygrp_{\vQ}  \subset \mrm{O}(\Physpc)$ generated by 
$\{ \vI, -\vI, \vQ\}$.  The maximal subgroup $\sygrp_{\mat{a}}  \subseteq \mrm{O}(\Physpc)$ 
under which such a tensor $\mat{a}$ is invariant
is termed the \emph{symmetry class} of the tensor $\mat{a}$.  It is obvious that any
even-order tensor is invariant under the inversion $-\vI$, i.e.\ the subgroup 
$\{\vI, -\vI\}$, so that always $\{\vI, -\vI\} \subseteq \sygrp_{\mat{a}}$.

\subsection{Hooke's law in Kelvin's matrix notation}  \label{SS:hooke-matrix}
Turning now to the case of linear elasticity, the 4th order elasticity tensor $\etns$ acts 
on the symmetric infinitesimal strain $\strnt$ to yield the Cauchy stress tensor 
\begin{equation} \label{eq:Hooke-1-p}
\strst = \etns:\strnt;
\end{equation} 
see also Appendix~\ref{SS:C-Kelvin} for details on the notation. 
The \feq{eq:Hooke-1-p} is Hooke's generalised law.  
Both stress and strain tensors lie in the space 
$\SymPh := \E{L}_s(\Physpc) = \sy(\Physpc)\subset \Physpc^{\otimes 2}$ 
of symmetric 2nd order tensors, i.e.\ 
$\strnt = \strnt^\trpos$, and $\strst = \strst^\trpos$; whereas the 
fourth order elasticity tensor 
$\etns \in \SymTen := \E{L}_s(\SymPh)\subset \Physpc^{\otimes 4}$ may be
seen as an SPD map on $\SymPh$.  

Choosing a particular basis in $\SymPh$, one may write Hooke's law in matrix-vector
notation, where the coefficient array of that basis is in the
representation space $\Strspc = \Rk$, with $k=\dim\SymPh$, thereby establishing a linear one-to-one
correspondence $\vrep: \SymPh \to \Strspc$ (see Appendix~\ref{SS:C-Kelvin}), which in 3D reads as
\begin{align}  \label{eq:Kv-vrep3-strs-1}
  \strsv &:= \vrep(\strst) := [\sigma_{11}, \sigma_{22}, \sigma_{33}, 
     \sqz\,\sigma_{23}, \sqz\,\sigma_{13}, \sqz\,\sigma_{12}]^\trpos \in  \Strspc \\
   \label{eq:Kv-vrep3-strn-1}
  \strnv &:= \vrep(\strnt) := [\vepsilon_{11}, \vepsilon_{22}, \vepsilon_{33}, 
     \sqz\,\vepsilon_{23}, \sqz\,\vepsilon_{13}, \sqz\,\vepsilon_{12}]^\trpos \in \Strspc .
\end{align}
This notation goes back to Kelvin \citep{kelvin1856, kelvin1878}, see also \citep{Helbig1996, helbig2013, mehrCow1990, Moakher2008}.  In 2D one has
\begin{align}  \label{eq:Kv-vrep2-strs-1}
  \strsv &:= \vrep(\strst) := [\sigma_{11}, \sigma_{22}, \sqz\,\sigma_{12}]^\trpos \in  \Strspc \\
   \label{eq:Kv-vrep2-strn-1}
  \strnv &:= \vrep(\strnt) := [\vepsilon_{11}, \vepsilon_{22}, \sqz\,\vepsilon_{12}]^\trpos \in\Strspc .
\end{align}
By a slight abuse of notation, the map $\vrep: \SymPh \to \Strspc$ is again denoted the
same as in 3D.

The linear relation in Hooke's law \feq{eq:Hooke-1-p}  in $\Strspc$ becomes
\begin{equation}  \label{eq:Hooke-vec-K}
  \strsv = \eTns \strnv ,
\end{equation}
with an SPD Kelvin matrix $\eTns \in \E{L}_s(\Strspc)$, thereby establishing a linear map
$\Vrep: \SymTen \to \E{L}_s(\Strspc)$, $\eTns = \Vrep(\etns)$.  In terms of the components 
of $\etns = (c_{i_1 i_2 i_3 i_4})$, in 3D the Kelvin matrix $\eTns$ and thus the map $\Vrep$
is given by
\begin{equation}  \label{eq:Kelvin-C-3D-1}
  \eTns  = \Vrep(\etns):= {\small \begin{bmatrix} 
   \mrm{c}_{1111} & \mrm{c}_{1122}& \mrm{c}_{1133} & 
   \sqz\,\mrm{c}_{1123} & \sqz\,\mrm{c}_{1113} & \sqz\,\mrm{c}_{1112} \\  
                  & \mrm{c}_{2222}& \mrm{c}_{2233} & 
   \sqz\,\mrm{c}_{2223} & \sqz\,\mrm{c}_{2213} & \sqz\,\mrm{c}_{2212} \\  
     &    & \mrm{c}_{3333} &  \sqz\,\mrm{c}_{3323} &  \sqz\,\mrm{c}_{3313} & \sqz\,\mrm{c}_{3312} \\  
     &    &   & \;\;\;2\,\mrm{c}_{2323} & \;\;\;2\,\mrm{c}_{2313} & \;\;\;2\,\mrm{c}_{2312} \\  
       & \mrm{SYM}  &   &  & \;\;\;2\,\mrm{c}_{1313} & \;\;\;2\,\mrm{c}_{1312} \\  
        & & & & & \;\;\;2\,\mrm{c}_{1212}  
   \end{bmatrix} } ,
\end{equation}
while in 2D one obtains
\begin{equation}  \label{eq:Kelvin-C-2D-1}
  \eTns  = \Vrep(\etns) := {\small \begin{bmatrix} 
   \mrm{c}_{1111} & \mrm{c}_{1122} &  \sqz\,\mrm{c}_{1112} \\  
             & \mrm{c}_{2222} &  \sqz\,\mrm{c}_{2212} \\  
   \mrm{SYM} &  &  \;\;\;2\,\mrm{c}_{1212}  
   \end{bmatrix} } .
\end{equation}

\subsection{Group reduction and local spectral decomposition} \label{SS:group-inv}
The elasticity class  \citep{nye_physical_1984, 
cowin_identification_1987, cowMehr1995, malgrange_symmetry_2014, bona_coordinate-free_2007}
of the material or the elasticity tensor $\etns \in \SymTen$ 
\citep{altenbach_tensor_2018, malyarenko_tensor-valued_2019} is determined
by rotating the coordinate system $\vhat{x} = \vQ \vx$, $\vQ\in \mrm{O}(\Physpc)$ 
in such a way as to reduce the number of constants in the 4th order elasticity 
tensor $\hat{\etns} = \vQ \star \etns \in\SymTen$ by introducing as many zeros 
as possible.  The tensor in this 
form then turns out to be invariant under a whole subgroup of rotations $\sycls 
\subseteq \mrm{O}(\Physpc)$, which is called the symmetry or invariance group of the
material resp.\ elasticity tensor.  For simplicity and without loss of generality, 
and as more is not needed in our modelling
approach, we restrict ourselves here to \emph{proper} rotations, i.e.\ the subgroup
$\SO(\Physpc) \subset \mrm{O}(\Physpc)$ of special orthogonal matrices, i.e.\ orthogonal
matrices $\vQ$ with $\det \vQ = 1$.  The symmetry or invariance group of a tensor will
therefore in the following also be only considered as a subgroup of $\SO(\Physpc)$.

\paragraph{Transformations of the Kelvin matrix:}
As just mentioned, a rotation $\vQ\in \SO(\Physpc)$ of the coordinate system in $\Physpc$ induces a 
transformation of the stress and strain tensors  
\citep{mehrCow1990, MehrabadiCowinJaric1995} and on the 4th order elasticity tensor 
$\hat{\etns} = \vQ \star \etns$, as described for the general case in \feq{eq:tens-Q-transform-q}.

In this context it is important to note that the linear map $\vrep: \SymPh \to \Strspc$
defined in the preceding \fsec{SS:hooke-matrix} in \feqs{eq:Kv-vrep3-strs-1}{eq:Kv-vrep3-strn-1}
in 3D, and in \feqs{eq:Kv-vrep2-strs-1}{eq:Kv-vrep2-strn-1} in 2D, is actually unitary, i.e.\
it preserves inner products; in fact this is here the advantage of the Kelvin notation over the
better known Voigt notation.  This also means that for a rotation $\vQ\in \SO(\Physpc)$ in
physical space, which one may recall preserves the inner product in $\Physpc$, 
there is a one-to-one correspondence $\Trep: \SO(\Physpc) \to \SO(\Strspc)$ 
\citep{mehrCow1990, MehrabadiCowinJaric1995} to a rotation 
$\tQ = \Trep(\vQ) \in \SO(\Strspc)$, which in turn preserves the inner product in $\Strspc$.
Concretely this means that when 
$\vhat{\sigma} = \vQ \star \strst = \vQ \strst \vQ^\trpos$ and   
$\vhat{\vepsilon} = \vQ \star \strnt = \vQ \strnt \vQ^\trpos$, and
\begin{equation}  \label{eq:unitarity-vrep-1}
\strsv = \vrep(\strst), \, \strnv = \vrep(\strnt), \, \That{s} = \vrep(\vhat{\sigma}), \,
   \That{e} = \vrep(\vhat{\vepsilon}), \;\text{ then }\;
   \That{e} = \tQ \strnv, \That{s} = \tQ \strsv,
\end{equation} 
as well as
\begin{equation}  \label{eq:unitarity-vrep-2}
\strst : \strnt = \vhat{\sigma} : \vhat{\vepsilon} = \strsv^\trpos \strnv = \That{s}^\trpos \That{e};
   \;\text{ and }\; \strsv^\trpos \strsv  = \That{s}^\trpos \That{s},\;
   \strnv^\trpos \strnv  = \That{e}^\trpos \That{e}. 
\end{equation}

It is not too difficult to see that $\Trep$ is actually an injective group homomorphism.
Thus the map $\Trep: \SO(\Physpc) \to  \SO(\Strspc)$ is a unitary representation of 
$\SO(\Physpc)$ on $\Strspc$, see e.g.\ \citep{faessStiefel1992, Jones1998, KlapperHahn2006} ---
and the precise definition of $\Trep$ may be found in Appendix~\ref{SS:A-rot-Strspc}.  Hence the 
image of $\SO(\Physpc)$ under $\Trep$, which will be denoted by 
$\ST := \im \Trep = \Trep(\SO(\Physpc)) \subseteq \SO(\Strspc)$, is a Lie subgroup
of $\SO(\Strspc)$.  

As is well known, e.g.\ see Appendix~\ref{SS:A-rot-Strspc}, the Lie algebra of a Lie group
is the tangent vector space at the identity, and it has usually the same notation
as the Lie group, but with small Fraktur letters; i.e.\ $\so(\Physpc)$ is the
Lie algebra of $\SO(\Physpc)$.  The Lie algebra of $\ST$ is denoted by $\st$.
The dimension of $\ST = \Trep(\SO(\Physpc))$, which is the dimension of its Lie 
algebra $\st$, is due to the injectivity of $\Trep$  that of $\SO(\Physpc)$ 
resp.\ $\so(\Physpc)$, i.e.\ $\dim(\so(\Physpc)) = \dim(\st)$.
The injective group homomorphism $\Trep: \SO(\Physpc) \to  \ST \subset \SO(\Strspc)$ 
induces an injective linear map on the corresponding
Lie algebras $\trep: \so(\Physpc) \to \so(\Strspc)$, such that $\st = \trep(\so(\Physpc))$, 
details of which may also be found in Appendix~\ref{SS:A-rot-Strspc}.

This implies that for a coordinate transform $\vQ\in \SO(\Physpc)$, which transforms the elasticity
tensor  $\etns \in \SymTen$ according to \feq{eq:tens-Q-transform-q} as 
$\Mhat{c} = \vQ \star \etns$, and whose induced transformation on $\Strspc$ is 
$\tQ = \Trep(\vQ)\in\ST$, the effect on the Kelvin matrix 
$\eTns = \Vrep(\etns) \in \E{L}_s(\Strspc)$  is given by
\citep{mehrCow1990, cowMehr1992, cowMehr1995, cowMehr1995a}
\begin{equation}  \label{eq:trans-Kv-mtrx}
\That{C} = \tQ \eTns \tQ^\trpos = \Vrep(\vQ \star \etns) = \Vrep(\Mhat{c}) .
\end{equation}

\paragraph{Group reduction:}
After this preparation one may address the group reduction,
see e.g.\ \citep{faessStiefel1992, Jones1998} and  \citep{nye_physical_1984, 
cowin_identification_1987, cowMehr1995, malgrange_symmetry_2014, bona_coordinate-free_2007,
altenbach_tensor_2018, malyarenko_tensor-valued_2019}.  
As already indicated, one seeks a rotation $\vQ\in \SO(\Physpc)$ 
such that $\mat{\rsymb{c}} = \vQ \star \etns$ has as many zeros as possible.
For the corresponding Kelvin matrix $\rsymb{\eTns} = \Vrep(\mat{\rsymb{c}})
\in \E{L}_s(\Strspc)$ this means (cf.\ \feq{eq:trans-Kv-mtrx}) that
\begin{equation}  \label{eq:group-dec}
\rsymb{\eTns} = \diag(\rsymb{\eTns}_j) = \tQ \eTns \tQ^\trpos   
\end{equation}
has block diagonal form with diagonal blocks $\rsymb{\eTns}_j$, see e.g.\ \citep{faessStiefel1992}.
This $\rsymb{\eTns}$ is called the group reduction of $\eTns = \tQ^\trpos \rsymb{\eTns} \tQ$, 
and the blocks $\rsymb{\eTns}_j$ on the diagonal indicate invariant subspaces of 
$\rsymb{\eTns} \in \Syp(\Strspc) \subset \sy(\Strspc)$.
The non-diagonal minimum non-zero pattern of
$\rsymb{\eTns}$ is characteristic for the elasticity class, 
and invariant under the rotations $\tV = \Trep(\vV) \in \syclS$ with 
$\vV \in \sycls \subseteq \SO(\Physpc)$.  In fact, the elasticity classes are commonly
discussed in this group reduced form \citep{nye_physical_1984, 
cowin_identification_1987, cowMehr1995, malgrange_symmetry_2014, bona_coordinate-free_2007,
altenbach_tensor_2018, malyarenko_tensor-valued_2019}.
When considering random choices resp.\ a stochastic model for $\eTns$, it may be
observed that the mean or expected value may be invariant under a larger
group $\syclsm$ for $\etns \in \SymPh$ resp.\ $\syclSm$ for $\eTns \in \sy(\Strspc)$
\citep{malyarenko_tensor-valued_2019}, i.e.\ $\sycls \subseteq \syclsm$ and 
$\syclS \subseteq \syclSm$.  Some possible choices for a mean will be
investigated in \fsecs{SS:Frechet-means}{SS:rep-requirem}.

Observe that a completely diagonal form with a diagonal $\rsymb{\eTns}$ would mean that
\feq{eq:group-dec} is an eigenvalue or spectral decomposition.
But it is usually not possible to achieve a completely diagonal form just with rotations
$\tQ \in \ST$, see also  \citep{KowalczykOstrowska2009, ItinHehl2015},
the general reason for which will be touched on now.

As already remarked in Appendix~\ref{SS:A-rot-Strspc}, the dimension of $\Physpc=\Rd$ is 
for $d>1$ smaller than that of $\Strspc$ --- $\dim\Physpc < \dim \Strspc = \dim \SymPh = 
\dim (\E{L}_s(\Physpc))$ --- which implies that the dimension of the Lie group $\SO(\Physpc)$, 
is also smaller than that of the Lie group $\SO(\Strspc)$, i.e.\
$\dim(\so(\Physpc)) < \dim (\so(\Strspc))$.    This in turn means that $\ST$ is a proper
subgroup of the full Lie group $\SO(\Strspc)$ of proper rotations on $\Strspc$, 
i.e.\ $\ST \subset \SO(\Strspc)$.  A similar relation obviously holds for its Lie algebra,
$\st \subset \so(\Strspc)$ and $\dim(\st) = \dim(\so(\Physpc)) < \dim (\so(\Strspc))$.

\paragraph{Spectral decomposition:}
The group reduction just discussed commonly produces a block-diagonal matrix $\rsymb{\eTns}$. 
In order for $\rsymb{\eTns}\in \E{L}_s(\Strspc)$ in \feq{eq:group-dec} to completely diagonalise 
$\eTns \in \E{L}_s(\Strspc)$ in general, $\tQ$ would need access to the whole group $\SO(\Strspc)$,
but one is only allowed $\tQ \in \ST \subset \SO(\Strspc)$ resulting from purely spatial rotations
$\tQ = \Trep(\vQ)$ with $\vQ \in \SO(\Physpc)$, which as just explained is only
a subgroup.  Still, \feq{eq:group-dec} is a good start, and
one can now proceed a step further by noting that $\rsymb{\eTns} \in \E{L}_s(\Strspc)$
is symmetric and thus has a spectral decomposition
\begin{equation}  \label{eq:group-eig-dec}
\rsymb{\eTns}  = \tV \tLbd \tV^\trpos   
\end{equation}
with $\tV \in \SO(\Strspc)$ and $\tLbd$ a diagonal matrix of eigenvalues of 
$\rsymb{\eTns}$.  As $\rsymb{\eTns}$ is block-diagonal, a similar structure will
appear in $\tV \in \SO(\Strspc)$, in fact \feq{eq:group-eig-dec} are separate
eigenproblems for each block $\rsymb{\eTns}_j$ from the group decomposition \feq{eq:group-dec}, 
see e.g.\ \citep{faessStiefel1992}.

As $\eTns \in \Syp(\Strspc)$
is SPD, so is its rotated form $\rsymb{\eTns} \in \Syp(\Strspc)$, and hence
$\tLbd \in \Dgp(\Strspc) \subset \Syp(\Strspc)$ is diagonally positive; such matrices
are denoted by $\Dgp(\Strspc)$.  $\tLbd \in \Dgp(\Strspc)$ is obviously also the matrix
of eigenvalues of $\eTns \in \Syp(\Strspc)$, and equating \feq{eq:group-dec} with
\feq{eq:group-eig-dec}, one obtains 
\begin{equation}  \label{eq:SpecDecSPD}
\tV \tLbd \tV^\trpos  = \tQ \eTns \tQ^\trpos \; \Rightarrow \; 
\eTns  = \tQ^\trpos \tV \tLbd \tV^\trpos \tQ = \tU \tLbd \tU^\trpos,
\end{equation}
where $\tV \in \SO(\Strspc)$ and $\tQ \in \ST \subset \SO(\Strspc)$.  But $\SO(\Strspc)$
is a group, and $\tQ^\trpos = \tQ^{-1} \in \ST \subset \SO(\Strspc)$, so that
$\tU := \tQ^\trpos \tV \in \SO(\Strspc)$, and hence \feq{eq:SpecDecSPD}
is a spectral or eigenvalue decomposition of $\eTns$;  this equation will later
form the basis for its representation.


\paragraph{Group invariance:}
Coming back to \feq{eq:group-dec}, and recollecting \feq{eq:tens-inv-rot},
invariance under some symmetry group $\sycls \subseteq \SO(\Physpc)$ for the
reduced elasticity tensor $\rsymb{\etns}$ means that
\begin{equation}  \label{eq:invrnc-elast-tens}
\forall\, \vW \in \sycls:\; \vW  \star \rsymb{\etns} = \rsymb{\etns}  . 
\end{equation}
For the Kelvin matrix  in the group reduced block diagonal form $\rsymb{\eTns}$ 
with $\vW \in \sycls$ and its associated $\tW = \Trep(\vW) \in \syclS = \Trep(\sycls)$, 
from \feq{eq:trans-Kv-mtrx} this translates to
$\rsymb{\eTns}= \tW \rsymb{\eTns} \tW^\trpos$, and hence \citep{faessStiefel1992}
\begin{equation}  \label{eq:invrnc-Kv-mtrx}
\forall\, \tW \in \syclS:\; \rsymb{\eTns} \tW = \tW \rsymb{\eTns}  , 
\end{equation}
where the mapped symmetry or invariance subgroup was designated as 
$\syclS = \Trep(\sycls) \subset \ST \subset \SO(\Strspc)$.
What \feq{eq:invrnc-Kv-mtrx} says is that $\rsymb{\eTns}$ and all $\tW \in \syclS$
commute, i.e. they are co-axial and can all be simultaneously
diagonalised with $\tV\in \SO(\Strspc)$, see also \feq{eq:group-eig-dec}.

It may be noted that the \feq{eq:invrnc-Kv-mtrx} is homogeneous linear in $\rsymb{\eTns}$.
As $\eTns \in \Syp(\Strspc) \subset \sy(\Strspc)$, the equations \feq{eq:invrnc-Kv-mtrx}
determine a subspace $\sy[\syclS] \subseteq \sy(\Strspc)$ in the vector space 
of all symmetric matrices. 
As already mentioned, $\Syp(\Strspc)$ is an open convex cone in $\sy(\Strspc)$; and thus
the Kelvin matrices which are invariant under $\syclS$, and hence belong to the same
elasticity class, are on the intersection of said subspace $\sy[\syclS]$ of $\sy(\Strspc)$
and the cone $\Syp(\Strspc)$, i.e.\ they all lie in the cone 
$\sy[\syclS] \cap \Syp(\Strspc) \subset \sy(\Strspc)$.



\subsection{Random tensor fields} \label{SS:rand-fields}
Let $\sptldom \subset \Rd$ be a bounded domain in a $d$-dimensional Euclidean space 
$\Rd \equiv \Physpc$ (here $d = 2$ or $d = 3$) where we want to model or generate a field of
elasticity tensors $\etns(\vx), \vx\in\sptldom$.  Rather than the elasticity tensor,
here we want equivalently to model the corresponding Kelvin matrix $\eTns = \Vrep(\etns)$.
As the exact value of $\eTns(\vx)$ at a location
$\vx \in \sptldom$ may be uncertain, either due to a highly heterogeneous material, or simply a 
lack of knowledge about the exact value, we assume that a probabilistic model is adopted, 
so that the Kelvin matrix field is modelled as a in a probability sense second order tensor 
valued random field (having first and second moments), i.e.\ a measurable mapping
\begin{equation}  \label{eq:rand-Kelvin-field}
\eTns(\vx, \evt) : \sptldom \times \smplspc \to \Syp(\Strspc) 
\end{equation}
on a probability space $(\smplspc, \sigalg, \prob)$, where $\smplspc$ represents the sample space 
containing all outcomes $\evt \in \smplspc$, $\sigalg$ is the $\sigma$-algebra of 
measurable subsets of $\smplspc$, $\prob$ a probability measure, and the corresponding
expectation operator is denoted by $\EXP{\cdot}$.  To avoid unnecessary 
clutter in the notation, the argument $\vx\in\sptldom$ will be omitted, unless it is necessary 
for understanding.  The same will often be true of the argument $\evt \in \smplspc$.

\paragraph{Mean field and covariance function:}
The first and second order probabilistic descrip\-tors of such a field would be the mean field,
which allows to split the field into the deterministic mean $\meansymb{\eTns}(\vx)$
and the zero mean random part $\Ttil{C}(\vx, \evt)$, and the covariance function
\begin{align}  \label{eq:mean-Kelvin-field}
\meansymb{\eTns}(\vx) &:=  \EXP{\eTns(\vx, \cdot)} = 
              \int_{\smplspc} \eTns(\vx, \evt)\; \prob(\di \evt), \\
       \label{eq:fluct-Kelvin-field}
      \Ttil{C}(\vx, \evt) &:= \eTns(\vx, \evt) - \meansymb{\eTns}(\vx), \\
  \label{eq:cov-Kelvin-field}
    \cov_{\eTns}(\vx,\vy) &:= \EXP{\Ttil{C}(\vx, \cdot) \otimes \Ttil{C}(\vy, \cdot)}.
\end{align}
The material symmetry-group properties are now encoded in the mean and covariance 
function \feqs{eq:mean-Kelvin-field}{eq:cov-Kelvin-field}, 
see e.g.\ \citep{malyarenko_tensor-valued_2019}.  Here one may see another
advantage of numerically working with the Kelvin matrix --- a 2nd order tensor --- which 
results in a covariance which is a 4th order valued tensor function 
on $\Strspc$,
although with some symmetries.  In case that one would form 
the analogue of \feq{eq:cov-Kelvin-field} for the elasticity tensor $\etns(\vx, \evt)$, i.e.\
\begin{equation}  \label{eq:cov-etns-field}
  \cov_{\etns}(\vx,\vy) = \EXP{\Mtil{c}(\vx, \cdot) \otimes \Mtil{c}(\vy, \cdot)},
\end{equation}
this would be an 8th order tensor on $\Physpc$,
although with many symmetries,
i.e.\ a much larger object.

In both cases this large number of covariance component functions 
encode simultaneously the spatial correlations \emph{and} the symmetry-group invariances.
Here we shall follow a different route, and, as will become clear in the 
following \fsec{S:modelling},
encode the symmetry-group properties of the random Kelvin matrix in a field of $n$
real parameters $\vz(\vx, \evt) \in\Rn$ and a map $\vz(\vx, \evt) \mapsto \eTns(\vx, \evt)$.
As will be seen later,
the number of parameters is at maximum $n=  \dim (\so(\Strspc)) + \dim (\dg(\Strspc))$.

The first and second order probabilistic descriptors of these parameters then are
\begin{align}  \label{eq:mean-param-field}
\meansymb{\vz}(\vx) &:=  \EXP{\vz(\vx, \cdot)} = \int_{\smplspc} \vz(\vx, \evt)\; \prob(\di \evt), \\
       \label{eq:fluct-param-field}
      \vtil{z}(\vx, \evt) &:= \vz(\vx, \evt) - \meansymb{\vz}(\vx), \\
\label{eq:cov-param-field}
\cov_{\vz}(\vx,\vy) &:= \EXP{\vtil{z}(\vx, \cdot) \otimes \vtil{z}(\vy, \cdot)},
\end{align}
and they then only carry the information about the spatial and probabilistic distribution.
The maximum number of covariance component functions in \feq{eq:cov-param-field}
is thus $n^2$, but again with the normal symmetry of a symmetric matrix, and thus
only $n(n+1)/2$.  This is actually the number of independent components in both
\feqs{eq:cov-Kelvin-field}{eq:cov-etns-field}.
Additionally, and more importantly,
as will be shown in  the following \fsec{S:modelling}, these parameters allow
one to separate the information about spatial orientation, stiffness or moduli, and eigenstrain
distribution through the Lie product representation, cf.\ \fsec{SS:Lie-Kelvin-represent}.

\paragraph{Spatial modelling of parameter fields:}
The parameters will be described in general terms in \fsec{SS:random-Kelvin}, and then in
detail for each elasticity class in \fsec{S:sepc_decomp}.  To arrive at a spatial model
for the random parameter field $\vz(\vx, \evt)$, one usually aims at a \emph{separated}
or tensor product representation in a sum or integral of $\vtil{z}(\vx, \evt)$, e.g.\ \citep{hgm07}.
\begin{equation} \label{eq:field-sep-rep}
   \vtil{z}(\vx, \evt) =  \sum_\ell \alpha_\ell \vr_\ell(\vx) \zeta_\ell(\evt), 
   \; \text{ or } \;
   \vtil{z}(\vx, \evt) = \int_{\Xi} \alpha(\xi) \vr(\vx,\xi) \zeta(\evt,\xi)\, \di\xi.
\end{equation}
\rvsi{Such separated representations actually follow from fairly general methods to analyse
and represent parametric objects like the fields in \feq{eq:field-sep-rep}, where the random
event $\evt$ may be regarded as the parameter, resulting
in tensor-product like expressions \citep{hgmRO-1-2018, hgmRO-3-2019}.}
Here $\zeta_\ell(\evt)$ resp.\ $\zeta(\evt,\xi)$ are usually zero mean unit variance random variables,
$\vr_\ell(\vx)$ resp.\ $\vr(\vx,\xi)$ are possibly normalised purely spatial functions, and
$\alpha_\ell$ resp.\ $\alpha(\xi)$ are some scaling coefficients.

Aiming for example at a \KL expansion of the spatial field $\vtil{z}(\vx,\evt)$ 
(see e.g.\ \citep{hgm07}), one has to solve the Fredholm integral eigenproblem with 
the covariance function $\cov_{\vz}(\vx,\vy)$  from \feq{eq:cov-param-field} as integral kernel
\begin{equation}  \label{eq:KLE-eigprob}
  \int_\sptldom \cov_{\vz}(\vx,\vy) \, \vr_\ell(\vy) \,\di \vy = 
    \mu_\ell \, \vr_\ell(\vx) ,
\end{equation}
for positive eigenvalues $\mu_\ell$ and eigenvector fields $\vr_\ell(\vx) \in \Rn$
and $\ell \in \D{N}$.  Here it was tacitly assumed that the integral operator in
\feq{eq:KLE-eigprob} is self-adjoint and positive and has a discrete spectrum, which would be usually 
the case with a bounded domain $\sptldom$ and continuous $\cov_{\vz}(\vx,\vy)$ for example.

It should be noted that spatial \rvsi{stochastic} homogeneity would be evident as usual 
from the fact that $\cov_{\vz}(\vx,\vy)$ would be a function of only the difference of
the arguments, $\vh := \vx - \vy$, \rvsi{say $\cov_{\vz}(\vx,\vy) = \vek{K}_{\vz}(\vx - \vy)
= \vek{K}_{\vz}(\vh)$}.  In that case the eigenproblem \feq{eq:KLE-eigprob} 
is a convolution equation, and it is well known how to solve those by Fourier
methods, as indeed the Fourier functions are (possibly generalised) eigenfunction
of the integral operator with kernel $\cov_{\vz}$ represented by \feq{eq:KLE-eigprob}.
\rvsi{Indeed, with the eigenfunctions known, the eigenvalues are obtained by a Fourier
transform --- continuous or in form of a series, depending on the integral operator 
\feq{eq:KLE-eigprob} and the domain $\sptldom$ --- of the matrix function $\vek{K}_{\vz}(\vh)$ above.
This Fourier transform $\vhat{K}_{\vz}(\vk)$, where $\vk$ is the vector of wave numbers resp.\
spatial frequencies dual to $\vh$, is the SPD \emph{spectral matrix}.  Locally diagonalised for
each $\vk$, it then displays the eigenvalues of the \KL eigenproblem \feq{eq:KLE-eigprob}.
This is actually the well know spectral description of stochastically homogeneous random fields.  
Further information may be found in \citep{hgm07} and the references
therein; see also the remarks in \fsec{SS:random-Kelvin}.  In the general --- 
non-stochastically homogeneous --- case, the \KL eigenproblem \feq{eq:KLE-eigprob} has to
be solved numerically, cf.\ e.g.\ \citep{Matthies2005, Matthies2008}.}

In any case, the orthonormal eigenfunctions $\vr_\ell(\vx)$ determine uncorrelated
mean zero unit variance
random variables $\zeta_\ell(\evt)$ through the projection of $\vtil{z}(\vx,\evt)$,
setting $\alpha_\ell = \sqrt{\mu_\ell}$,
onto the spatial eigenvector basis $\{ \vr_\ell(\vx) \}_{\ell \in \D{N}}$:
\begin{equation}  \label{eq:KLE-RV-proj}
      \zeta_\ell(\evt) = \frac{1}{\alpha_\ell}\;
  \int_\sptldom \vtil{z}(\vx,\evt)^\trpos  \, \vr_\ell(\vx) \,\di \vx .
\end{equation}
With this one has all the ingredients for the spatial representation in \feq{eq:field-sep-rep},
which is in this case called a \KL expansion, equivalent to a singular value decomposition
of the random field $\vtil{z}(\vx,\evt)$ viewed as an operator kernel, e.g.\ \citep{hgm07}.

\rvsi{As may be noted immediately, for Gaussian random fields all the uncorrelated random 
variables $\zeta_\ell$ in \feq{eq:KLE-RV-proj} are also Gaussian, and thus they are also 
independent.  For modelling purposes,
this is what is usually desired, a model in independent random variables.
As the parameter vector field $\vtil{z}$ effectively models the log of the SPD tensor field,
this partly explains the allure and pre-dominance of log-normal models for
such SPD tensors \citep{schwartzman_lognormal_2016}.  For cases where the log-parameters $\vtil{z}$
are not normal, the random variables $\zeta_\ell$ are merely uncorrelated, but not independent.
In this case it is possible to express these random variables in Norbert Wiener's 
\emph{polynomial chaos} expansion, cf.\ e.g.\ \citep{Matthies2005, hgm07, Matthies2008} 
and the references therein.
Thus one may represent $\zeta_\ell(\evt)$ by a series of multi-variate polynomials 
$\Psi_{\F{a}}$ --- $\F{a}$ is typically a multi-index --- 
in independent mean zero unit variance Gaussian random variables 
$\vek{\xi}(\evt) = (\xi_1(\evt),\dots,\xi_j(\evt),\dots)$ as
\begin{equation}  \label{eq:PCE-zeta}
\zeta_\ell(\evt) = \sum_{\F{a}} \beta^\ell_{\F{a}}\, \Psi_{\F{a}}(\vek{\xi}(\evt)),
\end{equation}
with some coefficients $\beta^\ell_{\F{a}}$.  This then yields the field $\vtil{z}$ in 
\feq{eq:field-sep-rep} as a function of \emph{independent} standard 
Gaussian variables $\vek{\xi}(\evt)$:
\begin{equation}  \label{eq:PCE-z}
  \vtil{z}(\vx, \evt) =  \sum_\ell \alpha_\ell \left(\sum_{\F{a}}  \beta^\ell_{\F{a}}\, 
  \Psi_{\F{a}}(\vek{\xi}(\evt)) \right)   \vr_\ell(\vx) = 
  \sum_{\F{a}} \Psi_{\F{a}}(\vek{\xi}(\evt)) \vg_{\F{a}}(\vx), 
\end{equation}
where the expression $\vg_{\F{a}}(\vx) := \sum_\ell \alpha_\ell \beta^\ell_{a}\,  \vr_\ell(\vx)$ 
may be seen as the polynomial chaos expansion coefficient.}

\rvsi{The foregoing covers the modelling aspect.} 
To perform the reverse task, \rvsi{the inverse problem of conditioning
\citep{mosegaard:2002:IP, tarantola:2005, AMStuart2010}}, one has to \rvsi{change}
the \rvsi{descriptions in \feqss{eq:field-sep-rep}{eq:PCE-zeta}{eq:PCE-z}}, e.g.\
spatial functions $\vr_\ell(\vx)$ and the random variables $\zeta_\ell(\evt)$
such that $\vz(\vx,\evt)$ has the proper statistics at each $\vx \in \sptldom$. 
\rvsi{This can range from finding the SPD tensor closest to measurements
\citep{FittEtAl2019, MoakherNorris2006, WeberGluegeBertram2019} to conditioning the
random SPD tensor field \citep{RosicKucerovaSykoraEtAl2013, RosicSykoraPajonkEtAl2016, 
MatthiesZanderEtAl2016, MatthiesZanderRosicEtAl2016, hgm18, NouySoize2014, SavvasPapai2020}.
One may note that some of the filtering methods --- e.g.\ the Gauss-Markov-Kalman filter --- 
for Bayesian conditioning work directly
on the polynomial chaos expansion coefficients $\vg_{\F{a}}(\vx)$ in \feq{eq:PCE-z}
\citep{PajonkRosicEtAl2012, RosicLitvinenkoEtAl2012, RosicSykoraPajonkEtAl2016, MatthiesZanderEtAl2016, 
MatthiesZanderRosicEtAl2016, hgm18, AI-HGM-EK-2020}.}


%
%
%
%
%
%
%
%
%
%
%
%
%
%


%
\section{Modelling of Elasticity Tensors}  \label{S:modelling}

%

%
%
%
%
%

The main idea from the preceding \fsec{SS:group-inv} for the modelling 
approach to be discussed here is contained in the
\feqss{eq:group-dec}{eq:group-eig-dec}{eq:SpecDecSPD}, which are collected here for easy
reference:
\begin{equation}   \label{eq:easy-ref}
  \rsymb{\eTns}  = \tQ \eTns \tQ^\trpos, \quad \rsymb{\eTns}  = \tV \tLbd \tV^\trpos \;\Rightarrow\;
  \eTns  = \tQ^\trpos \tV \tLbd \tV^\trpos \tQ = \tU \tLbd \tU^\trpos, \; \text{ with } \;
  \tU := \tQ^\trpos \tV .
\end{equation}
Here $\rsymb{\eTns} \in \Syp(\Strspc)$ is the group reduced form of the Kelvin matrix
$\eTns \in \Syp(\Strspc)$, and
$\tQ = \Trep(\vQ)\in\ST$ with $\vQ \in \SO(\Physpc)$, $\tV \in \SO(\Strspc)$ and hence
$\tU \in \SO(\Strspc)$ as well, and $\tLbd \in \Dgp(\Strspc)$.

\subsection{Lie representation} \label{SS:Lie-Kelvin-represent}
Thus, given $\rsymb{\eTns}$ in the form of $\tV \tLbd \tV^\trpos$, which means
in form of an eigenstrain distribution $\tV$ and eigenvalues $\tLbd$, plus an orientation
$\vQ\in\SO(\Physpc)$ in space which describes the rotation from $\eTns$ to $\rsymb{\eTns}$, 
one may form $\tQ=\Trep(\vQ)$ and $\tU = \tQ^\trpos \tV$ and then generate $\eTns$:
\begin{equation}   \label{eq:Lie-decomp}
  (\vQ, \tV, \tLbd) \mapsto (\tQ=\Trep(\vQ), \tV, \tLbd) \mapsto  (\tU  = \tQ^\trpos \tV, \tLbd)
  \mapsto \eTns = \tU \tLbd \tU^\trpos.
\end{equation}
This is a mapping from $\SO(\Physpc) \times \SO(\Strspc) \times \Dgp(\Strspc)$ to $\Syp(\Strspc)$,
resp.\ from $\ST \times \SO(\Strspc) \times \Dgp(\Strspc)$ to $\Syp(\Strspc)$

\paragraph{Lie group representation of the Kelvin matrix:}
The representation \feq{eq:Lie-decomp} will be called the Lie group representation of the Kelvin
matrix $\eTns \in \Syp(\Strspc)$, as $\vQ \in\SO(\Physpc)$ resp.\ $\tQ \in \ST$ 
and $\tV \in \SO(\Strspc)$ are elements
from matrix Lie groups.  But the diagonal SPD matrices $\Dgp(\Strspc)$ form a commutative
or Abelian Lie group under matrix multiplication, and hence $\tLbd \in \Dgp(\Strspc)$ is also from a 
Lie group.  The Lie representation in \feq{eq:Lie-decomp} is thus one on a 
product $\sS$ of Lie groups \citep{JungSchwartzmanGroisser2015,
schwartzman_lognormal_2016, GroisserJungSchwartzman2017, groissJungSchwman2017, shivaEtal2021}, 
which is again a Lie group:
\begin{equation}  \label{eq:prod-Liegrp}
  \sS := \ST \times \SO(\Strspc) \times \Dgp(\Strspc) .
\end{equation}

It should be noted here that the representation
in \feq{eq:Lie-decomp} is \emph{not unique}.  What is unique is the spectrum of $\eTns\in \Syp(\Strspc)$,
i.e.\ the set of diagonal elements of $\tLbd \in \Dgp(\Strspc)$.  They could be re-ordered, 
inducing a re-ordering of the columns of $\tV \in \SO(\Strspc)$.  This could be avoided by
insisting on an ordering, say by size, of the diagonal elements $\lambda_i$ of $\tLbd$.
But even more significantly, when degenerate or multiple eigenvalues occur, as is always
the case for the `higher' elasticity classes (cf.\ \fsec{S:sepc_decomp}), those eigenvalues
are associated with invariant subspaces of dimension two or higher.  This makes the quest for
an orthonormal eigenvector basis of such an invariant subspace --- the associated columns
of $\tV \in \SO(\Strspc)$ --- a matter of arbitrary choice.

As further detailed in Appendix~\ref{SS:metric}, for a general Lie group $G$, its
Lie algebra $\F{g}$, the tangent space at the identity, plays an important role
\citep{Carmo1992, Spivak-1-1999, Jost2005, AlexandrinoBettiol2015}.  
Specifically, it is possible to use the Lie exponential map --- for matrix Lie groups and algebras
this is the usual matrix exponential --- $\exp: \F{g} \to G$ to represent elements of the
group $G$ on a the free vector space $\F{g}$; the inverse of this map is the $\log: G \to \F{g}$.
Here we want to parametrise the product Lie group \feq{eq:prod-Liegrp} in this way
via its Lie algebra.

\paragraph{Exponential Lie algebra representation of the Kelvin matrix:}
The exponential and the logarithm allow one to shift the representation from the Lie group $\sS$
in \feq{eq:prod-Liegrp} to its Lie algebra $\sSl$, which is a combination of the Lie algebras
of the individual factors:
\begin{equation}  \label{eq:prod-Liealg}
 \sSl = \st \times \so(\Strspc) \times \dg(\Strspc). 
\end{equation}
Here $\dg(\Strspc) \subset \sy(\Strspc)$ is the vector space of diagonal matrices,
the Lie algebra of $\Dgp(\Strspc)$.  
With the exponential map, this finally yields
a representation of the Kelvin matrix $\eTns$ on the vector space $\sSl$ in \feq{eq:prod-Liealg}.
It allows one also to separate the effects of spatial orientation,
represented by $\vQ \in \SO(\Physpc)$, strain distribution, as 
represented by $\tV \in \SO(\Strspc)$, and stiffness, as given by 
$\tLbd \in \Syp(\Strspc)$.  In \fsec{SS:Frechet-means} and Appendix~\ref{SS:metric} 
this is used further to define the `elasticity metric', 
which is used in the sequel for Fréchet means.  

It should be noted that to represent $\etns\in\SymTen$ resp.\ $\eTns\in \Syp(\Strspc)$, 
usually not all of $\SO(\Physpc)$ resp.\ $\ST$ is required, as one has invariance 
w.r.t.\ rotations $\vW\in\sycls$ resp.\ $\tW=\Trep(\vW)\in\syclS$ 
in the symmetry group of the material.  
We shall not introduce the group theoretic complication which might arise from a transition
to factor groups here 
\citep{malyarenko_random_2017, altenbach_tensor_2018, malyarenko_tensor-valued_2019},
and allow possibly
\emph{wrapped} representations, as they occur also often in circular statistics, cf.\ e.g.\
\citep{jammalamadaka_topics_2001, JammalamadakaKozubowski2017}.  A similar `wrapping problem'
occurs in the representation of spatial rotations $\SO(n)$ on its Lie algebra 
$\so(n)$ via the exponential map,
see also the remarks at the end of Appendix~\ref{SS:A-rot-2-3}.

As is easily seen from \feq{eq:SpecDecSPD} and \feq{eq:group-dec} and the above discussion, 
the group invariance properties of the symmetric tensor 
\begin{equation}  \label{eq:log-eTns}
\tH = \log \eTns = \tU \log(\tLbd) \tU^\trpos =
   \tQ^\trpos \tV \log(\tLbd) \tV^\trpos \tQ \in \sy(\Strspc)
\end{equation}
and its exponential $\eTns = \exp \tH$ are the same, a fact which was also used in
\citep{GuilleminotSoize2013a, GuilleminotSoize2014}.  And as we are using the
Lie algebra $\dg(\Strspc)$ of the Lie group $\Dgp(\Strspc)$, we are essentially
modelling  $\tH = \log \eTns$ in \feq{eq:log-eTns}.

Note that one should actually
apply some scaling prior to using the logarithm in \feq{eq:log-eTns}, as the
quantities involved may have physical dimensions.  One could use in \feq{eq:log-eTns}
a diagonal scaling matrix $\tD\in \Dgp(\Strspc)$ --- often this is just a multiple
of the identity, $\tD = \alpha \tI$ for some $\alpha>0$ --- of the same physical dimensions as
$\tLbd$ and then compute with 
\begin{equation}  \label{eq:log-eTnsDscl}
\tH = \log \eTns = \tU \log(\tD^{-1}\tLbd) \tU^\trpos =
   \tQ^\trpos \tV \log(\tD^{-1}\tLbd) \tV^\trpos \tQ \in \sy(\Strspc).
\end{equation}
Or one could use a reference matrix
$\whsymb{\eTns}\in \Syp(\Strspc)$ of the same physical dimensions as $\eTns$ --- often 
$\whsymb{\eTns} = \tD \in \Dgp(\Strspc)$ is a diagonal matrix, or even just a multiple 
of the identity $\whsymb{\eTns} = \alpha \tI$ as above --- and define the logarithm as 
\begin{equation}  \label{eq:log-eTnsscl}
\tH = \log(\whsymb{\eTns}^{-1/2} \eTns \whsymb{\eTns}^{-1/2}) \in \sy(\Strspc),
\end{equation}
see also e.g.\ \citep{shivaEtal2021}.  It is tacitly assumed that some such transformation to 
non-dimensional quantities has been carried out, so there is no need to clutter the notation with it.

Together with the exponential Rodrigues formulas in Appendix~\ref{S:appx-Rotations},
this means that all the invariance constraints
in the Lie group 
representation \feq{eq:Lie-decomp} 
may be applied on the Lie algebra level in \feq{eq:prod-Liealg}, but 
a little refinement is in order here.  As in the product Lie group $\sS$ the
first factor $\ST \subset \SO(\Strspc)$ is contained in the second one, it is clear that in
the  Lie algebra $\sSl = \st \times \so(\Strspc) \times \dg(\Strspc)$
one has $\st \subset \so(\Strspc)$, and so the second factor can be reduced to 
$\st^\perp \subset \so(\Strspc)$, the orthogonal complement of $\st$ in $\so(\Strspc)$.
Hence we may use for the representation the direct sum of Lie algebras
\begin{equation}  \label{eq:prod-Liealg2}
  \rpal = \st \oplus \st^\perp \oplus \dg(\Strspc) .
\end{equation}

Hence in order that one may represent the Kelvin matrix later in detail
in \fsec{S:sepc_decomp}, the following general picture emerges:
starting with a triple $(\tR,\tP,\tM) \in \rpal$
one first builds the logarithm $\tH$ in factored form, and then $\eTns$:
\begin{align}  \label{eq:exp-Kelvin-chain}
  (\tR,\tP,\tM) \quad &\longmapsto\quad (\exp\tR,\exp\tP,\tM)\quad \longmapsto\quad 
  \tH = \tQ^\trpos \tV \tM \tV^\trpos \tQ = \log \eTns \\  \label{eq:exp-Kelvin-chain2}
  &\longmapsto \quad \eTns = \tU \exp(\tM) \tU^\trpos =  \tU \tLbd \tU^\trpos = \exp \tH ,
  \phantom{Hhhhhhhhhhhhhh}
\end{align}
where, as before, $\tQ = \exp\tR$, $\tV = \exp\tP$, $\tU = \tQ^\trpos \tV$, and
$\tLbd = \exp \tM$; in between one has $\tH = \log \eTns \in \sy(\Strspc)$, and finally 
$\eTns = \exp\tH \in \Syp(\Strspc)$.  This exponential representation 
$\exp: \sy(\Strspc) \to \Syp(\Strspc)$ ensures positive definiteness of $\eTns$.

As what we finally seek
is a measurable map $\eTns: \sptldom \times \smplspc \to \Syp(\Strspc)$ to give the
elasticity Kelvin matrix $\eTns(\vx,\evt)$ for a location $\vx \in \sptldom$ and event
$\evt \in \smplspc$, from what was just outlined in \feq{eq:exp-Kelvin-chain}, this will be done
in several steps.  One of them is to produce random tensor fields with values in
$\rpal$ in \feq{eq:prod-Liealg2}, i.e.\ random triples 
$\sptldom \times \smplspc \to (\tR,\tP,\tM) \in \rpal$.
From this, at each location $\vx$  and event $\evt$, the log tensor $\tH = \log \eTns$ is built according to \feq{eq:exp-Kelvin-chain}, and then
finally $\eTns =  \exp \tH$ as just described in \feq{eq:exp-Kelvin-chain2}.

In order to put the choice of random
elements into proper perspective, in the following \fsecs{SS:Frechet-means}{SS:rep-requirem}
some consideration on how to to choose a suitable kind of mean or expectation operator
on $\Syp$ will be given, together with some requirements for such a choice.  Then 
we can continue on how to generate random Kelvin matrices in general in \fsec{SS:random-Kelvin}.

\subsection{Fr\'echet means on manifolds} \label{SS:Frechet-means}
As the set of SPD maps $\Syp(\Strspc)$ is not a flat vector space, but a manifold
which can be accessed in different ways, this leads
to  the question of what kind of average or mean to use for
elements of $\Syp(\Strspc) \subset \sy(\Strspc)$, and this is
tied to the question of how distances are measured on this set.
As the ambient space
$\sy(\Strspc)$ is a vector space, it appears at first natural to 
calculate averages and means as one would in a vector space, using the common arithmetic
or Euclidean mean.  But, as mentioned already in \fsec{S:intro}, it has been found that by 
using this kind of mean one introduces undesirable effects 
\cite{PennecFillardAyache2004, AndoLiMathias2004, Moakher2005, ArsignyFillardPennecEtAl2006, 
ArsignyFillardPennecEtAl2007, drydenEtal2009, DrydenEtal2010, WangSalehianEtAl2014, FujiiSeo2015,
FeragenFuster2017}.   As further described there in \fsec{S:intro} and also in 
Appendix~\ref{S:appx-metric}, these are the so-called the 
\emph{swelling, fattening, and shrinking} effects 
\cite{schwartzman_random_nodate, JungSchwartzmanGroisser2015, 
schwartzman_lognormal_2016, GroisserJungSchwartzman2017, groissJungSchwman2017, FeragenFuster2017}.

\paragraph{Variational characterisation of the mean:}
In this context where we want to describe shortly how to generalise the arithmetic or Euclidean mean. 
One may note first that the mean has a variational
characterisation, cf.\ also Appendix~\ref{SS:frechet} and \citep{Oller_1995, 
PennecFillardAyache2004, Pennec2006, Moakher2006, Moakher2008, MoakherZerai2011, FeragenFuster2017, 
ThanwerdasPennec2019, SommerFletcherPennec2020, Fletcher2020, shivaEtal2021, GuiguiEtal2023} 
for further discussion, which is the key for further generalisation.
This characterisation is reproduced in Appendix~\ref{S:appx-metric} for better reference,
see \feeqs{eq:var-def-EuclMean}{eq:x-basedvar-P}, and it is based
on a minimisation: 
\begin{equation}  \label{eq:var-def-EuclMean-p}
  \olsi{\tX}^{(\vtheta)} := \arg \min_{\tX \in\sy(\Strspc)}  \Psi_{\vtheta}(\tX),
\end{equation}
where the $\tX$-based variance $\Psi_{\vtheta}(\tX)$ (cf.\ also \feq{eq:x-basedvar-P})
is the so-called `loss function', here based on the general distance function 
resp.\ metric $\vtheta$.  In the simple case of a finite collection of items
$\{ \tX_j\}$ to be averaged this would be 
$\Psi_{\vtheta}(\tX) = \sum_j w_j \vtheta(\tX,\tX_j)^2$,
with weights $w_j \ge 0$, $\sum_j w_j = 1$.
The usual arithmetic resp.\ Euclidean mean is obtained
by use of the Euclidean distance $\vtheta_2$ in $\Psi_{\vtheta}(\tX)$,
see Appendix~\ref{S:appx-metric}.
But a Euclidean distance on $\Syp(\Strspc)$ is not necessarily natural, and it is what seems to 
be responsible for the undesirable effects just mentioned.  So one has to look for other 
distance functions $\vtheta$, cf.\ also Appendix~\ref{SS:metric}, which to use in
\feq{eq:var-def-EuclMean-p} in place of the Euclidean metric 
$\vtheta_2$ in the variational characterisation.

\paragraph{Possible alternative means:}
There are several possible choices, cf.\ Appendix~\ref{SS:metric} and \citep{Oller_1995, 
PennecFillardAyache2004, Pennec2006, Moakher2006, Moakher2008, MoakherZerai2011, FeragenFuster2017, 
ThanwerdasPennec2019, SommerFletcherPennec2020, Fletcher2020, shivaEtal2021, GuiguiEtal2023},
as well as the references therein.  The connection with \fsec{SS:Lie-Kelvin-represent} is as follows.
It has been mentioned here before that the set of positive definite matrices $\Syp(\Strspc)
\subset \sy(\Strspc)$ is geometrically an open convex cone in $\sy(\Strspc)$.  Hence $\Syp(\Strspc)$
can be equipped with the structure of a differentiable manifold.  

The representation of $\Syp(\Strspc)$ on the product Lie group $\sS$ in \feq{eq:prod-Liegrp}, and
via the maps \feq{eq:exp-Kelvin-chain} finally on the vector space $\sSl$ in \feq{eq:prod-Liealg},
allows one to map a Euclidean structure on the Lie algebra $\sSl$ onto a Riemannian structure on
the product Lie group $\sS$, each factor separately, see Appendix~\ref{S:appx-metric} for more details.
The Riemannian structure enables the definition of geodesics --- locally shortest paths --- and
thus allows to define a distance between two elements of each of the factor groups 
as the length of the shortest geodesic between those two elements.  

From this one may define a product metric on the product Lie group $\sS$
 \citep{WangSalehianEtAl2014, JungSchwartzmanGroisser2015, schwartzman_lognormal_2016, 
GroisserJungSchwartzman2017, groissJungSchwman2017, FeragenFuster2017, shivaEtal2021}, 
and take this as the basis for a metric on $\Syp(\Strspc)$.
The product Lie group $\sS$ used here has a finer structure than that used for 
$\Syp(\Physpc)$ in \citep{shivaEtal2021}, or general SPD matrices in 
\citep{JungSchwartzmanGroisser2015, schwartzman_lognormal_2016, GroisserJungSchwartzman2017,
groissJungSchwman2017, FeragenFuster2017}, which only uses the spectral 
decomposition \feq{eq:SpecDecSPD}, which would correspond to a
representation on only $\SO(\Strspc)\times \Dgp(\Strspc)$.
Here we want to use this finer structure given by the representation in \feq{eq:Lie-decomp},
which includes the invariance resp.\ symmetry group reduction, in the definition of a new metric,
which is explained in detail in Appendix~\ref{SS:metric}.
It is proposed to call this metric the `elasticity metric' or `elasticity distance' $\vtheta_E$
in \feq{eq:elast-metric2}.  It allows
a fine control to measure differences in spatial orientation $\tQ\in\ST$ in \feq{eq:Lie-decomp},
differences in strain distributors $\tV \in \SO(\Strspc)$, and differences in Kelvin moduli
$\tLbd \in \Dgp(\Strspc)$, and combine them with different weights (cf.\ Appendix~\ref{SS:metric}).

It is based on the product distance on $\sS$ (see \feq{eq:elast-dist} in
Appendix~\ref{SS:metric}).  
For two elements $\eTns_1, \eTns_2 \in \Syp(\Strspc)$ with representations
$\eTns_j = \tQ_j^\trpos \tV_j \tLbd_j \tV_j^\trpos \tQ_j$, $j=1,2$ based on
$(\tR_j,\tP_j,\tM_j) \in \rpal$ as in \feq{eq:exp-Kelvin-chain},
the squared product distance $\tilde{\vtheta}_E^2$ is given by
\begin{equation}  \label{eq:elast-dist-p}
   \tilde{\vtheta}_E(\eTns_1, \eTns_2)^2 = \vtheta_L(\tLbd_1, \tLbd_2)^2 +
   c_{\Physpc} \vtheta_R(\vQ_1, \vQ_2)^2 + c_{\Strspc} \vtheta_R(\tV_1, \tV_2)^2 ,
\end{equation}
with two positive tuning constants $c_{\Physpc}, c_{\Strspc}$.  The distances on the
component Lie groups are: 
for the rotation groups the standard Riemannian metric (cf.\ \feq{eq:SO-dist}) 
$\vtheta_R(\vQ_1, \vQ_2)= \nd{\log (\vQ_1 \vQ_2^\trpos)}_F$ with $\vQ_j \in \SO(\Physpc)$,
and completely analogous also for $\vtheta_R(\tV_1, \tV_2)$ with $\tV_j \in \SO(\Strspc)$;
and for $\Dgp(\Strspc)$ also the logarithmic Riemannian metric (cf.\ \feq{eq:Diag-dist})
$\vtheta_L(\vLbd_1, \vLbd_2)    := \nd{\log \vLbd_1 - \log \vLbd_2}_F =\nd{\tM_1 - \tM_2}_F$.
Additionally, one has now to worry about the non-uniqueness of the product representation,
to obtain from $\tilde{\vtheta}_E$ in \feq{eq:elast-dist-p}
the final elasticity distance $\vtheta_E$ in \feq{eq:elast-metric2}.
Here, as shown in Appendix~\ref{SS:metric}, one minimises over all possible
Lie product group representations $\eTns = \tQ^\trpos \tV \tLbd \tV^\trpos \tQ$,
and, as the geodesic is only locally a shortest path, over all `intermediate stops';
but for small variations \feq{eq:elast-dist-p} is all what is needed.

To recap quickly Appendix~\ref{S:appx-metric} here, with
this elasticity metric one may go back to \feq{eq:var-def-EuclMean-p}, employing the
variational characterisation with metric $\vtheta$, and replace the Euclidean metric 
$\vtheta=\vtheta_2$ with the elasticity metric $\vtheta=\vtheta_E$
mentioned in \feq{eq:elast-metric2}.
At the end of Appendix~\ref{S:appx-metric}, an elasticity Fr\'echet mean or 
expected value $\D{E}_{\vtheta_E}$ is thus defined in \feq{eq:elast-mean}
for a random elasticity tensor, utilising as loss function
the $\Tbar{C}$-based variance $\Psi_{\vtheta_E}(\Tbar{C})$ in \feq{eq:elast-var-C}, which is
based on the elasticity metric $\vtheta_E$ in \feq{eq:elast-metric}.

\subsection{Requirements for representation and mean} \label{SS:rep-requirem}
While the introduction of a new metric and mean where motivated in the preceding
\fsec{SS:Frechet-means} by geometrical considerations, it might be actually worthwhile
to consider what one may require from a metric for elasticity tensors resp.\ Kelvin matrices,
and even for any kind of numerical representation in general.

As the goal is to be able to define, model, and represent random tensor fields $\eTns(\vx,\evt)$
on some spatial domain $\sptldom \subset \Rd$, $d=2$ or $d=3$ with typical point
$\vx \in \sptldom$, and $\evt \in \smplspc$ an event in some appropriate probability space
$(\smplspc,\sigalg,\prob)$ with $\sigma$-algebra $\sigalg$ and probability measure $\prob$,
i.e.\ a measurable map $\eTns: \sptldom \times \smplspc \to \Syp(\Strspc)$, here we formulate
some requirements which in our view have to be satisfied for the modelling, so that
the random tensor field $\eTns(\vx,\evt)$ is usable in e.g.\ stochastic FEM calculations
and uncertainty quantification (e.g.\ \citep{hgm07, Soize2017} and the references therein).  See also
\citep{shivaEtal2021} for similar remarks regarding second-order SPD random tensor fields.

\paragraph{Representation requirements:}
The requirements we suggest should be met is that
at almost all points $\vx \in \sptldom$ and realisations $\evt \in \smplspc$,
\begin{enumerate}
\item even under numerical approximations like truncation of series resp.\ numerical quadrature,
  the tensor $\eTns(\vx,\evt)$ has to be SPD,  i.e.\ 
  $\eTns(\vx,\evt) = \eTns(\vx,\evt)^\trpos \in \sy(\Strspc)$, and
  for all $\tx \neq \tnb{0} \in \Strspc :   \tx^\trpos \eTns(\vx,\evt) \tx > 0 $;
\item  the tensor $\eTns(\vx,\evt)$ has to be invariant 
  under some group of transformations $\syclS \subseteq \ST $, 
  the invariance or symmetry class of each realisation;
\item the mean $\olsi{\eTns}^{(\vtheta_D)}(\vx) := 
  \D{E}_{\vtheta_D}(\eTns(\vx,\cdot)) \in \Syp(\Strspc)$ based on some desirable metric 
  $\vtheta_D$ has to be invariant under a possibly larger group of transformations 
  $\syclSm$, with $\ST \supseteq \syclsm \supset \syclS$, 
  the invariance or symmetry class of the mean. 
\end{enumerate}
The first item is a necessity in order to be able to successfully calculate with the tensor field
$\eTns(\vx,\evt)$.  The second item makes sure that each realisation has the proper spatial
invariance resp.\ symmetry or elasticity class.  The third item allows to control the elasticity
class of the mean resp.\ expected value.  For the most part the concern here is the analysis,
modelling, and representation of $\eTns(\vx,\evt)$ at one point $\vx \in \sptldom$ and 
sample $\evt \in \smplspc$, so that for the sake of brevity these arguments will be dropped 
in the sequel, just as they have been in most of the description above,
up to the point in \fsec{SS:random-Kelvin} when spatial random fields will finally be considered.

\paragraph{Desiderata for a Fréchet mean:}
Turning now to the mean and metric addressed in the third item above,
first note that a change of physical units or to non-dimensional quantities 
is a diagonal scaling like \feq{eq:log-eTnsDscl}, and the mean
should scale the same way, i.e.\ first scaling and the averaging should give the
same result as first averaging and then scaling.  Secondly, a change of coordinate system by some 
$\tQ = \Trep(\vQ) \in \ST$ for any $\vQ \in \SO(\Physpc)$ should affect
the same change in the mean, i.e. commutativity of averaging and coordinate transform.  
And lastly, one may note that Hooke's law
--- in vector notation in \feq{eq:Hooke-vec} --- can just as well be formulated
in terms of compliances, i.e.\
\begin{equation}  \label{eq:Hooke-vec-inv}
   \strnv = \eTns^{-1} \strsv = \tS \strsv; \quad \text{ with compliance } \tS = \eTns^{-1}.
\end{equation}
Hence it seems desirable to require the `mean material law' to be independent of the arbitrariness
of formulating the material law, e.g.\ \feq{eq:Hooke-vec} or \feq{eq:Hooke-vec-inv},
and therefore demand an appropriate invariance under inversion, i.e.\
averaging and then inverting should give the same result as first inverting and then averaging.  
In simplest terms, given $n$ Kelvin matrices $\{\eTns_j \}_{j=1}^n \subset \Syp(\Strspc)$, 
let $\olsi{\eTns_j}^{(\vtheta_D)}$ denote their Fréchet mean based on some 
desirable distance $\vtheta_D$,  cf.\ Appendix~\ref{S:appx-metric}, and let $\tQ=\Trep(\vQ)\in\ST$,
$\vQ \in \SO(\Physpc)$ be an arbitrary spatially induced rotation.  
The requirements for the mean then are:
\begin{enumerate}
\item invariance under scaling: $\forall\; \alpha > 0$:
     $\olsi{(\alpha\eTns_j)}^{(\vtheta_D)} = \alpha\,\olsi{\eTns_j}^{(\vtheta_D)}$;
\item invariance under orthogonal transformations:
     $\olsi{(\tQ\eTns_j\tQ^\trpos)}^{(\vtheta_D)} = \tQ \olsi{\eTns_j}^{(\vtheta_D)} \tQ^\trpos$;
\item invariance under inversion: let $\{\tS_j = \eTns_j^{-1} \}_{j=1}^n \subset \Syp(\Strspc)$ 
      with mean $\olsi{\tS_j}^{(\vtheta_D)}$ be the inverses to the Kelvin matrices 
      $\{\eTns_j \}$, it should hold that 
     $ \olsi{\tS_j}^{(\vtheta_D)} = (\olsi{\eTns_j}^{(\vtheta_D)})^{-1}$.
\end{enumerate}

\paragraph{Desiderata for a metric to define a Fréchet mean:}
When looking at the variational definition of the Fréchet mean 
\feq{eq:var-def-EuclMean-p}, but with another metric $\vtheta_D$ in place of the Euclidean 
metric, or the more general formulas in
Appendix~\ref{S:appx-metric}, it is quite obvious that the Fréchet mean inherits the
required properties  from the underlying metric $\vtheta_D$ on $\Syp(\Strspc)$.  
The requirements for a desired metric $\vtheta_D$, and hence for a Fréchet mean based on it, 
are thus to satisfy for all $\eTns_1, \eTns_2 \in \Syp(\Strspc), \alpha > 0$, 
and all $\tQ \in \ST$:
\begin{enumerate}
\item quasi-invariance under scaling: $\vtheta_D(\alpha \eTns_1,\alpha \eTns_2) =
   f(\alpha) \vtheta_D(\eTns_1, \eTns_2)$, \\ with a monotone $f(\alpha)>0$;
\item invariance under orthogonal transformations: $\vtheta_D(\tQ\eTns_1\tQ^\trpos,
   \tQ\eTns_2\tQ^\trpos) = \vtheta_D(\eTns_1, \eTns_2)$;
\item invariance under inversion: $\vtheta_D(\eTns_1^{−1}, \eTns_2^{−1}) = 
   \vtheta_D(\eTns_1, \eTns_2)$;
\end{enumerate}
The Euclidean distance $\vtheta_2$ satisfies only the first two items, 
and so the arithmetic or Euclidean mean only satisfies the first two requirements for means.
The elasticity metric $\vtheta_E$ proposed here, see Appendix~\ref{S:appx-metric}, 
satisfies all three requirements, and so does the associated mean.  There are some other
metrics which satisfy all three requirements, see the discussion in 
\citep{FeragenFuster2017, shivaEtal2021},
but the elasticity metric seems to deal particularly well with the afore mentioned
swelling, fattening, and shrinking effects.

\subsection{Generation of Random Kelvin Matrices}  \label{SS:random-Kelvin}
At first assume that $\vx \in \sptldom$ is fixed, i.e.\  consider
the generation of random Kelvin matrices at one location.  The spatial modelling
will be described later, so the argument $\vx$ will be suppressed.  
Similarly to the situation in \fsec{SS:Lie-Kelvin-represent}
when introducing $\tH = \log \eTns$, where a `reference matrix' $\whsymb{\eTns}\in \Syp(\Strspc)$ 
was used in \feq{eq:log-eTnsscl}, this concept is useful also here.  
One may actually think of the
reference matrix $\whsymb{\eTns}$ as an appropriate mean or expected value.  
It will be pointed out under what conditions this holds.  Start with the Lie product group
decomposition of the deterministic reference matrix $\whsymb{\eTns}$ as in \feq{eq:Lie-decomp}:
\begin{align}   \label{eq:Lie-ref-dec}
   \whsymb{\eTns} &= \tQ_0^\trpos \tV_0 \tLbd_0 \tV_0^\trpos \tQ_0, \quad \text{ with } \quad  
   (\tQ_0,\tV_0) = (\exp\tR_0,\exp\tP_0), \\  \label{eq:exp-Kelvin-ch-wh}
   \rpal \ni (\tR_0,\tP_0,\tM_0) &\mapsto (\tQ_0,\tV_0,\tM_0) \mapsto
  \whsymb{\tH} := \tQ_0^\trpos \tV_0 \tM_0 \tV_0^\trpos \tQ_0 = \log \whsymb{\eTns} \in \sy(\Strspc);
\end{align}
from where $\whsymb{\eTns} = \exp\whsymb{\tH}\in \Syp(\Strspc)$ can be recovered as in 
\feq{eq:exp-Kelvin-chain2}.  Here one can then choose an elasticity invariance class
according what is appropriate for the mean as indicated in \fsec{SS:group-inv}, 
i.e.\ invariance under $\syclsm$ resp.\ $\syclSm$.

How to pick a specific elasticity invariance class either for the mean --- i.e.\ $\syclsm$ 
resp.\ $\syclSm$ --- or for the random part --- i.e.\ $\sycls$ resp.\ $\syclS$ --- will
be given in detail for each class in the next \fsec{S:sepc_decomp}.

As a second step, in \feq{eq:exp-Kelvin-chain2} each component $(\tQ,\tV,\tLbd) \in \sS$
is written as a product
\begin{alignat}{2}   \label{eq:prod-Q}
   \tQ(\evt) &=  \tQ_1(\evt)\tQ_0,\quad&\tQ_1(\evt)&=\exp(\theta(\evt)\tR_1(\evt)); \\  
   \label{eq:prod-V}
   \tV(\evt) &= \tV_1(\evt)\tV_0,\quad&\tV_1(\evt)&=\exp(\tP_1(\evt)); \\  \label{eq:prod-Lbd}
   \tLbd(\evt) &=  \tLbd_1(\evt)\tLbd_0,\quad&\tLbd_1(\evt)&=\exp(\tM_1(\evt)) , 
\end{alignat}
computed from random elements $(\theta(\evt)\tR_1(\evt),\tP_1(\evt),\tM_1(\evt)) \in \rpal$.  The random
variable equivalent to \feq{eq:Lie-decomp} resp.\ \feqs{eq:exp-Kelvin-chain}{eq:exp-Kelvin-chain2} 
then is
\begin{multline}  \label{eq:Lie-rand-dec}
   \eTns(\evt) = \tQ^\trpos(\evt) \tV(\evt) \tLbd(\evt) \tV^\trpos(\evt) \tQ(\evt) \\ =
   (\tQ_1(\evt)\tQ_0)^\trpos (\tV_1(\evt)\tV_0) (\tLbd_0\tLbd_1(\evt)) 
   (\tV_1(\evt)\tV_0)^\trpos (\tQ_1(\evt)\tQ_0),
\end{multline}
where $\tQ_0 = \Trep(\vQ_0) \in \ST$ and $\tQ_1(\evt) = \Trep(\vQ_1(\evt)) \in \ST$ 
originate from spatial rotations $\vQ_0, \vQ_1(\evt) \in \SO(\Physpc)$.

\paragraph{Random rotations in physical space $\Physpc$:}
The purpose of the rotation in \feq{eq:prod-Q}, $\tQ(\evt) = \Trep(\vQ) \in \ST$,
resulting from a spatial rotation $\vQ \in \SO(\Physpc)$, 
is to transforms the desired final $\eTns \in \Syp(\Strspc)$ to 
the special form $\rsymb{\eTns} \in \Syp(\Strspc)$ --- that is why in \feq{eq:Lie-rand-dec}
its inverse $\tQ^{-1}(\evt)= \tQ^\trpos(\evt)$ is used for the back-transformation. 
The special forms to be addressed below for each elasticity class  have
at least one distinguished rotation axis --- except for tri-clinic material, which has
none, and isotropic material, which does not need a spatial rotation --- 
which as is usual taken to be the 3-axis.
Additionally, there may be other distinguished rotation axes in the plane orthogonal to
the 3-axis, i.e.\ the 1-2-plane.  In $\mrm{O}(\Physpc)$ there also would be reflections,
but as here we confine ourselves to proper rotations $\vQ \in \SO(\Physpc)$, reflections are
not considered any further.

As recalled in Appendix~\ref{SS:A-rot-2-3}, a rotation $\vQ\in \SO(\Physpc)$ can be 
represented by a normalised skew matrix $\vR \in \so(\Physpc)$ and a rotation angle 
$\theta$ through the exponential map given by the Rodrigues formulas, 
\feq{eq:rodrigues} in 3D and \feq{eq:exp-Q2} in 2D.

In 2D the skew matrix $\vR$ is constant, cf.\ \feq{eq:rodrigues-P2}, whereas in 3D
it is conveniently specified by a vector $\vq \in \D{R}^3, \vq = \theta\,\vr$.  Here
$\vr$ is a unit vector, and the length of $\vq$ is equal to the absolute value of $\theta$,
and from that one obtains $\vR = \skwr(\vr)$ (cf.\ \feq{eq:rodrigues-R}).

In 2D one may take the parameter $q = \theta$ (a 1D vector), and in 3D the 
3D-vector $\vq \in \D{R}^3$ ($\vq = \theta\,\vr$).  
Thus the deterministic part $\vQ_0$ of $\vQ(\evt) = \vQ_1(\evt)\vQ_0$
is always described by a deterministic vector $\vq_0\in \D{R}^{m_Q}$, $m_Q \le \dim\so(\Physpc)$ 
(so in 2D $\dim\so(\Physpc) =1$ and in 3D $\dim\so(\Physpc) =3$),
and the random part $\vQ_1(\evt)$ through a random vector $\vq_1(\evt)\in \D{R}^{m_Q}$.
Finally one obtains a correspondence for \feq{eq:Lie-rand-dec}, 
\begin{align}   \label{eq:q0-Q0-map}
 \D{R}^{m_Q} \ni \vq_0 &\longmapsto \tQ_0 = \Trep(\vQ_0) \in \ST, \\  \label{eq:q1-Q1-map}
 \D{R}^{m_Q} \ni \vq_1(\evt) &\longmapsto \tQ_1(\evt) = \Trep(\vQ_1(\evt)) \in \ST,
\end{align}
check Appendix~\ref{SS:A-rot-Strspc} for further details.


One way to think of this combined rotation --- a concept which is employed also for the 
factors $\tV(\evt)$ and $\tLbd(\evt)$ in \feq{eq:Lie-rand-dec} --- is that $\tQ_0 \in\ST$
does the `main part', i.e.\ it transforms the block diagonal form $\rsymb{\eTns}$
into the average spatial orientation.  The additional factor 
$\tQ_1(\evt) = \Trep(\vQ_1(\evt)) \in \ST$ introduces a random component on top of this.
This will be achieved if the expected value of this random rotation $\vQ_1(\evt)$ --- for the 
$\vtheta_R$-metric on $\SO(\Physpc)$ in \feq{eq:SO-dist} --- is the identity.
The total rotation then has mean $\vQ_0$.  Thus, in light of \feqs{eq:q0-Q0-map}{eq:q1-Q1-map},
one may in effect say that the random orientation can be written as 
$\tQ(\evt) = \tQ_1(\vq_1(\evt)) \tQ_0(\vq_0)$.

The identity $\vI = \exp(\theta \vR)$ corresponds to $\theta = 0$, i.e.\ to $\vq = \vek{0}$.
Thus, for a random realisation, generate a random
spatial rotation vector $\vq_1(\evt) \in \D{R}^m$ with mean at the origin.
Note that from \feqs{eq:SO-dist}{eq:prod-Q} one gleans that the $\vtheta_R$-distance 
(cf.\ Appendix~\ref{S:appx-metric})  between $\vQ_1(\evt)\vQ_0$ and $\vQ_0$ is
\begin{equation}   \label{eq:dist-Qs-evt}
   \vtheta_R(\vQ_1(\evt)\vQ_0,\vQ_0) = \nd{\log (\vQ_1(\evt)\vQ_0 \vQ_0^{-1})}_F =
     \nd{\log \vQ_1(\evt)}_F = \nd{\theta(\evt)\vR_1(\evt)}_F .
\end{equation}
As this shows,
one possibility when this will make the corresponding $\vQ_1(\evt)$ have mean $\vI$
in the $\vtheta_R$ metric is when $\theta(\evt)\vR_1(\evt) = \skwr(\vq_1(\evt))$ 
resp.\ $\vq_1(\evt)$ has a distribution which is symmetric about the origin.
Observe that if in addition to the mean (the origin) also the variance of the  
vector $\vq_1(\evt) \in\D{R}^{m_Q}$ is specified, the \emph{maximum entropy distribution}
(e.g.\ \citep{raoJammaSengup2001}) for $\vq_1(\evt)$ is the centred normal or 
Gaussian distribution, which obviously is symmetric to its mean, the origin.

\paragraph{Random transformations in strain space $\C{T}$:} 
This is in many ways very similar to random rotations in physical space as just described,
only that $\tV(\evt) = \tV_1(\evt)\tV_0$ with $\tV_1(\evt)=\exp(\tP_1(\evt))$ in \feq{eq:prod-V}
is in $\SO(\Strspc)$, and the randomness enters here in a slightly different way.

As a general spatial rotation $\tQ = \Trep(\vQ) \in\ST \subset\SO(\Strspc)$ with 
$\vQ \in \SO(\Physpc)$ can not diagonalise the Kelvin matrix $\eTns$ for any elasticity 
invariance class, a further rotation $\tV \in\SO(\Strspc)$ in strain representer
space $\Strspc$ is needed for this.  Thus these orthogonal $\tV_0 \exp \tP_0, 
\tV_1(\evt) = \exp \tP_1(\evt)$,
described by $\tP_0, \tP_1(\evt) \in \st^\perp \subset \so(\Strspc)$,
are part of the material description, i.e.\ the definition of the block diagonal
group-reduced form of the Kelvin matrix $\rsymb{\eTns}$, as its columns specify 
the Kelvin eigen-strain representers resp.\ the eigenvectors of the elasticity matrix
in the invariant subspace that was found through the reduction of the group representation.
So, although it is an abstract rotation one is talking about, this is not a rotation
in physical space.

One chooses again, analogous as before for the spatial rotation, a deterministic 
$\tP_0  \in \st^\perp \subset \so(\Strspc)$ to define the reference 
or mean eigen-strain, again defined by a parameter vector $\vp_0 \in \D{R}^{m_V}$,
where $m_V \le \dim \st^\perp$,
and a random $\tP_1(\evt)  \in \st^\perp$, also  defined by a parameter 
vector $\vp_1(\evt) \in \D{R}^{m_V}$.
As previously for the spatial rotation in \feqs{eq:q0-Q0-map}{eq:q1-Q1-map}, one obtains
\begin{align}   \label{eq:p0-V0-map}
 \D{R}^{m_V} \ni \vp_0 &\mapsto \tP_0 = \skwr(\vp_0) \in \so(\Strspc) \mapsto 
     \tV_0 = \exp \vP_0 \in \SO(\Strspc), \\  \label{eq:p1-V1-map}
 \D{R}^{m_V} \ni \vp_1(\evt) &\mapsto \tP_1(\evt) = \skwr(\vp_1(\evt)) \in \so(\Strspc) \mapsto
    \tV_1(\evt) = \exp \vP_1(\evt) \in \SO(\Strspc).
\end{align}
Thus again the final result may be written as $\tV(\evt) = \tV_1(\vp_1(\evt)) \tV_0(\vp_0)$.
Details for these versions of the map $\vp_0 \mapsto \tV_0$, as well as $\skwr$ will be
given for each elasticity class in \fsec{S:sepc_decomp}.  

In case $\vp_1(\evt)$ has a
distribution which is symmetric about the origin, and as for spatial rotations the
$\vtheta_R$-distance (cf.\ Appendix~\ref{S:appx-metric}) is used on the 
Lie product factor $\SO(\Strspc)$, similar considerations as before show that then
$\tV_1(\evt)$ has the $\vtheta_R$-mean equal to the identity $\tI$ and thus
$\tV_0$ is the $\vtheta_R$-mean of $\tV(\evt) = \tV_0 \tV_1(\evt)$.  
And again, in case that additionally also the variance of the  
vector $\vp_1(\evt) \in\D{R}^{m_V}$ is specified, the \emph{maximum entropy distribution}
(e.g.\ \citep{raoJammaSengup2001}) for $\vp_1(\evt)$ is the centred normal or 
Gaussian distribution.

\paragraph{Random Kelvin moduli:}
The diagonal eigenvalue matrix we want to split according to \feq{eq:prod-Lbd}, similarly as
for the spatial rotations and the rotations in strain space $\Strspc$, i.e.\
$\tLbd(\evt) = \tLbd_1(\evt) \tLbd_0 \in \Dgp(\Strspc)$.  One may think of  $\tLbd_0$ as 
the deterministic or mean part, and  $\tLbd_1(\evt)$ as random with mean equal to unity.  
Translating this into logarithmic space, one obtains a modelling on $\dg(\Strspc)$ by
taking $\tM_0 \in \dg(\Strspc)$ as fixed such that $\tLbd_0 = \exp \tM_0$ and 
$\tM_1(\evt) \in \dg(\Strspc)$ as a zero mean random part such that
\begin{equation}  \label{eq:log-stiff}
  \tLbd(\evt) = \tLbd_1(\evt) \tLbd_0 = (\exp \tM_1(\evt))(\exp \tM_0) = \exp(\tM_1(\evt)+\tM_0),
\end{equation}
the last equality being true as $\Dgp(\Strspc)$ is an Abelian Lie group, and therefore
$\tLbd_0$ and $\tLbd_1$ commute.

One may point out that in case the distribution of $\tM_1(\evt)$ is symmetric to
the origin, then the expected value of $\tLbd_1(\evt)$ according to the logarithmic distance 
$\vtheta_L$ (see \feq{eq:Diag-dist} in Appendix~\ref{S:appx-metric}) will be the unit matrix $\tI$.  
And additionally it may be observed that the \emph{maximum entropy distribution} 
for given mean and variance on $\tM_1(\evt)$ is the centred
the Gaussian or normal distribution.  

To obtain a concise representation in real parameters, one chooses a deterministic
vector $\vek{\mu}_0 = [\mu_{0,1},\dots,\mu_{0,m_{\Lambda}}] \in \D{R}^{m_\Lambda}$, 
where $m_\Lambda$ is the number of different log-eigenvalues,
of the diagonal of $\tM_0 = \diag_{\#} \vek{\mu}_0$, and the subscript \# means that the
entries of $\vek{\mu}_0$ are distributed at the right positions of the diagonal.
Additionally, the zero-mean random part $\tM_1(\evt)$ is parametrised by
a random vector $\vek{\mu}_1(\evt) \in \D{R}^{m_{\Lambda}}$,  which is then used in 
$\tM_1(\evt) = \diag_{\#}\vek{\mu}_1(\evt)$.
Finally one obtains (cf.\ \refeq{eq:log-stiff})
\begin{align}  \label{eq:gen-Kelvin}
  \tM(\evt) &=  \tM_1(\vek{\mu}_1(\evt)) + \tM_0(\vek{\mu}_0);\; \text{ and }
  \\   \label{eq:gen-Kelvin2}
  \tLbd(\evt) &=  \tLbd_1(\vek{\mu}_1(\evt))  \tLbd_0(\vek{\mu}_0) =
  \exp \tM(\evt) =  \diag_{\#}(\ee^{(\mu_{1,k}(\evt) + \mu_{0,k})}).
\end{align}

Sometimes one wants a guarantee that for each realisation $\evt \in \smplspc$ one has
$\lambda_{1,k}(\evt) > \lambda_{1,k+1}(\evt)$.
This can be achieved by not generating $\lambda_{1,k}$ and $\lambda_{1,k+1}$ through their logarithms
as just explained, but rather to realise that $\lambda_{1,k} = \rho_k + \lambda_{1,k+1}$, where
$\rho_k$ has to be positive.  So one generates the free variables $\mu_{1,k+1}$ --- the logarithm
of $\lambda_{1,k+1}$ --- and $\tau_k$ --- the logarithm of the difference 
$\lambda_{1,k}-\lambda_{1,k+1}$ --- and sets $\lambda_{1,k+1} = \ee^{\mu_{1,k+1}}$, 
$\rho_k = \ee^{\tau_k}$, and then $\lambda_{1,k} = \lambda_{1,k+1} + \rho_k$.  
As in every realisation one has that
$\rho_k > 0$, this guarantees that $\lambda_{1,k}(\evt) > \lambda_{1,k+1}(\evt)$ for
each realisation $\evt \in \smplspc$.

\paragraph{Probability distributions of the parameters:}
As was described in the previous sections, the random parameters $\vq_1, \vp_1, \vek{\mu}_1$
determine the stochastic behaviour of the generated Kelvin matrices $\eTns$.  Therefore it should be
mentioned what influence the character of the probability distributions for those random
parameters has on the solvability of e.g.\ the elliptic boundary value problem.
 
In the normal $\Hp^1 - \Lp_2$ setting, as will be described in a bit more detail further
below, the Kelvin matrix $\eTns$ and its inverse has to be essentially bounded independent
of spatial position or random event for the boundary value problem to be well-posed. 

The parameters $\vq_1$ and $\vp_1$ are mapped through the Lie algebra $\so(\Physpc)$ 
resp.\ $\so(\Strspc)$ into the compact Lie groups $\SO(\Physpc)$ resp.\ $\SO(\Strspc)$, 
and thus have no influence on the question of boundedness.
If their distributions are unbounded, they are wrapped on the compact Lie group to 
produce so-called circular statistics, e.g.\ see \citep{jammalamadaka_topics_2001} 
--- just think of the 2D case, where the Lie algebra $\so(2)$ is the real line, which 
is mapped via $\alpha \mapsto (\cos(\alpha), \sin(\alpha))$ onto the circle and 
in this way onto the corresponding rotation matrix.  
Thus the mapped probability distributions of the rotation matrices $\tQ$ and $\tV$ 
are always bounded, and hence can not cause the Kelvin matrix $\eTns$ or 
its inverse to be unbounded.

This is different for the eigenvalues, the Kelvin moduli or their logarithms $\vek{\mu}_1$.  
In case the distribution of $\vek{\mu}_1$ is bounded, so are the distributions of
$\eTns$ or its inverse.  But in case it is only known that the $\vek{\mu}_1$ have finite
variance, the maximum entropy distribution \citep{Jaynes-2003} is  
Gaussian, and hence unbounded.  Thus it is also unbounded for $\eTns$ or 
its inverse.  Further below this will be discussed further.

\rvsi{
In \fsec{SS:rand-fields} it was advocated to model random fields in a separated fashion like
in \feq{eq:field-sep-rep}, as this separates the probabilistic content from the spatial variation
in a tensor-product like manner.  Given the spatial functions, the attached random variables
will be given through some kind of projection like \feq{eq:KLE-RV-proj}.  One possibility
how to express these random variables through a polynomial chaos expansion in
independent standard Gaussian random variables was already
sketched at the end of \fsec{SS:rand-fields}.  This can be taken as one example of the so-called
NORTA (NORmal To Anything) approach, cf.\ e.g.\ \citep{chenNORTA2001, GhoshHendNORTA03}, where
groups of random variables with known correlation and marginal distributions are determined
as functions of standard normal or Gaussian random variables.
}

\rvsi{
Especially in the modelling of the Kelvin moduli, observe that what is proposed is to model
the log-moduli, and then essentially take the exponential, thereby guaranteeing positivity.
This means that the point-wise exponential transforms the probability distribution of the
log-moduli into the one of the Kelvin moduli themselves.  If, as a realistic example, the
log-moduli $\mu$ were modelled as a Gaussian field, the Kelvin moduli $\lambda = \exp \mu$ 
would be log-normal.  One may interpret the exponential here also in a different way:  Let
$F_g$ be the Gaussian cumulative distribution function of the log-moduli $\mu$, and $F_\ell$ the
log-normal cumulative distribution function of the Kelvin moduli $\lambda$, 
then $F_\ell^{-1} \circ F_g = \exp$.  Now it easy to see, that if some other probability distribution
of the Kelvin moduli $\lambda$ were desired, say a probability distribution with
cumulative distribution function $F_o$, then one could simply take the point transformation
$\lambda = F_o^{-1} \circ F_g(\mu)$.  The thing to watch out for in this case is not to
make any approximations for $F_o^{-1} \circ F_g$ which could numerically destroy positivity.
}

\paragraph{Spatial random parametrisation:}
Putting everything together, one may now define with $n = m_Q + m_V + m_{\Lambda}$ a 
parameter vector space $\Rn \equiv \D{R}^{m_Q} \times \D{R}^{m_V} \times \D{R}^{m_\Lambda}$ and
in it a deterministic vector $\meansymb{\vz} := (\vq_0, \vp_0, \vek{\mu}_0)$ to define the deterministic
(cf.\ \feqs{eq:exp-Kelvin-chain}{eq:exp-Kelvin-chain2}) part 
$(\tR_0(\vq_0),\tP_0(\vp_0),\tM_0(\vek{\mu}_0)) \in \rpal$, and also the zero mean random vector
\begin{align}  \label{eq:rand-vec-param}
  &\vtil{z}(\evt) := (\vq_1(\evt), \vp_1(\evt), \vek{\mu}_1(\evt)), \quad \text{ to define } \\
  &(\tR_1(\vq_1(\evt)), \tP_1(\vp_1(\evt)), \tM_1(\vek{\mu}_1(\evt))) \in \rpal.
\end{align}
Now we set $\vz(\evt) := \meansymb{\vz} + \vtil{z}(\evt)$, and following 
\feeqs{eq:prod-Q}{eq:Lie-rand-dec}
one thus arrives at the construction of the desired random Kelvin matrices 
$\eTns(\evt) = \eTns(\vz(\evt))$ at any spatial point $\vx \in \sptldom$.

To extend this to the construction of random spatial fields of Kelvin matrices, one may
recall \fsec{SS:rand-fields}.  Thus to extend the above modelling to random fields,
one observes that the above description of a parameter vector has to be extended to spatial
parameter vector fields.  Thus, by defining a deterministic spatial field
$\meansymb{\vz}(\vx) = (\vq_0(\vx), \vp_0(\vx), \vek{\mu}_0(\vx))\in \Rn$ for 
$\vx \in \sptldom$,  and a zero mean random field
$\vtil{z}(\vx,\evt) = (\vq_1(\vx,\evt), \vp_1(\vx,\evt), \vek{\mu}_1(\vx,\evt))\in \Rn$, one has
everything that is needed to define random fields of Kelvin matrices with all the desired
properties.

The random parameter field $\vz(\vx,\evt)$ is up to stochastic second order described by its
mean $\meansymb{\vz}(\vx)$ and covariance $\cov_z(\vx,\vy)\in \D{R}^{n\times n}$, 
see \feqs{eq:mean-param-field}{eq:cov-param-field} in \fsec{SS:rand-fields}.  
The covariance now takes the more detailed form of a joint correlation matrix function:
\begin{multline}  \label{eq:JointCorr}
  \cov_z(\vx,\vy)  =  \EXP{
    (\vq_1(\vx,\cdot), \vp_1(\vx,\cdot), \vek{\mu}_1(\vx,\cdot)) \otimes  
    (\vq_1(\vy,\cdot), \vp_1(\vy,\cdot), \vek{\mu}_1(\vy,\cdot)) } \\ 
  =  \EXP{  \begin{matrix} 
     \vq_1(\vx,\cdot) \otimes \vq_1(\vy,\cdot) & \vq_1(\vx,\cdot) \otimes \vp_1(\vy,\cdot) 
      & \vq_1(\vx,\cdot) \otimes \vek{\mu}_1(\vy,\cdot) \\
     \vp_1(\vx,\cdot) \otimes \vq_1(\vy,\cdot) & \vp_1(\vx,\cdot) \otimes \vp_1(\vy,\cdot) 
      & \vp_1(\vx,\cdot) \otimes \vek{\mu}_1(\vy,\cdot) \\
     \vek{\mu}_1(\vx,\cdot) \otimes \vq_1(\vy,\cdot) & \vek{\mu}_1(\vx,\cdot) \otimes \vp_1(\vy,\cdot) 
      & \vek{\mu}_1(\vx,\cdot) \otimes \vek{\mu}_1(\vy,\cdot)  
     \end{matrix}  }.
\end{multline}
This form shows that it contains all the correlation structures between spatial 
orientation, encoded in $\vq_1$,
eigen-strain distribution, encoded in $\vp_1$, and stiffness, encoded in $\vek{\mu}_1$.

Heterogeneous materials and their statistical description
may be found in \citep{torquato:2001} and references therein.  
The family of Matérn covariance functions
(e.g.\ \citep{Matern_1986, Cressie1993, GneitingEtAl2010}) are an often used family 
of covariance functions in this connection.  They allow a separate control of
smoothness and correlation length of the random field, as well as
efficient approximation  via low-rank tensor methods \citep{Litvinenko_Keyes2019}, which
is important in the solution of the Fredholm integral equation \feq{eq:KLE-eigprob}
and field expansion \feq{eq:field-sep-rep}.
\rvsi{This kind of covariance function is used to describe stochastically homogeneous random 
fields as already described at the end of \fsec{SS:rand-fields}, where the \KL-eigenproblem
\feq{eq:KLE-eigprob} becomes a convolution equation.  This means that the --- possibly generalised
--- eigenfunctions are the Fourier functions.  Given this, the spectrum of the operator
in \feq{eq:KLE-eigprob} is computed via Fourier transform of the covariance function,
 cf.\ \citep{hgm07}.
In addition to the versatility of this family of covariance functions this (continuous) spectrum 
in the eigenproblem \feq{eq:KLE-eigprob} is analytically known.}  
Unfortunately, neither this covariance nor its 
spectrum are compactly supported --- an indication that the integral operator in 
\feq{eq:KLE-eigprob} is not compact.  One example of a covariance function with compact support is
given in \citep{Gneiting2002}.  This renders the integral operator in \feq{eq:KLE-eigprob}
compact, hence it has a discrete spectrum and a proper \KL{} series expansion
\feq{eq:field-sep-rep} --- whereas in the case of a continuous spectrum this becomes an integral.

\paragraph{Boundary value problems:}
Let us return to the question of regularity, resp.\ boundedness required from the
field of Kelvin matrices $\eTns(\vx)$.  The solution space for an elasticity boundary 
value problem with a deterministic field $\eTns(\vx)$ as coefficients for the elastic displacement
field $\vu(\vx)$ would in the simplest instance be a closed subspace of the Hilbert-Sobolev 
space $\Hp^1(\sptldom)$, incorporating the essential boundary conditions, 
say for simplicity $\vu(\vx) \in \Ho^1(\sptldom)$, i.e.\ $\vu(\vx)$
vanishes on the boundary $\partial \sptldom$.  The right-hand side of distributed loads
then typically has to be in the dual space $\Hp^{-1}(\sptldom)$.
Thus the Kelvin matrix $\eTns(\vx)$ as coefficient field has to be in $\Lp_\infty(\sptldom)$, 
i.e.\ essentially bounded, in order to have a well-posed problem, 
e.g.\ \citep{Ciarlet1978, hgm07, OdenDemko2010, OdenReddy2012, Ciarlet2013}.   
For that, one would also want the inverse of the Kelvin matrix, $\eTns^{-1}(\vx)$,
to be essentially bounded.  This is well known standard elliptic theory,
e.g.\ \citep{Ciarlet1978, segalKunze78, yosida-fa-1980, 
OdenDemko2010, OdenReddy2012, Ciarlet2013}.

Considering the stochastic extension of the boundary value problem with a random 
Kelvin matrix $\eTns(\vx,\evt)$, one approach is to require finite second moments 
for the random solution displacement field $\vu(\vx,\evt)$, i.e.\ to
be in the tensor product Hilbert space $\Ho^1(\sptldom) \otimes \Lp_2(\smplspc)$,
and the right-hand side of random distributed loads to be in its dual space
$\Hp^{-1}(\sptldom) \otimes \Lp_2(\smplspc)$.
The formulation of a well-posed problem in this case requires
the random Kelvin matrix $\eTns(\vx,\evt) \in 
\Lp_\infty(\sptldom \times \smplspc)$ to be essentially bounded, as well as its inverse,
e.g.\ see \citep{Matthies2005, soize_2006, BabuskaNobileTemp07, hgm07, Matthies2008, 
LordPowell2014, DungSchwabX2022}. 

This may be regarded as the standard case, and, as mentioned above, it requires the 
probability densities of the log-eigenvalues $\vek{\mu}_1$ to be essentially bounded.
One possibility is the quite versatile beta distribution, see e.g.\ \citep{Grigoriu2016}.
But, as already mentioned, the maximum entropy distribution for the
log-eigenvalues $\vek{\mu}_1$ with known finite second moments is the normal
or Gaussian distribution \citep{Jaynes-2003}, making this a highly desired choice
\citep{schwartzman_lognormal_2016}.
In this case the eigenvalues and thus the Kelvin matrix $\eTns(\vx,\evt)$ have a 
lognormal distribution and are thus unbounded.  It is possible
to proceed also in this case, with different techniques, which either make the solution
space smaller, or change the solution concept slightly; see e.g.\
\citep{EspigHackbuschEtAl2014, LordPowell2014, HoangSchwab2014,
BachmayrCohenDeVoreEtAl2017, BachmayrCohenMigliorati2018, HerrmannSchwab2019, DungSchwabX2022}
for more details.

The numerical discretisation and solution of such stochastic boundary value problems resp.\
stochastic partial differential equations (SPDEs) is beyond the scope of this paper, 
and the interested reader
may find more material in e.g.\ \cite{Matthies2005, BabuskaNobileTemp07, hgm07, Matthies2008, 
grigoriu2012, EspigHackbuschEtAl2014, HoangSchwab2014, LordPowell2014, HerrmannSchwab2019, 
DungSchwabX2022} and the references therein.
  The numerical solution of such SPDEs is
also necessary when one wants to identify the properties of the tensor field from measurements
resp.\ observations.

\paragraph{Identification of the parameters:}
\rvsi{The tasks of identification, calibration, and conditioning have already been briefly
touched upon at the end of \fsec{SS:rand-fields}.  Mathematically speaking, these are all
so-called inverse problems.  In our view, general techniques for such identification 
resp.\ inverse problems are the Bayesian methods, see e.g.\
\citep{mosegaard:2002:IP, tarantola:2005, marzouk:2007:JCP, marzouk:2009:CompPhys, AMStuart2010, 
RosicLitvinenkoEtAl2012, PajonkRosicEtAl2012, RosicKucerovaSykoraEtAl2013, 
MatthiesZanderEtAl2016, MatthiesZanderRosicEtAl2016, RosicSykoraPajonkEtAl2016, hgm18}
and the references therein.  These probabilistic methods may be used for many purposes,
e.g.\ for anisotropic materials \citep{FittEtAl2019, SavvasPapai2020}.  But the detailed description
of the} identification  of the symmetry classes of materials and of the numerical values
of the properties, i.e.\ the eigenvalues and orientations, is a topic beyond the scope 
of this work, so we offer only some remarks.
The identification of symmetry classes may be based e.g.\ on the knowledge of the
crystal structure of the material \citep{nye_physical_1984}, or on its fabrication
\citep{Cowin2013}, or of its natural genesis.  
Otherwise one must rely on measurements or observations.  For the identification
of elasticity classes, see \citep{CowinMehr1989, FrancoisEtAl1998, 
DanekEtAl2015, DanekSlawinski2015, GierlachDanek2018}.
For the kind of stochastic modelling proposed here, some pointers are given in 
section 4.2 of \citep{Grigoriu2016}.

With  informative measurements or tests
quite complicated material behaviour can be identified and calibrated, see e.g.\
\citep{RosicKucerovaSykoraEtAl2013, MatthiesZanderEtAl2016, MatthiesZanderRosicEtAl2016, RosicSykoraPajonkEtAl2016}, and also e.g.\
\citep{DobrillaEtAl2023, AIbrahimbEtAl_SN-2022, AI-HGM-EK-2020}
for highly nonlinear and irreversible behaviour.
Whether the tests or experiments are informative may be tested computationally beforehand.
Thus the test or measurement and updating may be performed first virtually in the computer,
and in this way help in the design of informative experiments.



%
%
%
%
%
%
%
%
%
%
%
%
%
%


%
\section{Detailed Lie Representation} \label{S:sepc_decomp}

%

%
%
%
%
%
%

According to \feq{eq:rand-vec-param} in \fsec{SS:random-Kelvin}, the random field of Kelvin
matrices can be described by a $n$-dimensional random field of parameters $\vz \in \Rn$,
where $n = m_Q + m_V + m_{\Lambda}$, and 
$\vz = (\vq, \vp, \vek{\mu})$ is such that $\vq \in \D{R}^{m_Q}$ describes the
spatial rotations, $\vp \in \D{R}^{m_V}$ describes the eigen-strain distribution, and
$\vek{\mu} \in \D{R}^{m_\Lambda}$ describes the eigenvalues or Kelvin moduli.
These parameters will now be given for all the elasticity classes.  As this is simpler in 2D,
this is where we start in \fsec{SS:el_cl_2D}, and then move to the 3D case in \fsec{SS:el_cl_3D}.
As we will accept possible `wrapping' in the parametrisation of rotations, for the sake of
simplicity it is refrained from giving the formal group-theoretic descriptions of the 
material symmetry resp.\ invariance groups.  This will be described only verbally by 
pointing out familiar geometrical objects with the same symmetry.

Here the special form of the Kelvin matrix $\rsymb{\eTns}$ (see the development 
\feqs{eq:group-dec}{eq:SpecDecSPD} in \fsec{SS:group-inv}) 
will be given in its group reduced form, i.e.\ reduced to block-diagonal form.
In cases where one can not achieve two or more diagonal blocks --- e.g.\ for 
tri-clinic materials --- one at least wants to have a form which has as many 
zeros in $\rsymb{\eTns}$ as possible.  
The special form $\rsymb{\eTns}$ can then be spatially rotated --- the reverse of 
\feq{eq:group-dec} --- to the final form $\eTns = \tQ^\trpos \rsymb{\eTns} \tQ$,
this is all contained in the relations \feq{eq:easy-ref} at the beginning of \fsec{S:modelling}.
How to build the orthogonal $\tQ \in \SO(\Strspc)$ from a spatial rotation 
$\vQ \in \SO(\Physpc)$ is recalled for the sake of completeness from the literature
in Appendix~\ref{S:appx-Rotations}.

\subsection{Elasticity classes in 2D}  \label{SS:el_cl_2D}
In 2D there are four symmetry classes \citep{blin1996}, to be shortly and informally described 
by pointing out a geometric figure with the same symmetry.    The Hasse diagram  --- drawn 
horizontally in \ffig{F:Hasse2} --- shows the relation of the lattice of elasticity symmetry or 
invariance groups as subgroups of the full orthogonal group, where the hooked arrow means 
a subgroup relation, e.g.\ the symmetry group of the orthotropic class is a subgroup of the
symmetry group of the tetragonal class.  The largest group is on the left,
the \emph{isotropic} class with the symmetries of a circle (i.e.\ the whole 2D rotation group), 
then the next smaller one is the \emph{tetragonal} class with the symmetry of a square, then
the \emph{orthotropic} class with the symmetry of a general rectangle, and finally the 
\emph{tri-clinic} class with the symmetry of a parallelogram.

\begin{figure}[htb!]
\begin{tikzcd}  
      \text{\phantom{tri}} &
      \text{[isotropic]}  & \arrow[l, hookrightarrow] \text{[tetragonal]} & \arrow[l, hookrightarrow] 
     \text{[orthotropic]} & \arrow[l, hookrightarrow] \text{[tri-clinic]} 
\end{tikzcd}
    \caption{Hasse diagram of symmetry subgroups of 4th order tensors in 2D}  \label{F:Hasse2}
\end{figure}

\paragraph{Tri-clinic material in 2D:}
This section covers not only tri-clinic, but also mono-clinic materials, as there is no
difference in 2D.
As mentioned before, these materials have in 2D the symmetry of a parallelogram.
The Kelvin matrix $\eTns$ can not be simplified through a group reduction with a 
spatial rotation $\tQ \in \ST$
to a block-diagonal form beyond the general form \feq{eq:Kelvin-C-2D-1}, where
one may note that there are $n=6 =\dim\,\SymTen = \dim\,\SymPh$ free parameters.
But, as in 2D one has $\dim\,\ST = \dim\,\st = 1$, one parameter 
$\vq = (\theta) \in \C{Q} \equiv \D{R}^{m_Q}$ ($m_Q = 1$) may be used
for a spatial rotation, see \feqs{eq:nVoigt-orth2D}{eq:q-to-Q-in-23} in 
Appendix~\ref{S:appx-Rotations}, to induce one zero in the upper right corner \citep{cowMehr1995},
this echos the first of the relations in \feq{eq:easy-ref}:
\begin{equation}  \label{eq:Kelvin-C-2D-2}
  \rsymb{\eTns}_{\mrm{tri-2D}}  = {\small \begin{bmatrix} 
   \rsymb{\mrm{c}}_{1111} & \rsymb{\mrm{c}}_{1122} &  0 \\  
             & \rsymb{\mrm{c}}_{2222} &  \sqz\,\rsymb{\mrm{c}}_{2212} \\  
   \mrm{SYM} &  &  \;\;\;2\,\rsymb{\mrm{c}}_{1212}  
   \end{bmatrix} } .
\end{equation}
The spatial rotations $\vQ \in \SO(\Physpc)$ resp.\ $\tQ \in \ST$
decide on whether the spatial orientation of the material axes changes or not.  
Hence one wants to keep this possibility in the stochastic modelling process.

What is needed next is to use the spectral decomposition of $\rsymb{\eTns}$, the second of the
relations in \feq{eq:easy-ref}.
The matrix in \feq{eq:Kelvin-C-2D-2} has generally three distinct positive eigenvalues 
$\lambda_1, \lambda_2$, and $\lambda_3$.  
For the three distinct eigenvalues one thus needs $m_{\Lambda} = 3$ log-eigenvalue
parameters $\D{R}^{m_{\Lambda}} \ni \vek{\mu} = (\mu_1, \mu_2, \mu_3) \mapsto 
\tM = \diag(\vek{\mu})$, such that 
$\tLbd(\vek{\mu}) = \diag(\lambda_1, \lambda_2, \lambda_3) = \exp \tM$.

The Kelvin matrix $\rsymb{\eTns}$ in \feq{eq:Kelvin-C-2D-2} can of course be diagonalised 
as in \feq{eq:group-eig-dec}, but the corresponding matrix of eigenvectors 
$\tV\in\SO(\Strspc)$ is not in $\ST$, i.e.\ not induced by a spatial rotation.

In \fsec{SS:Lie-Kelvin-represent} it was explained how to keep the spatial rotation 
described by $\C{Q} \ni \vq \mapsto \vQ \mapsto \tQ $ --- cf.\ Appendix~\ref{SS:A-rot-Strspc} 
--- from interfering with the `strain distributor' part $\tV \in \SO(\Strspc)$
by introducing the orthogonal split in the corresponding Lie algebra
already mentioned in \feq{eq:prod-Liealg2}: $\so(\Strspc) = \st \oplus \st^\perp$.  
As just mentioned, in 2D one has $\dim\,\st = 1 = m_Q$, and this is needed for the rotation 
angle parameter $\vq = (\theta) \in \C{Q}$ to describe the spatial rotation. 
This means $m_V = n - m_Q - m_{\Lambda} = 6 - 1 - 3 = 2 = \dim\,\st^{\perp}$ angle-like 
parameters $\vp = (s_1, s_2) \in \C{Q}^\perp \equiv \D{R}^{m_V}$ can now be used to describe the 
eigenvectors of the Kelvin matrix in \feq{eq:Kelvin-C-2D-2}, i.e.\ the columns 
of $\tV \in \SO(\Strspc)$ in \feq{eq:group-eig-dec}.

There are at least two ways of generating such a $\tV \in \SO(\Strspc)$: one is to use the 
\feeqs{eq:Q-perp}{eq:s-to-P-in-23} in Appendix~\ref{SS:A-rot-Strspc}, where the
parameters $\vp \in \C{Q}^\perp$ are directly used for a map 
$\C{Q}^\perp \to \st^\perp \to \SO(\Strspc)$ to generate a 
$\tV(\vp) = \tV(s_1, s_2) \in \SO(\Strspc)$.
Another way is to use two 2D rotation matrices
--- cf.\ \feq{eq:2Drotmat} --- embedded in 3D depending on angles $\vp = (s_1, s_2)$:
\begin{equation}  \label{eq:2D-rot-subm-V12}
  \tV_i(s_1) := \begin{bmatrix} 1 & 0 & 0 \\ 0 & \cos s_1 & -\sin s_1  \\
                  0 & \sin s_1 & \phantom{-}\cos s_1  \end{bmatrix};  \qquad
  \tV_{ii}(s_2) := \begin{bmatrix} \cos s_2 & -\sin s_2 & 0  \\
                 \sin s_2 & \phantom{-}\cos s_2 & 0 \\ 0 & 0 & 1  \end{bmatrix}.
\end{equation}
Here the explicit formulation of the skew matrices $\tP_i, \tP_{ii} \in \st$ and the
explicit use of the exponential map in the mapping chain $\vp \mapsto \tP \mapsto \tV$
according to Appendix~\ref{SS:A-rot-2-3} has been shortened by giving just the final result.
Then with $\tV(\vp) = \tV(s_1,s_2) = \tV_i(s_1)\tV_{ii}(s_2)$ it is easily seen 
that (cf.\ \feq{eq:group-eig-dec})
\begin{equation}  \label{eq:Kelvin-C-2D-tri}
\rsymb{\eTns}(\vek{\mu}, \vp) = \tV(\vp) \tLbd(\vek{\mu}) \tV^\trpos(\vp)
\end{equation}
has the form \feq{eq:Kelvin-C-2D-2}, which corresponds to the spectral decomposition as the
second of the relations in \feq{eq:easy-ref}.  In both cases the parameters $\vp = (s_1, s_2)$
are purely a description of the distribution of eigen-strains for the material specification.
Note that the first column of the matrix $\tv_1$ of $\tV = [\tv_1 \tv_2 \tv_3]$ is the
eigenvector or eigen-strain for the eigenvalue $\lambda_1$, $\tv_2$ the eigenvector of $\lambda_2$,
and so on.  If one wants to distinguish the bulk or volumetric response from the shear response,
e.g.\ by assigning the largest of the eigenvalues to this response, then one may
choose the eigenvector $\tv_j$ which has the largest projection onto the volumetric strain
$\tn$ (\feq{eq:n-rep2D} in Appendix~\ref{SS:C-Kelvin}).  By assigning the largest eigenvalue
to that eigenvector --- it was described previously how to generate random eigenvalues where one
is always larger than the others --- one can ascertain that the bulk or volumetric response
is always stiffer than the shear response.

By construction, the form in \feq{eq:Kelvin-C-2D-tri} is invariant under rotations 
in the symmetry group $\syclS = \Trep(\sycls) = [\text{tri-clinic}]$.
Finally, the elasticity matrix for a tri-clinic material in 2D with the proper spatial
orientation $\vq = (\theta)$ may be generated from \feq{eq:Kelvin-C-2D-tri}
in a 2-step rotation procedure via
\begin{align}   \label{eq:gen-C-mat-tri-clin-2D}
  \eTns_{\mrm{tri-2D}}(\vq,\vek{\mu},\vp) &= \tQ^{\trpos}(\vq)\rsymb{\eTns}(\vek{\mu}, \vp)
  \tQ(\vq) ,
  \quad \text{ or even in more detail } \\   \nonumber
  &= \tQ^\trpos_0 \tQ_1^\trpos(\vq) \tV_1(\vp) \tV_0 
  \tLbd_0 \tLbd_1(\vek{\mu}) \tV_0^\trpos \tV_1^\trpos(\vp) \tQ_1(\vq) \tQ_0 .
\end{align}
The last equation in \feq{eq:gen-C-mat-tri-clin-2D} --- the third of the relations
in \feq{eq:easy-ref} --- is the concrete version of
\feq{eq:Lie-rand-dec} in \fsec{SS:random-Kelvin}.  One may recall that
$\whsymb{\eTns} = \tQ^\trpos_0 \tV_0 \tLbd_0 \tV_0^\trpos \tQ_0$ is the constant reference matrix ---
e.g.\ the scaling-rotation mean --- which may be actually invariant under a larger 
symmetry group $\syclSm = \Trep(\syclsm) \supset \syclS$, i.e.\ further to the left 
in \ffig{F:Hasse2}.
In case one of the eigenvectors is a pure shear, i.e.\ proportional to $\ty$ in \feq{eq:y-rep2D} 
in Appendix~\ref{SS:C-Kelvin}, then the material is in fact \emph{orthotropic}, to be
described next.

\paragraph{Orthotropic material in 2D:}
As already indicated, such materials have the symmetry of a general rectangle in 2D.
By the group reduction process, i.e.\ a spatial rotation $\tQ \in \ST$ given by one parameter 
$\vq = (\theta) \in \C{Q} \equiv \D{R}^{m_Q}$ ($m_Q = 1$), see 
\feqs{eq:nVoigt-orth2D}{eq:q-to-Q-in-23} in Appendix~\ref{S:appx-Rotations},
the Kelvin matrix for these materials may be brought into the following block-diagonal
form \citep{cowMehr1995}, with four free parameters:
\begin{equation}  \label{eq:Kelvin-C-2D-ortho}
  \rsymb{\eTns}_{\mrm{ortho-2D}}  = {\small \begin{bmatrix} 
   \rsymb{\mrm{c}}_{1111} & \rsymb{\mrm{c}}_{1122} &  0 \\  
             & \rsymb{\mrm{c}}_{2222} &  0 \\  
   \mrm{SYM} &  &  \;\;\;2\,\rsymb{\mrm{c}}_{1212}  
   \end{bmatrix} } ,
\end{equation}
which has one  obvious eigen-stiffness or Kelvin shear modulus $\lambda_3=2\,\rsymb{\mrm{c}}_{1212}$ 
with associated eigenvector or eigen-strain distributor
\begin{equation}  \label{eq:v2-ortho2}
   \tv_3 = [0, 0, 1]^{\trpos},
\end{equation}
a simple shear.  The other two come from the diagonalisation of the upper $2 \times 2$
upper left sub-matrix of \feq{eq:Kelvin-C-2D-ortho}, and have the general form of a rotation
of $\tn, \ty$ in \feqs{eq:n-rep2D}{eq:y-rep2D}
about the fixed vector $\tv_3$ in \feq{eq:v2-ortho2}in the strain-representer space $\Strspc$:
\begin{equation}  \label{eq:v1-ortho2}
   \tv_1(s) = [\cos s, \sin s, 0]^{\trpos}\quad\text{ and }\quad
   \tv_2(s) = [-\sin s, \cos s, 0]^{\trpos} ,
\end{equation}
associated with the other two Kelvin moduli $\lambda_1, \lambda_2$.
Observe that the single parameter $\vp = (s)  \in \C{Q}^\perp \equiv \D{R}^{m_V}$ ($m_V = 1$) 
is not an angle in a spatial sense, but is one in the two-dimensional
strain representer subspace  formed by the first two components,
characterising the material, or more precisely, the Kelvin eigen-strains.
Here, as just before, the mapping chain $\vp \mapsto \tP \mapsto \tV$ according to 
Appendix~\ref{SS:A-rot-2-3} has been shortened to display just the final result $\tV(\vp)$.

The previous remarks made for tri-clinic material regarding the
possible choice of the largest Kelvin module to
be associated with the volumetric response apply here as well. 
Again, for the three generally distinct eigenvalues one needs $m_{\Lambda} = 3$ log-eigenvalue
parameters $\D{R}^{m_{\Lambda}} \ni \vek{\mu} = (\mu_1, \mu_2, \mu_3) \mapsto 
\tM = \diag(\vek{\mu})$, such that 
$\tLbd(\vek{\mu}) = \diag(\lambda_1, \lambda_2, \lambda_3) = \exp \tM$.
Additionally, one has with \feqs{eq:v2-ortho2}{eq:v1-ortho2} 
$\tV(\vp) = \tV(s) = [\tv_1(s) \tv_2(s) \tv_3]$, and with that
formally the same kind of spectral representation as in \feq{eq:Kelvin-C-2D-tri}
for $\rsymb{\eTns}_{\mrm{ortho-2D}}(\vek{\mu},\vp)$.
The Kelvin elasticity matrix $\eTns_{\mrm{ortho-2D}}(\vq,\vek{\mu},\vp)$ for 
an orthotropic material in 2D with the proper spatial orientation 
$\vq = (\theta)$ may then again be generated as in \feq{eq:Lie-rand-dec} in \fsec{SS:random-Kelvin},
or in \feq{eq:gen-C-mat-tri-clin-2D}
in a 2-step rotation procedure, see also the remarks there.   
The total number of parameters $(\vq, \vek{\mu}, \vp)$
is $n= m_Q + m_{\Lambda} + m_V = 1 + 3 + 1 = 5$.
In case $s=\uppi/4$ in \feq{eq:v1-ortho2}, i.e.\ in case the first two eigenvectors 
are $\tv_1 = \tn$ and $\tv_2 = \ty$, the material is actually \emph{tetragonal}.

\paragraph{Tetragonal material in 2D :}
As already indicated, such materials have the symmetry of a square in 2D.
Again, by the group reduction process and a spatial rotation given by one parameter 
$\vq = (\theta) \in \C{Q} \equiv \D{R}^{m_Q}$ ($m_Q = 1$),
the Kelvin matrix for these materials may be brought into the following block-diagonal
form \citep{cowMehr1995}, with three free parameters:
\begin{equation}  \label{eq:Kelvin-C-2D-tetra}
  \rsymb{\eTns}_{\mrm{tetra-2D}}  = {\small \begin{bmatrix} 
   \rsymb{\mrm{c}}_{1111} & \rsymb{\mrm{c}}_{1122} &  0 \\  
             & \rsymb{\mrm{c}}_{1111} &  0 \\  
   \mrm{SYM} &  &  \;\;\;2\,\rsymb{\mrm{c}}_{1212}  
   \end{bmatrix} } ,
\end{equation}
completely determining the spectral decomposition, with, as before in the 2D orthotropic case
with the obvious Kelvin shear modulus $\lambda_3=2\,\rsymb{\mrm{c}}_{1212}$ and eigenvector
$\tv_3$ as in \feq{eq:v2-ortho2}, and the two additional generally distinct
Kelvin moduli ($m_{\Lambda}=3$): $\lambda_1 = \rsymb{\mrm{c}}_{1111} + \rsymb{\mrm{c}}_{1122}$ and
$\lambda_2 = \rsymb{\mrm{c}_{1111}} - \rsymb{\mrm{c}}_{1122}$, 
i.e.\ $\lambda_1 = \lambda_2 + 2\, \rsymb{\mrm{c}}_{1122}$.

The associated eigenvectors are fixed ($m_V=0$), 
corresponding to $s = \uppi/4$ in \feq{eq:v1-ortho2}: 
$\tv_1 = \tn$ and $\tv_2 = \ty$, cf.\ \feqs{eq:n-rep2D}{eq:y-rep2D}, hence
$\tV =  [\tv_1 \tv_2 \tv_3]$ is a fixed matrix.  
As is easily observed, $\lambda_1$ is a volumetric Kelvin modulus,
whereas $\lambda_2$ is a shear Kelvin modulus; and the previous remarks made for 
tri-clinic material regarding the possible choice of the largest Kelvin module to
be associated with the volumetric response apply here as well.
 
Here again, for the three generally distinct eigenvalues one needs 
$m_{\Lambda} = 3$ log-eigenvalue parameters $
\vek{\mu} = (\mu_1, \mu_2, \mu_3)$.  With the above fixed $\tV$
one obtains formally the same kind of spectral representation as in \feq{eq:Kelvin-C-2D-tri}
for $\rsymb{\eTns}_{\mrm{tetra-2D}}(\vek{\mu})$ in \feq{eq:Kelvin-C-2D-tetra}.
The tetragonal 2D Kelvin elasticity matrix $\eTns_{\mrm{tetra-2D}}(\vq,\vek{\mu})$ 
with the proper spatial orientation 
$\vq = (\theta)$ may then again be generated as in \feq{eq:Lie-rand-dec} in \fsec{SS:random-Kelvin},
or in \feq{eq:gen-C-mat-tri-clin-2D}
in a 2-step rotation procedure.   The total number of parameters $(\vq, \vek{\mu})$
is $n= m_Q + m_{\Lambda} + m_V = 1 + 3 + 0 = 4$.
In case the two shear Kelvin moduli coincide, i.e.\ $\lambda_2 =\lambda_3$, 
the material is actually \emph{isotropic}.

\paragraph{Isotropic material in 2D:}
As already mentioned, such a material has the symmetry group of a circle in 2D.
This means that no spatial rotation is needed ($m_Q = 0$) to bring the Kelvin matrix to
the block-diagonal form \feq{eq:Kelvin-C-2D-tetra} with $\lambda_3 = 2\,\rsymb{\mrm{c}}_{1212}
= \rsymb{\mrm{c}}_{1111} - \rsymb{\mrm{c}}_{1122} = \lambda_2$, 
the Kelvin matrix has this form in any spatial orientation.  Thus there are only two
($n = m_Q + m_{\Lambda} + m_V = 0+2+0 = 2$) free parameters for the two
log-eigenvalues $\mu_1, \mu_2=\mu_3$.  The spectral decomposition  as in \feq{eq:Kelvin-C-2D-tri}
with the matrix $\tV$ as in the tetragonal case is the final product 
$\rsymb{\eTns}_{\mrm{iso-2D}}$,  no spatial rotations
as in \feq{eq:Lie-rand-dec} in \fsec{SS:random-Kelvin}, or in \feq{eq:gen-C-mat-tri-clin-2D}
are necessary, as the  Kelvin matrix $\rsymb{\eTns}_{\mrm{iso-2D}}$ is invariant
under such rotations.

\subsection{Elasticity classes in 3D}  \label{SS:el_cl_3D}
\begin{figure}[htb!]
\begin{center}
\begin{tikzcd}
    & { \text{[isotropic]} } & \\
    {\text{[cubic]} }   \arrow[ur, hookrightarrow] & & 
    {\text{[transv.\ isotropic]} } \arrow[ul, hookrightarrow] \\
    {\text{[tetragonal]} }   \arrow[u, hookrightarrow]  \arrow[urr, hookrightarrow]& & 
    {\text{[trigonal]}}   \arrow[u, hookrightarrow] \\
    {\text{[orthotropic]} }   \arrow[u, hookrightarrow] & & \\
    & {\text{[mono-clinic]} }   \arrow[ul, hookrightarrow] \arrow[uur, hookrightarrow]  & \\
    & {\text{[tri-clinic]} }   \arrow[u, hookrightarrow] &
\end{tikzcd}
\end{center}
    \caption{Hasse diagram of symmetry subgroups of 4th order tensors in 3D}  \label{F:Hasse}
\end{figure}

In 3D there are eight elasticity classes \citep{cowMehr1995, forteVianel1996, 
newn2005, slaw2007, tinder2008, oliveAuffray2013, auffrayEtal2014, malyarenko_random_2017, 
malyarenko_tensor-valued_2019}.  The Hasse diagram for 3D in \ffig{F:Hasse} shows 
the relation of the lattice of elasticity symmetry or invariance groups
as subgroups of the full orthogonal group in 3D.  They will be again described 
informally by pointing out geometric bodies with the same symmetry.  On top is 
the \emph{isotropic} class with the symmetries of a sphere,
i.e.\ the whole orthogonal group, with the sub-classes --- descending in the Hasse
diagram in \ffig{F:Hasse} along the left branch --- the \emph{cubic} class 
with the symmetries of a cube, the \emph{tetragonal} class 
with the symmetries of a square prism (a right prism with a square base), the 
\emph{orthotropic} class with the symmetries of a right prism (a right prism with a 
general rectangle as base),  the \emph{mono-clinic} class with the symmetries of 
a right prism with a parallelogram as base, and the \emph{tri-clinic} class 
with the symmetries of a parallelepiped (an oblique prism with a parallelogram as base).
On the right branch of the Hasse diagram for 3D in \ffig{F:Hasse} are the 
\emph{transversely isotropic} class with the symmetries of a circular cylinder and the 
\emph{trigonal} class with the symmetries of a triangular prism (a right prism with a 
equilateral triangle as base).

\paragraph{Tri-clinic material in 3D:}
As mentioned before, these materials have in 3D the symmetry of a parallelepiped.
As in the 2D case for tri-clinic materials,
the Kelvin matrix $\eTns$ can not be simplified through a group reduction with a 
spatial rotation $\tQ=\Trep(\vQ) \in \ST$ to a block-diagonal form beyond the general form 
\feq{eq:Kelvin-C-3D-1}, and there are $n=21$ free parameters.
But, as in 3D one has $\dim\,\ST = \dim\,\st = 3$, three parameters 
$\vq = (q_1,q_2,q_3) \in \C{Q} \equiv \D{R}^{m_Q} \subset \D{R}^{15}$ ($m_Q = 3$) may be used
for a spatial rotation, the direction of the vector $\vq$ actually being the rotation axis 
of $\vQ(\vq) \in \SO(3)$, see \feqs{eq:so-6-R-15}{eq:rodrigues6} in 
Appendix~\ref{S:appx-Rotations}, to induce three zeros in the right of
the top row \citep{cowMehr1995}, this echos the first of the relations in \feq{eq:easy-ref}:
\begin{equation}  \label{eq:tri-clinic-3D}
  \rsymb{\eTns}_{\text{tri-3D}}  = {\small \begin{bmatrix} 
   \rsymb{\mrm{c}}_{1111} & \rsymb{\mrm{c}}_{1122}& \rsymb{\mrm{c}}_{1133} & 0 & 0 & 0 \\  
                  & \rsymb{\mrm{c}}_{2222}& \rsymb{\mrm{c}}_{2233} & 
   \sqz\,\rsymb{\mrm{c}}_{2223} & \sqz\,\rsymb{\mrm{c}}_{2213} & \sqz\,\rsymb{\mrm{c}}_{2212} \\  
     &    & \rsymb{\mrm{c}}_{3333} &  \sqz\,\rsymb{\mrm{c}}_{3323} &  \sqz\,\rsymb{\mrm{c}}_{3313} 
             & \sqz\,\rsymb{\mrm{c}}_{3312} \\  
     &    &   & \;\;\;2\,\rsymb{\mrm{c}}_{2323} & \;\;\;2\,\rsymb{\mrm{c}}_{2313} 
          & \;\;\;2\,\rsymb{\mrm{c}}_{2312} \\  
       & \mrm{SYM}  &   &  & \;\;\;2\,\rsymb{\mrm{c}}_{1313} & \;\;\;2\,\rsymb{\mrm{c}}_{1312} \\  
        & & & & & \;\;\;2\,\rsymb{\mrm{c}}_{1212}  
   \end{bmatrix} } .
\end{equation}
As before in 2D, the spatial rotations $\vQ \in \SO(\Physpc)$ resp.\ $\tQ \in \ST$
decide on whether the spatial orientation of the material axes changes or not, 
and thus one wants to keep this possibility in the stochastic modelling process.

The matrix in \feq{eq:tri-clinic-3D} has generally six distinct positive eigenvalues 
$\lambda_1, \dots, \lambda_6$, for which one needs $m_{\Lambda} = 6$ log-eigenvalue
parameters to generate the diagonal eigenvalue matrix:
$\D{R}^{m_{\Lambda}} \ni \vek{\mu} = (\mu_1, \dots, \mu_6) \mapsto 
\tM = \diag(\vek{\mu}) \mapsto \tLbd(\vek{\mu}) = \diag(\lambda_j) = \exp \tM$.

Next one wants to use the spectral decomposition of $\rsymb{\eTns}_{\text{tri-3D}}$, the second 
of the relations in \feq{eq:easy-ref}.  As a $\tQ \in \ST$ was used to generate the special form
\feq{eq:tri-clinic-3D}, this matrix can be  diagonalised 
as in \feq{eq:group-eig-dec}, but the corresponding matrix of eigenvectors 
$\tV\in\SO(\Strspc)$ is not induced by a spatial rotation, and thus generated by 
a $\tP \in \st^\perp$.  See \fsec{SS:Lie-Kelvin-represent} on how to keep the spatial rotation 
$\tQ$ from interfering with the `strain distributor' part $\tV \in \SO(\Strspc)$
by introducing the orthogonal split in the corresponding Lie algebra
already mentioned in \feq{eq:prod-Liealg2}: $\so(\Strspc) = \st \oplus \st^\perp$.  
As $\dim\,\st^\perp = m_V = n - m_Q - m_{\Lambda} = 12$, all that is needed now is
a basis of the space $\C{Q}^\perp$, given in \feq{eq:span-V6} in Appendix~\ref{SS:A-rot-Strspc}.
With that basis $\{\vk_1,\dots,\vk_{12}\}$ and 12 parameters $(s_1,\dots,s_{12})$
one may form the vector $\vp = \sum_{j=1}^{12} s_j \vk_j \in \C{Q}^\perp$, and then map
this on to $\vp \mapsto \tP(\vp) \in \st^\perp$ according to \feq{eq:so-6-R-15}. 
This is finally mapped to $\tV(\vp) = \exp \tP(\vp)$, to obtain 
$\rsymb{\eTns}_{\text{tri-3D}}(\vek{\mu}, \vp)$
as in \feq{eq:Kelvin-C-2D-tri}, which corresponds to the spectral decomposition as the
second of the relations in \feq{eq:easy-ref}.

The columns of the matrix $\tV = [\tv_1,\dots,\tv_6]$ are the eigenvectors corresponding 
to the eigenvalues $\lambda_1, \dots, \lambda_6$.
 As before, in case one wants to distinguish the bulk or volumetric response from the shear 
 response, e.g.\ by assigning the largest of the eigenvalues to this response, then one may
choose the eigenvector $\tv_j$ which has the largest projection onto the volumetric strain
$\tn$ given in \feq{eq:n-rep3D} in Appendix~\ref{SS:C-Kelvin}, and assign the largest eigenvalue
to that eigenvector.

Finally, as before, the 2-step rotation procedure with a spatial rotation $\tQ(\vq)$
given by $\vq \in \C{Q}$ may be used as in \feq{eq:gen-C-mat-tri-clin-2D}
to rotate the Kelvin matrix to the desired coordinate system.
As pointed out before, the last equation in \feq{eq:gen-C-mat-tri-clin-2D} giving
$\eTns_{\text{tri-3D}}(\vq,\vek{\mu},\vp)$ here --- the third relation in \feq{eq:easy-ref} --- 
is the concrete version of \feq{eq:Lie-rand-dec} in \fsec{SS:random-Kelvin}, cf.\ the 
remarks for tri-clinic materials in 2D in \fsec{SS:el_cl_2D}.
In case two of the eigenvectors are pure shears, i.e.\ proportional to 
$\ty$ in \feq{eq:y-rep3D} and $\tz$ in \feq{eq:z-rep3D} in Appendix~\ref{SS:C-Kelvin}, 
or some linear combination thereof, then the material is in fact \emph{mono-clinic}, to be
described next.

\paragraph{Mono-clinic material in 3D:}
As indicated previously, these materials have in 3D the symmetry of a right prism 
with a parallelogram as base.
In this case  \citep{cowMehr1995} the Kelvin matrix $\eTns$ can be simplified through 
a group reduction with a spatial rotation $\tQ \in \ST$ to a block-diagonal form 
with three diagonal blocks \feq{eq:mono-clinic-3D}, and in this form there are $12$ 
free parameters.  An additional $m_Q = \dim\,\ST = \dim\,\st = 3$ parameters 
$\vq = (q_1,q_2,q_3) \in \C{Q} \equiv \D{R}^{m_Q}$ are used
for the spatial rotation $\tQ(\vq)$ just referenced for the group reduction process,
giving the total number of parameters for mono-clinic material as $n=15$; 
see \feqs{eq:so-6-R-15}{eq:rodrigues6} in Appendix~\ref{S:appx-Rotations}, 
this is the first relation in \feq{eq:easy-ref}:
\begin{equation}  \label{eq:mono-clinic-3D}
  \rsymb{\eTns}_{\text{mono-3D}}  = {\small \begin{bmatrix} 
   \rsymb{\mrm{c}}_{1111} & \rsymb{\mrm{c}}_{1122}& \rsymb{\mrm{c}}_{1133} & 
         \sqz\,\rsymb{\mrm{c}}_{1123} & 0 & 0 \\  
                  & \rsymb{\mrm{c}}_{2222}& \rsymb{\mrm{c}}_{2233} & 
   \sqz\,\rsymb{\mrm{c}}_{2223} &   0 & 0 \\
     &    & \rsymb{\mrm{c}}_{3333} &  \sqz\,\rsymb{\mrm{c}}_{3323} &   0 & 0 \\
     &    &   & \;\;\;2\,\rsymb{\mrm{c}}_{2323} &  0 & 0 \\
       & \mrm{SYM}  &   &  & \;\;\;2\,\rsymb{\mrm{c}}_{1313} &  0 \\
        & & & & & \;\;\;2\,\rsymb{\mrm{c}}_{1212}  
   \end{bmatrix} } .  
\end{equation}
The matrix in \feq{eq:mono-clinic-3D} has generally six distinct positive eigenvalues 
$\lambda_1, \dots, \lambda_6$, for which one needs, as before, $m_{\Lambda} = 6$ log-eigenvalue
parameters to generate the diagonal eigenvalue matrix:
$\vek{\mu} = (\mu_1, \dots, \mu_6) \mapsto 
\tM = \diag(\vek{\mu}) \mapsto \tLbd(\vek{\mu}) = \diag(\lambda_j) = \exp \tM$.
Two eigenvalues are immediately recognisable from \feq{eq:mono-clinic-3D}, namely
the Kelvin moduli
$\lambda_5 = 2\,\rsymb{\mrm{c}}_{1313}$ and $\lambda_6 = 2\,\rsymb{\mrm{c}}_{1212}$
with corresponding eigenvectors $\tv_5=[0,0,0,0,1,0]^\trpos$ and $\tv_5=[0,0,0,0,0,1]^\trpos$.

The other four eigenvectors in $\tV = [\tv_1,\dots,\tv_4,\tv_5,\tv_6]^\trpos$ come
from the diagonalisation of the upper left $4 \times 4$ block  $\check{\eTns}$ in 
\feq{eq:mono-clinic-3D}.  This has a spectral decomposition $\check{\eTns} = \check{\tV}
\check{\tLbd}\check{\tV}^\trpos$ with $\check{\tV} \in \SO(4)$.  As $\dim\,\SO(4) = 
\dim\,\so(4) = 6$, one needs $m_V = 6 = n - m_{\Lambda} - m_Q$ additional parameters
$\vp = [s_1, \dots, s_6]^\trpos \in \C{Q}^\perp \equiv \D{R}^{m_V}$ to model this.
Taking the 4-dimensional version of \feq{eq:so-6-R-15} in Appendix~\ref{SS:A-rot-Strspc},
this is mapped to $\check{\tP}(\vp) \in \so(4)$, and this on to
$\check{\tV}(\vp) = \exp \check{\tP}(\vp) \in \SO(4)$, yielding the eigen-strains 
of the eigenvalues resp.\  Kelvin moduli of $\check{\eTns}$.  
Each column $\check{\tv}_j\in \D{R}^4,
j=1,\dots,4$ in $\check{\tV}=[\check{\tv}_1,\check{\tv}_2,\check{\tv}_3,\check{\tv}_4]$ 
is then extended into $\D{R}^6$ by padding with two zeros: 
$\tv_j(\vp) = [\check{\tv}_j^\trpos(\vp),0,0]^\trpos\in \D{R}^6, j=1,\dots,4$ to
complete $\tV(\vp)$.

From this eigen-strain matrix $\tV(\vp)$ and the diagonal $\tLbd(\vek{\mu})$ one then
obtains a random mono-clinic Kelvin matrix of the form \feq{eq:mono-clinic-3D}:
$\rsymb{\eTns}_{\text{mono-3D}}(\vek{\mu},\vp) = \tV(\vp) \tLbd(\vek{\mu}) \tV^\trpos(\vp)$,
as in \feq{eq:Kelvin-C-2D-tri}, corresponding to the spectral decomposition as the
second of the relations in \feq{eq:easy-ref}.
The by now familiar 2-step rotation procedure with a spatial rotation $\tQ(\vq)$
given by $\vq \in \C{Q}$ may be used as in \feq{eq:gen-C-mat-tri-clin-2D}
to rotate the Kelvin matrix to the desired coordinate system, and again
the last equation in \feq{eq:gen-C-mat-tri-clin-2D} yielding
$\eTns_{\text{mono-3D}}(\vq,\vek{\mu},\vp)$  --- the third relation in \feq{eq:easy-ref} --- 
is the concrete version of \feq{eq:Lie-rand-dec} in \fsec{SS:random-Kelvin}.
In case $\tv_4 = [0,0,0,1,0,0]^\trpos$, the material is in fact \emph{orthotropic}.

\paragraph{Orthotropic material in 3D:}
It was already mentioned that these materials have in 3D the symmetry of a right prism 
--- a prism with a general rectangle as base.  As in all the previous cases, the group
reduction process with a rotation $\tQ(\vq)=\Trep(\vQ(\vq)) \in \ST$ leads to a block-diagonal 
form \feq{eq:ortho-3D} with three rotation axis parameters 
$\vq = (q_1,q_2,q_3) \in \C{Q} \equiv \D{R}^{m_Q}$ ($m_Q = 3$), 
see \feqs{eq:so-6-R-15}{eq:rodrigues6} in Appendix~\ref{S:appx-Rotations}.
The resulting Kelvin matrix of the form \citep{cowMehr1995}:
\begin{equation}  \label{eq:ortho-3D}
  \rsymb{\eTns}_{\text{ortho-3D}}  = {\small \begin{bmatrix} 
   \rsymb{\mrm{c}}_{1111} & \rsymb{\mrm{c}}_{1122}& \rsymb{\mrm{c}}_{1133} & 0 & 0 & 0 \\  
                  & \rsymb{\mrm{c}}_{2222}& \rsymb{\mrm{c}}_{2233} &  0 &   0 & 0 \\
     &    & \rsymb{\mrm{c}}_{3333} &  0 &   0 & 0 \\
     &    &   & \;\;\;2\,\rsymb{\mrm{c}}_{2323} &  0 & 0 \\
       & \mrm{SYM}  &   &  & \;\;\;2\,\rsymb{\mrm{c}}_{1313} &  0 \\
        & & & & & \;\;\;2\,\rsymb{\mrm{c}}_{1212}  
   \end{bmatrix} }   
\end{equation}
has $m_{\Lambda} + m_V =9$ free parameters, so that together with the rotation axis 
$\vq$ with $m_Q = 3$ parameters, in total there are $n = m_Q + m_{\Lambda} + m_V = 12$ parameters.

The matrix in \feq{eq:ortho-3D} has generally six distinct positive eigenvalues 
$\lambda_1, \dots, \lambda_6$, for which one needs, as before, $m_{\Lambda} = 6$ log-eigenvalue
parameters to generate the diagonal eigenvalue matrix $\vek{\mu} = (\mu_1, \dots, \mu_6) \mapsto 
\tM = \diag(\vek{\mu}) \mapsto \tLbd(\vek{\mu}) = \diag(\lambda_j) = \exp \tM$.
Three eigenvalues are obvious from \feq{eq:ortho-3D}, namely
the Kelvin moduli $\lambda_4 = 2\,\rsymb{\mrm{c}}_{2323}$, 
$\lambda_5 = 2\,\rsymb{\mrm{c}}_{1313}$, and $\lambda_6 = 2\,\rsymb{\mrm{c}}_{1212}$
with corresponding eigenvectors $\tv_4=[0,0,0,1,0,0]^\trpos$, $\tv_5=[0,0,0,0,1,0]^\trpos$, 
and $\tv_6=[0,0,0,0,0,1]^\trpos$.

The other three eigenvectors come
from the diagonalisation of the upper left $3 \times 3$ block  $\check{\eTns}$ in 
\feq{eq:ortho-3D}.  Its spectral decomposition is $\check{\eTns} = \check{\tV}
\check{\tLbd}\check{\tV}^\trpos$ with $\check{\tV} \in \SO(3)$.  Therefore, 
$m_V = 3 = n - m_{\Lambda} - m_Q$ additional parameters
$\vp = [s_1, s_2, s_3]^\trpos \in \C{Q}^\perp \equiv \D{R}^{m_V}$ to model this.
As $m_V = 3$, one may actually use Rodrigues original formula
\feeqs{eq:rodrigues-P}{eq:rodrigues} in Appendix~\ref{SS:A-rot-2-3}, such
that $\vp$ is the rotation vector, but now in the strain representer space $\Strspc$,
and not in physical space $\Physpc.$  This produces a rotation matrix
$\check{\tV}(\vp) = \exp \check{\tP}(\vp) \in \SO(3)$, yielding the eigen-strains 
of the eigenvalues resp.\  Kelvin moduli of $\check{\eTns}$.  
Each column $\check{\tv}_j\in \D{R}^3$ in 
$\check{\tV}=[\check{\tv}_1,\check{\tv}_2,\check{\tv}_3]$ 
is then extended into $\D{R}^6$ by padding with three zeros: 
$\tv_j(\vp) = [\check{\tv}_j^\trpos(\vp),0,0,0]^\trpos\in \D{R}^6, j=1,\dots,3$ to
complete $\tV(\vp)$.
The familiar 2-step rotation procedure with a spatial rotation $\tQ(\vq)$
given by $\vq \in \C{Q}$ may be used as in \feq{eq:gen-C-mat-tri-clin-2D}
to rotate the Kelvin matrix to the desired coordinate system.

\paragraph{Trigonal material in 3D:}
It was already mentioned that these materials have in 3D the symmetry of a triangular prism. 
As in all the previous cases, the Kelvin matrix can be simplified in the group
reduction process with a rotation $\tQ(\vq)=\Trep(\vQ(\vq)) \in \ST$  with three 
rotation axis parameters $\vq = (q_1,q_2,q_3) \in \C{Q} \equiv \D{R}^{m_Q}$ ($m_Q = 3$)
--- a so-called \emph{natural} coordinate system with the spatial 3-axis equal to the
three-fold rotation axis --- into a form \citep{cowMehr1995} with $n-m_Q=6$ free parameters
(i.e.\ the total number of parameters is $n=9$):
\begin{equation}  \label{eq:trigon-3D}
  \rsymb{\eTns}_{\text{trigon-3D}}  = {\small \begin{bmatrix} 
   \rsymb{\mrm{c}}_{1111} & \rsymb{\mrm{c}}_{1122}& \rsymb{\mrm{c}}_{1133} & 
                     \phantom{-}\sqz\,\rsymb{\mrm{c}}_{1123} & 0 & 0 \\  
     & \rsymb{\mrm{c}}_{1111}& \rsymb{\mrm{c}}_{1133} &  
                      -\sqz\,\rsymb{\mrm{c}}_{1123} & 0 & 0 \\
     &    & \rsymb{\mrm{c}}_{3333} &  0 &   0 & 0 \\
     &    &   & \;\;\;2\,\rsymb{\mrm{c}}_{2323} &  0 & 0 \\
       & \mrm{SYM}  &   &  & \;\;\;2\,\rsymb{\mrm{c}}_{2323} &  \;\;\;2\,\rsymb{\mrm{c}}_{1123} \\
        & & & & & \rsymb{\mrm{c}}_{1111} - \rsymb{\mrm{c}}_{1122}  
   \end{bmatrix} }   
\end{equation}
The matrix in \feq{eq:trigon-3D} has generally \citep{sutcl1992} only four distinct 
positive eigenvalues $\lambda_1, \dots, \lambda_4$, as two of them are degenerate 
double eigenvalues.  For this one needs $m_{\Lambda} = 4$ log-eigenvalue
parameters to generate the diagonal eigenvalue matrix $\vek{\mu} = (\mu_1, \dots, \mu_4) \mapsto 
\tM = \diag_{\sharp}(\vek{\mu}) \mapsto \exp \tM = \tLbd(\vek{\mu})$, where $\diag_{\sharp}$
means that the four distinct eigenvalues are put in the right six locations on the diagonal,
doubling included.

Now $m_V = n - m_Q - m_{\Lambda} = 9 - 3 - 4 = 2$ more parameters $\vp = [s_1, s_2]^\trpos$
are needed to fully describe the material via the eigenvector matrix $\tV(\vp) = 
[\tv_1, \tv_2, \tv_3, \tv_4, \tv_5, \tv_6]$.  The two simple eigenvectors for 
$\lambda_1, \lambda_2$ have the general form  \cite{sutcl1992}:
\begin{equation}  \label{eq:v1-tri}
   \tv_{1} = \frac{1}{\sqz}[ \cos \alpha, \cos \alpha,\sqz \sin \alpha, 0, 0, 0]^{\trpos},\quad
   \tv_{2} = \frac{1}{\sqz}[ \sin \alpha, \sin \alpha, -\sqz \cos \alpha, 0, 0, 0]^{\trpos}.
\end{equation}
The angle $\alpha$ in strain representer space signifies a rotation about the shear 
$\ty$ in \feq{eq:y-rep3D}, which means that $\tv_1, \tv_2$ are rotated images of $\tn$
in \feq{eq:n-rep3D} and $\tz$ \feq{eq:z-rep3D}.  In case $\tan \alpha = \sqz$, i.e.\
for the angle $\alpha_n = \arctan \sqz$, the vectors $\tv_1, \tv_2$ coincide with $\tn, \tz$.
therefore we set in \feq{eq:v1-tri} $\alpha = \alpha_n + s_1$, so that $s_1 = 0$ corresponds
to the special case.

The two two-dimensional eigen-subspaces with repeated eigenvalues $\lambda_3 = \lambda_4$ 
and $\lambda_5=\lambda_6$ have as possible bases \citep{sutcl1992} vectors 
of the general form --- pure shears:
\begin{align}  \label{eq:v2-tri}
   \tv_{3} &= \frac{1}{\sqz}[ \cos s_2, -\cos s_2, 0, -\sqz \sin s_2, 0, 0]^{\trpos},\;
   \tv_{4} = [ 0, 0, 0, 0, -\sin s_2, \cos s_2]^{\trpos}, \text{ and } \\ \label{eq:v3-tri}
   \tv_{5} &= \frac{1}{\sqz}[ \sin s_2, -\sin s_2, 0, \phantom{-}\sqz \cos s_2, 0, 0]^{\trpos},
   \; \tv_{6} = [ 0, 0 , 0, 0, \phantom{-}\cos s_2, \sin s_2]^{\trpos} .
\end{align}
Again the parameter $s_2$ can be interpreted as an angle, and note that for $s_2 = 0$
these eigenvectors all become simple shears.
With this the description of the $n=9$ parameters $(\vq, \vek{\mu}, \vp)$ is complete.
The last step is the 2-step rotation procedure as before
to rotate the Kelvin matrix to the desired coordinate system.

\paragraph{Tetragonal material in 3D:}
As mentioned, these materials have in 3D the symmetry of a square prism. 
As in the previous cases, the Kelvin matrix can be simplified by group
reduction with a rotation $\tQ(\vq)=\Trep(\vQ(\vq)) \in \ST$  with three 
rotation axis parameters $\vq = (q_1,q_2,q_3) \in \C{Q} \equiv \D{R}^{m_Q}$ ($m_Q = 3$)
--- such that the spatial 3-axis is equal to the
four-fold rotation axis --- \citep{cowMehr1995} with $n-m_Q=6$ free parameters
(i.e.\ the total number of parameters is $n=9$) into the form:
\begin{equation}  \label{eq:tetra-3D}
  \rsymb{\eTns}_{\text{tetra-3D}}  = {\small \begin{bmatrix} 
   \rsymb{\mrm{c}}_{1111} & \rsymb{\mrm{c}}_{1122}& \rsymb{\mrm{c}}_{1133} & 0 & 0 & 0 \\  
     & \rsymb{\mrm{c}}_{1111}& \rsymb{\mrm{c}}_{1133} & 0 & 0 & 0 \\
     &    & \rsymb{\mrm{c}}_{3333} &  0 &   0 & 0 \\
     &    &   & \;\;\;2\,\rsymb{\mrm{c}}_{2323} &  0 & 0 \\
       & \mrm{SYM}  &   &  & \;\;\;2\,\rsymb{\mrm{c}}_{2323} &  0 \\
        & & & & & 2\,\rsymb{\mrm{c}}_{1212}  
   \end{bmatrix} }   ,
\end{equation}
which has generally \citep{sutcl1992} five independent eigenvalues resp.\ Kelvin moduli.
Obviously $\lambda_4 = \lambda_5 = 2\,\rsymb{\mrm{c}}_{2323}$ is a double eigenvalue with
the two-dimensional eigen-space spanned by the vectors
$\tv_4=[0,0,0,1,0,0]^\trpos$ and $\tv_5=[0,0,0,0,1,0]^\trpos$.
The other obvious simple eigenvalue is $\lambda_6 = 2\,\rsymb{\mrm{c}}_{1212}$ with
eigenvector $\tv_6=[0,0,0,0,0,1]^\trpos$.
Thus one needs $m_{\Lambda} = 5$ log-eigenvalue
parameters to generate the diagonal eigenvalue matrix: $\vek{\mu} = (\mu_1, \dots, \mu_5) \mapsto 
\tM = \diag_{\sharp}(\vek{\mu}) \mapsto \exp \tM = \tLbd(\vek{\mu})$, where again $\diag_{\sharp}$
means that the five distinct eigenvalues are put in the right six locations on the diagonal.

The other three eigenvectors come
from the diagonalisation of the upper left $3 \times 3$ block  $\check{\eTns}$ in 
\feq{eq:tetra-3D}.  By simple inspection \citep{sutcl1992} it can be seen that 
$\check{\tv}_3 = \frk{1}{\sqz}[-1,1,0] = \ty$ (cf.\ \feq{eq:y-rep2D}) is a fixed 
eigenvector of $\check{\eTns}$ with
eigenvalue $\lambda_3 = \rsymb{\mrm{c}}_{1111} - \rsymb{\mrm{c}}_{1122}$.  Padding
$\check{\tv}_3$ with 3 zeros in the by now familiar fashion gives the fixed eigenvector 
$\tv_3 = \ty$ (cf.\ \feq{eq:y-rep3D}) of $\rsymb{\eTns}_{\text{tetra-3D}}$ in \feq{eq:tetra-3D}.
The other eigenvectors for the two simple eigenvalues $\lambda_1, \lambda_2$ of 
$\check{\eTns}$ depend on an angle-like strain-distribution parameter $\vp = [\alpha]$,
and have the general form  \cite{sutcl1992} (already  padded with zeros):
\begin{equation}  \label{eq:v12-tetra}
   \tv_1 = \frac{1}{\sqz}[ \cos \alpha, \cos \alpha,\sqz \sin \alpha, 0, 0, 0]^{\trpos},\;
   \tv_2 = \frac{1}{\sqz}[ \sin \alpha, \sin \alpha, -\sqz \cos \alpha, 0, 0, 0]^{\trpos}.
\end{equation}
This then finally gives the eigenvector matrix $\tV(\vp) = 
[\tv_1(\vp), \tv_2(\vp), \tv_3, \tv_4, \tv_5, \tv_6]$.
Thus a tetragonal material including its orientation in the physical space $\Physpc$
is completely described by the $n=9$ parameters $(\vq, \vek{\mu}, \vp)$.
Finally, the 2-step rotation procedure rotates the Kelvin matrix to the desired coordinate system.

\paragraph{Cubic material in 3D:}
As mentioned, these materials have in 3D the symmetry of a cube. 
The Kelvin matrix can be simplified by the group
reduction with a rotation $\tQ(\vq)=\Trep(\vQ(\vq)) \in \ST$  with three 
rotation axis parameters $\vq = (q_1,q_2,q_3) \in \C{Q} \equiv \D{R}^{m_Q}$ ($m_Q = 3$)
--- such that the sides of the cube align with the axes --- 
\citep{cowMehr1995} with $n-m_Q=3$ free parameters
(i.e.\ the total number of parameters is $n=6$) into the form:
\begin{equation}  \label{eq:cubic-3D}
  \rsymb{\eTns}_{\text{cubic-3D}}  = {\small \begin{bmatrix} 
   \rsymb{\mrm{c}}_{1111} & \rsymb{\mrm{c}}_{1122}& \rsymb{\mrm{c}}_{1122} & 0 & 0 & 0 \\  
     & \rsymb{\mrm{c}}_{1111}& \rsymb{\mrm{c}}_{1122} & 0 & 0 & 0 \\
     &    & \rsymb{\mrm{c}}_{1111} &  0 &   0 & 0 \\
     &    &   & \;\;\;2\,\rsymb{\mrm{c}}_{2323} &  0 & 0 \\
       & \mrm{SYM}  &   &  & \;\;\;2\,\rsymb{\mrm{c}}_{2323} &  0 \\
        & & & & & 2\,\rsymb{\mrm{c}}_{2323}  
   \end{bmatrix} }   .
\end{equation}
It has generally \citep{sutcl1992} only three independent eigenvalues resp.\ Kelvin moduli,
one repeated eigenvalue of multiplicity two, and one repeated eigenvalue of multiplicity three, 
as $\lambda_4 = \lambda_5 = \lambda_6 = 2\,\rsymb{\mrm{c}}_{2323}$ is a triple eigenvalue with
the three-dimensional eigen-space spanned by the vectors
$\tv_4=[0,0,0,1,0,0]^\trpos, \tv_5=[0,0,0,0,1,0]^\trpos$, and $\tv_6=[0,0,0,0,0,1]^\trpos$.
Thus one needs $m_{\Lambda} = 3$ log-eigenvalue
parameters to generate the diagonal eigenvalue matrix: $\vek{\mu} = (\mu_1, \mu_2, \mu_3) \mapsto 
\tM = \diag_{\sharp}(\vek{\mu}) \mapsto \exp \tM = \tLbd(\vek{\mu})$; here again $\diag_{\sharp}$
means that the three distinct eigenvalues are put in the right six locations on the diagonal.

As $m_V = n - m_Q - m_{\Lambda} = 6 - 3 - 3 = 0$, the other three eigenvectors may be chosen 
as fixed.
They are found to be $\lambda_1 = \rsymb{\mrm{c}}_{1111} + 2\,\rsymb{\mrm{c}}_{1122}$ with
eigenvector $\tv_1 = \tn$ (cf.\ \feq{eq:n-rep3D}) and $\lambda_2 = \lambda_3 = 
\rsymb{\mrm{c}}_{1111} - \rsymb{\mrm{c}}_{1122}$ with the two-dimensional eigen-space
spanned by the shear-vectors $\tv_2 = \ty, \tv_3 = \tz$, cf.\ \feqs{eq:y-rep3D}{eq:z-rep3D}.
The fixed matrix of eigenvectors is thus $\tV = [\tv_1,\tv_2,\tv_3,\tv_4,\tv_5,\tv_6]$,
and the six parameters to describe the material are $(\vq, \vek{\mu})$.
Finally, the usual 2-step rotation procedure rotates the Kelvin matrix 
to the desired coordinate system.

\paragraph{Transversely isotropic material in 3D:}
As indicated, these materials have in 3D the symmetry of a circular cylinder. 
To simplify the Kelvin matrix by the group
reduction, a rotation $\tQ(\vq)=\Trep(\vQ(\vq)) \in \ST$  with only \emph{two} 
rotation axis parameters $\vq = (q_1,q_2) \in \C{Q} \equiv \D{R}^{m_Q}$ is
required ($m_Q = 2$, as no rotation around the cylinder axis is necessary)
--- such that the cylinder axis aligns with the 3-axis --- 
\citep{cowMehr1995} with $n-m_Q=5$ free parameters
(i.e.\ the total number of parameters is $n=7$) into the form:
\begin{equation}  \label{eq:trans-3D}
  \rsymb{\eTns}_{\text{tr-iso-3D}}  = {\small \begin{bmatrix} 
   \rsymb{\mrm{c}}_{1111} & \rsymb{\mrm{c}}_{1122}& \rsymb{\mrm{c}}_{1133} & 0 & 0 & 0 \\  
     & \rsymb{\mrm{c}}_{1111}& \rsymb{\mrm{c}}_{1133} & 0 & 0 & 0 \\
     &    & \rsymb{\mrm{c}}_{3333} &  0 &   0 & 0 \\
     &    &   & \;\;\;2\,\rsymb{\mrm{c}}_{2323} &  0 & 0 \\
       & \mrm{SYM}  &   &  & \;\;\;2\,\rsymb{\mrm{c}}_{2323} &  0 \\
        & & & & & \rsymb{\mrm{c}}_{1111} - \rsymb{\mrm{c}}_{1122}  
   \end{bmatrix} }   .
\end{equation}
This Kelvin matrix has generally \citep{sutcl1992} four independent eigenvalues resp.\
Kelvin moduli; and there are two eigenvalues of multiplicity two.  One is obviously
$\lambda_4 = \lambda_5 = 2\,\rsymb{\mrm{c}}_{2323}$ with the fixed two-dimensional eigen-space
spanned by the vectors $\tv_4=[0,0,0,1,0,0]^\trpos$ and $\tv_5=[0,0,0,0,1,0]^\trpos$.
The other double eigenvalue is $\lambda_3 = \lambda_6 = \rsymb{\mrm{c}}_{1111} - 
\rsymb{\mrm{c}}_{1122}$ with the fixed two-dimensional eigen-space
spanned by the vectors $\tv_3=\ty$ (cf.\ \feq{eq:y-rep3D}) and $\tv_6=[0,0,0,0,0,1]^\trpos$.
Thus one needs $m_{\Lambda} = 4$ log-eigenvalue
parameters to generate the diagonal eigenvalue matrix: $\vek{\mu} = (\mu_1, \dots, \mu_4) \mapsto 
\tM = \diag_{\sharp}(\vek{\mu}) \mapsto \exp \tM = \tLbd(\vek{\mu})$, where again $\diag_{\sharp}$
means that the four distinct eigenvalues are put in the proper six locations on the diagonal.

As for tetragonal material above, the other three eigenvectors come
from the diagonalisation of the upper left $3 \times 3$ block  $\check{\eTns}$ in 
\feq{eq:trans-3D}, which is identical to that block for tetragonal material. 
The eigenvalue $\lambda_3$ and its eigenvector which can be found by simple
inspection \citep{sutcl1992} was already noted above.
The other eigenvectors for the two simple eigenvalues $\lambda_1, \lambda_2$ of 
$\check{\eTns}$ depend on an angle-like strain-distribution parameter $\vp = [\alpha]$,
and have the same general form  \cite{sutcl1992} as for tetragonal material \feq{eq:v12-tetra},
giving $\tv_1(\vp)$ and $\tv_2(\vp)$.   This finally yields the eigenvector matrix $\tV(\vp) = 
[\tv_1(\vp), \tv_2(\vp), \tv_3, \tv_4, \tv_5, \tv_6]$.
Thus a transversely isotropic material including its orientation in the physical space $\Physpc$
is completely described by the $n=7$ parameters $(\vq, \vek{\mu}, \vp)$, and
the final 2-step rotation procedure rotates the Kelvin matrix to the desired coordinate system.

\paragraph{Isotropic material in 3D:}
As is well known, these materials have in 3D the symmetry of a sphere. 
No simplification of the Kelvin matrix by the group reduction process is possible, and
no spatial rotations are required ($m_Q = 0$).  The Kelvin matrix has the same form
in any coordinate system with \citep{cowMehr1995} $n = 2$ free parameters:
\begin{equation}  \label{eq:iso-3D}
  \rsymb{\eTns}_{\text{iso-3D}}  = {\small \begin{bmatrix} 
   \rsymb{\mrm{c}}_{1111} & \rsymb{\mrm{c}}_{1122}& \rsymb{\mrm{c}}_{1122} & 0 & 0 & 0 \\  
     & \rsymb{\mrm{c}}_{1111}& \rsymb{\mrm{c}}_{1122} & 0 & 0 & 0 \\
     &    & \rsymb{\mrm{c}}_{1111} &  0 &   0 & 0 \\
     &    &   & \rsymb{\mrm{c}}_{1111} - \rsymb{\mrm{c}}_{1122} &  0 & 0 \\
       & \mrm{SYM}  &   &  & \rsymb{\mrm{c}}_{1111} - \rsymb{\mrm{c}}_{1122} &  0 \\
        & & & & & \rsymb{\mrm{c}}_{1111} - \rsymb{\mrm{c}}_{1122}  
   \end{bmatrix} }   .
\end{equation}
There is one eigenvalue with multiplicity five, $\lambda_j = 
\rsymb{\mrm{c}}_{1111} - \rsymb{\mrm{c}}_{1122} = 2\,G, (j=2 \dots 6)$, with $G$ the well-known
\emph{shear modulus}, and its fixed five-dimensional eigen-space of shears spanned by
\begin{align}  \label{eq:v2-iso}
   \tv_2 &=  \ty,\; \tv_3 = \tz  \quad\text{cf. \feqs{eq:y-rep3D}{eq:z-rep3D}, and} \\  \nonumber
   \tv_4 &= [0, 0 , 0, 1, 0 , 0]^\trpos,\;
   \tv_5 = [0, 0 , 0, 0, 1 , 0]^\trpos,\;
   \tv_6 = [0, 0 , 0, 0, 0 , 1]^\trpos.
\end{align}
The simple Kelvin modulus $\lambda_1 = \rsymb{\mrm{c}}_{1111} + 2\,\rsymb{\mrm{c}}_{1122} = 3\,K$, 
where $K$ is the well-known \emph{bulk modulus}, has the eigenvector
$\tv_1 =  \tn$ cf.\  \feq{eq:n-rep3D}, i.e.\ volume change.
To generate the diagonal eigenvalue matrix: $\vek{\mu} = (\mu_1,  \mu_2) \mapsto 
\tM = \diag_{\sharp}(\vek{\mu}) \mapsto \exp \tM = \tLbd(\vek{\mu})$, where again $\diag_{\sharp}$
means that the two distinct eigenvalues are put in the proper six locations on the diagonal.
Thus an isotropic material is completely described by the $n= m_{\Lambda} =2$ parameters 
$(\vek{\mu})$ --- the log-eigenvalue parameters --- to generate the diagonal eigenvalue matrix: 
$\vek{\mu} = (\mu_1, \mu_2) \mapsto \tM = \diag_{\sharp}(\vek{\mu}) 
\mapsto \exp \tM = \tLbd(\vek{\mu})$.  The eigenvector matrix $\tV$ has the fixed ($m_V = 0$)
columns $\tv_j, (j=1 \dots 6)$, and no final spatial rotation is needed.

\rvsn{\subsection{Example}}
\rvsn{We consider the average orthotropic elastic properties of human cortical femoral bone in 3D, obtained from measurements on 60 specimens, by \cite{ashman_continuous_1984}. The coefficients are listed in \ftbl{Table:mat_param_orth}. 
\begin{table}[h!]
	\centering
	{\begin{tabular}{lll}
			\specialrule{1pt}{1pt}{1pt}
			\begin{tabular}[c]{@{}l@{}}
				Young's \\ modulus (GPa) 
			\end{tabular} & 
			\begin{tabular}[c]{@{}l@{}} 
				Poisson's \\ ratio 
			\end{tabular} & 
			\begin{tabular}[c]{@{}l@{}} 
				Shear\\ modulus (GPa) 
			\end{tabular} \\ \midrule
			$Y_1 = 12$ & $\nu_{21} = 0.422$, $\nu_{12} = 0.376$ & $G_{12} = 4.53$\\ 
			$Y_2 = 13.4$ & $\nu_{31} = 0.371$, $\nu_{13} = 0.222$ & $G_{13} = 5.61$\\
            $Y_3 = 20$ & $\nu_{32} = 0.35$, $\nu_{23} = 0.235$ & $G_{23} = 6.23$\\
			\specialrule{1pt}{1pt}{1pt}
	\end{tabular}}
	\caption{Orthotropic material parameters of cortical femoral bone\label{Table:mat_param_orth}}
\end{table}
Following this, the compliance matrix---the inverse of the elasticity matrix---in Voigt notation is constructed, as shown in Eq.(6) in \cite{ashman_continuous_1984}. The subsequent inversion of the compliance matrix yields the  elasticity matrix. Finally, by multiplying a factor of 2 to the lower-right $3\times3$ block of diagonal elements, we derive the Kelvin matrix of an orthotropic material in a block diagonal form $\rsymb{\eTns}_{\text{ortho-3D}}$, as depicted in \feq{eq:ortho-3D}.}

\rvsn{A key objective of this study is to decouple the modelling of scaling, eigen-strain, and orientation parameters in the Kelvin matrix $\eTns$---see \feq{eq:SpecDecSPD} for the spectral decomposition. To represent variations in these aspects, we model a one-dimensional field of the orthotropic elasticity matrices over the spatial domain $x\in[0, 1]$.}

\rvsn{To this, the SPD matrix at the left end of the domain at $x=0$ remains the previously defined $\rsymb{\eTns}_{\text{ortho-3D}}$, which for simplicity is referred to as $\aTns$. On the other end at $x=1$, the Kelvin matrix denoted as $\bTns$ is defined separately for each variation. For scaling, the eigenvalues $\lambda_5$ and $\lambda_6$ of $\aTns$ are multiplied by a factor of 5. For rotation in physical space, the matrix $\bTns$ is obtained by rotating $\aTns$ by an angle $\theta = 60^{\circ}$ or $\pi/3$ in radians in X-Z plane in clockwise direction i.e, the rotation vector is $\vek{q} = (0, \theta, 0)$. For eigen-strain variation, the upper-left $3\times3$ block matrix of $\aTns$ undergoes a similar rotation, where the rotation-like vector reads $\vek{p} = (0, \theta, 0)$.}
\begin{figure}[!h]
	\begin{subfigure}[b]{\textwidth}
		\centering
		\includegraphics[width=0.75\linewidth]{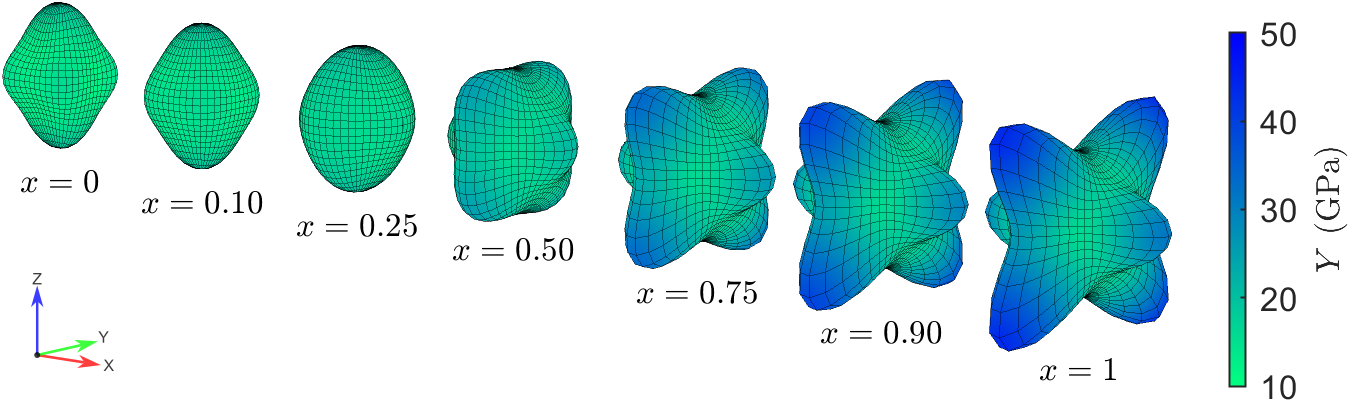}
		\caption{Riemannian metric for scaling ($\vtheta_L(\tLbd_1, \tLbd_2)$)}
		\label{}
	\end{subfigure}
	\begin{subfigure}[b]{\textwidth}
		\centering
		\includegraphics[width=0.75\linewidth]{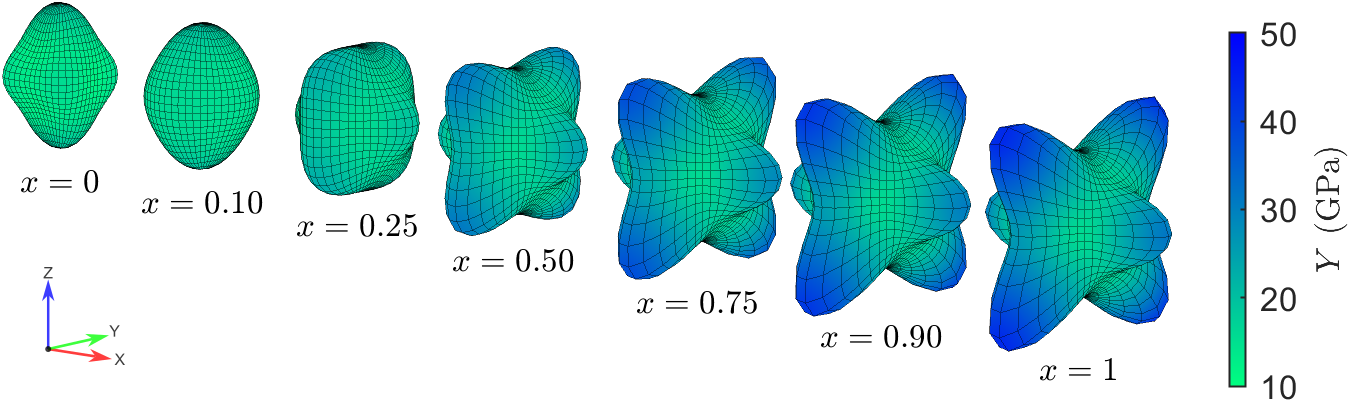}
		\caption{Euclidean metric ($\vtheta_2(\eTns_1, \eTns_2)$)}
		\label{}
	\end{subfigure}
	\caption{Visualization of Young's modulus for the interpolation of Kelvin matrices under varying scaling parameters}
	\label{fig:scaling}
\end{figure}

\rvsn{The spatial field is built by interpolating from $\aTns$ to $\bTns$ using the Elasticity metric $\tilde{\vtheta}_E$ as defined in \feq{eq:elast-dist-p} and Appendix~\ref{SS:metric}. Meaning that, for scaling-based interpolation, the metric reduces to log-Euclidean  $\vtheta_L(\tLbd_1, \tLbd_2)$, whereas, for rotations and eigen-strain-based fluctuations, the interpolation is dependent on Riemannian distance for special orthogonal matrices i.e., $\vtheta_R(\vQ_1, \vQ_2)$ and $\vtheta_R(\tV_1, \tV_2)$, respectively. Additionally, for comparison, we also interpolate the SPD matrices using the usual Euclidean metric $\vtheta_2(\eTns_1, \eTns_2)$, detailed in Appendix~\ref{SS:frechet}.}

\begin{figure}[!h]
	\begin{subfigure}[b]{\textwidth}
		\centering
		\includegraphics[width=0.75\linewidth]{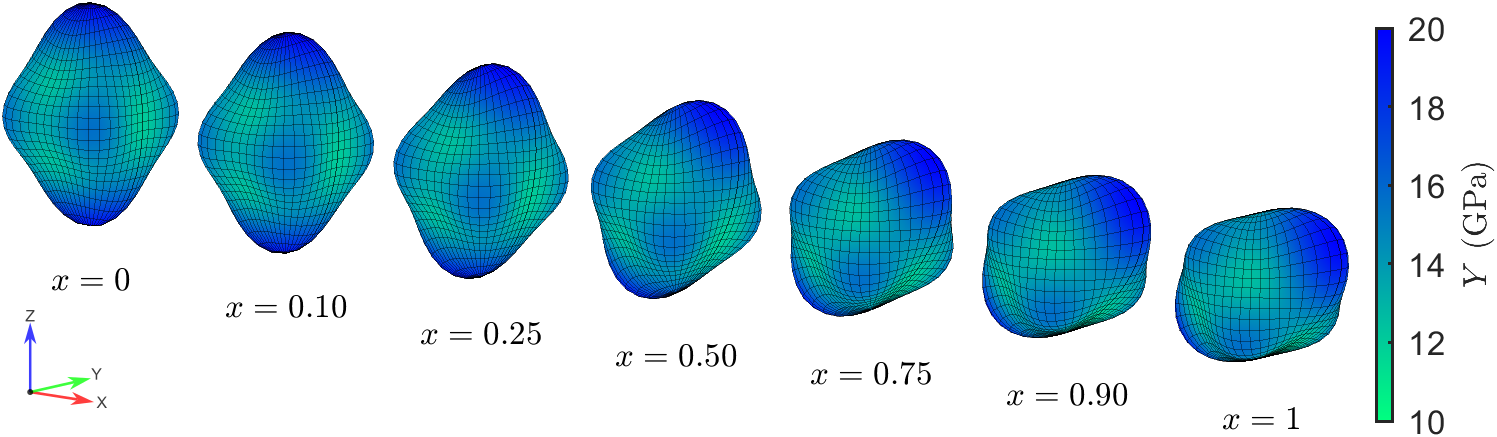}
		\caption{Riemannian metric for rotation ($\vtheta_R(\vQ_1, \vQ_2)$)}
		\label{}
	\end{subfigure}
	\begin{subfigure}[b]{\textwidth}
		\centering
		\includegraphics[width=0.75\linewidth]{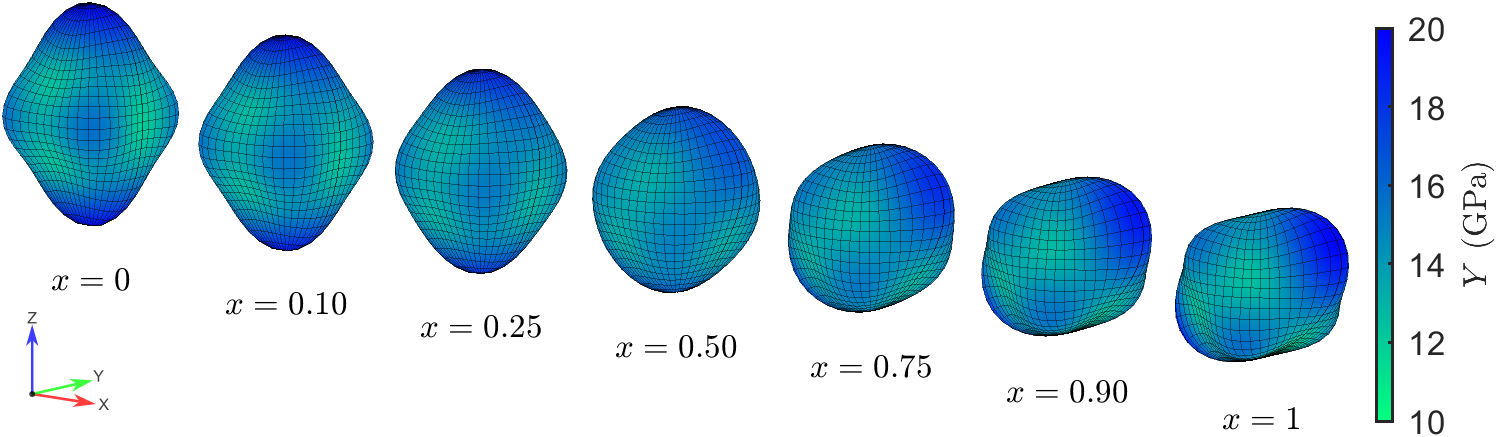}
		\caption{Euclidean metric ($\vtheta_2(\eTns_1, \eTns_2)$)}
		\label{}
	\end{subfigure}
	\caption{Visualization of Young's modulus for the interpolation of Kelvin matrices under varying orientation parameters}
	\label{fig:rotation}
\end{figure}

\rvsn{The visualization of interpolation of Kelvin matrices is done by analysing one of the characteristic elastic parameters such as Young's modulus ($Y$) for a given loading direction, as shown in \cite{nordmann_visualising_2018}. \ffig{fig:scaling} illustrates the interpolation under scaling variations at given spatial points $x$, while \ffigs{fig:rotation}{fig:eigenstrain} depict the effects of variations in orientation and eigen-strain, respectively. Additionally, each figure demonstrates the interpolation between $\aTns$ and $\bTns$ using both the Riemannian (non-Euclidean) and Euclidean metric. }
\begin{figure}[!h]
	\begin{subfigure}[b]{\textwidth}
		\centering
		\includegraphics[width=0.75\linewidth]{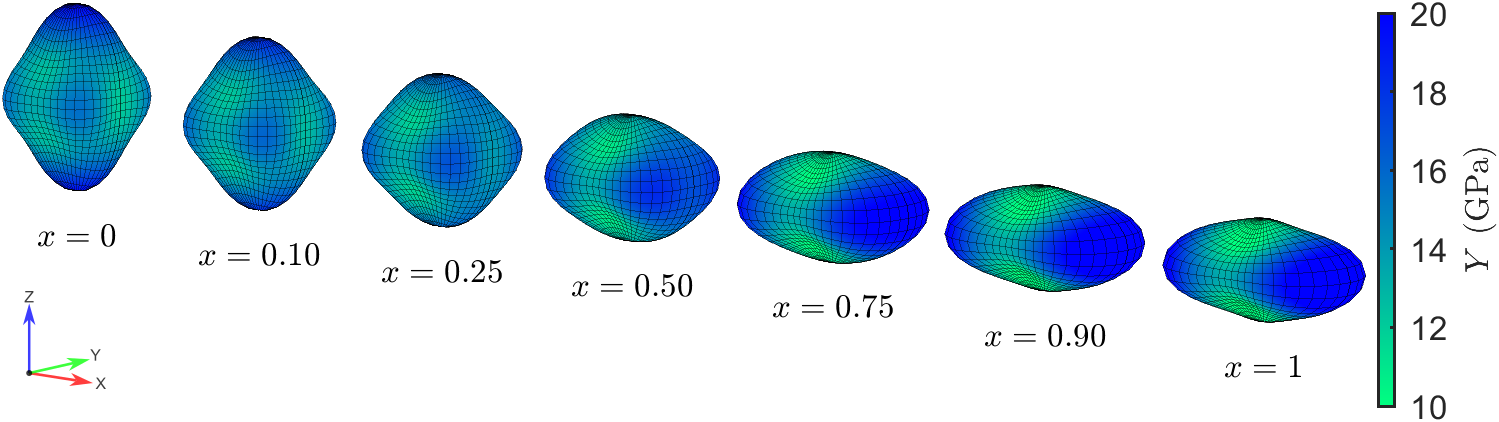}
		\caption{Riemannian metric for eigen-strain ($\vtheta_R(\tV_1, \tV_2)$)}
		\label{}
	\end{subfigure}
	\begin{subfigure}[b]{\textwidth}
		\centering
		\includegraphics[width=0.75\linewidth]{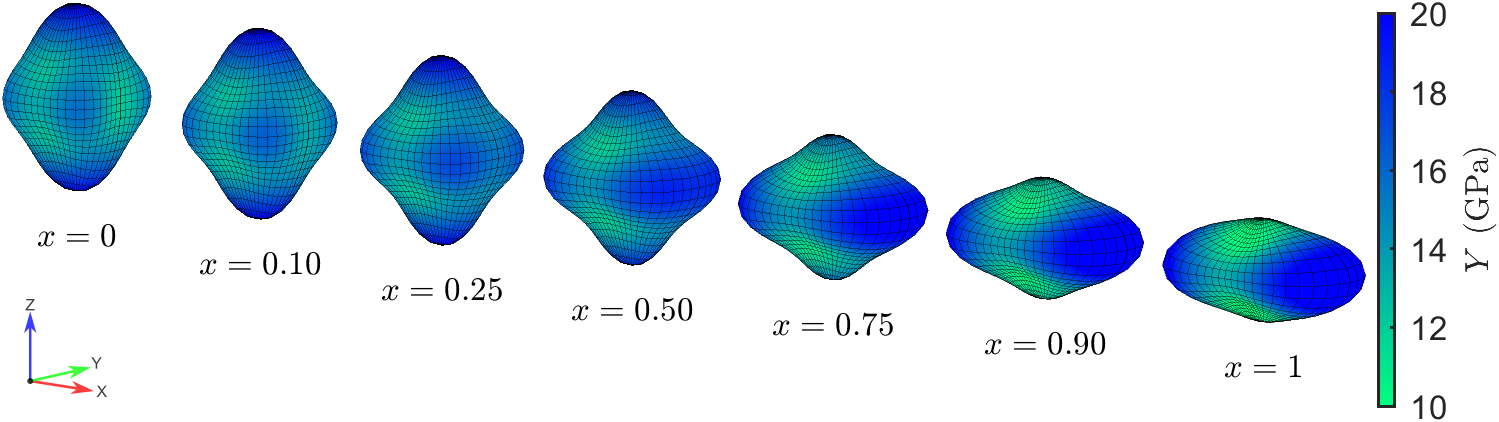}
		\caption{Euclidean metric ($\vtheta_2(\eTns_1, \eTns_2)$)}
		\label{}
	\end{subfigure}
	\caption{Visualization of Young's modulus for the interpolation of Kelvin matrices under varying eigen-strain parameters}
	\label{fig:eigenstrain}
\end{figure}

\rvsn{In all three cases, we observe notable differences between Euclidean and non-Euclidean interpolation. To quantify these differences, we compute the determinant of the Kelvin matrices $\eTns$, which serves as a volumetric measure. The comparison plots for the three variations—scaling, rotation, and eigen-strain—are presented in \ffig{fig:determinant}.} 
\begin{figure}[h]
    \begin{subfigure}{0.32\textwidth}
        \centering
%
%
\definecolor{mycolor1}{rgb}{0.49000,0.18000,0.56000}%
\definecolor{mycolor2}{rgb}{0.30000,0.75000,0.93000}%
\begin{tikzpicture}
\footnotesize
\begin{axis}[%
width=4cm,
height=4cm,
at={(0,0)},
scale only axis,
xmin=1,
xmax=7,
tick label style={font=\scriptsize},
xtick={1,2,3,4,5,6,7},
xticklabels={{0},{0.1},{0.25},{0.5},{0.75},{0.9},{1}},
xlabel style={font=\color{white!15!black}},
xlabel={$x$},
ymin=0,
ymax=180000000,
ylabel style={font=\color{white!15!black}, yshift=-10pt},
ylabel={$\det(C)$},
axis background/.style={fill=white},
legend style={at={(0.0,1.4)}, anchor=north west, legend cell align=left, align=left, draw=none}
]
\addplot [color=mycolor1, line width=1.3pt, mark size=1.3pt, mark=*, mark options={solid, fill=mycolor1, mycolor1}]
  table[row sep=crcr]{%
1	6681886.34258271\\
2	9219196.78137396\\
3	14941152.0799424\\
4	33409431.7129135\\
5	74705760.3997119\\
6	121072383.402742\\
7	167047158.564568\\
};
\addlegendentry{Riemannian ($\vtheta_L(\tLbd_1, \tLbd_2)$)}

\addplot [color=mycolor2, dashed, line width=1.3pt, mark size=1.3pt, mark=*, mark options={solid, fill=mycolor2, mycolor2}]
  table[row sep=crcr]{%
1	6681886.34258271\\
2	13096497.2314621\\
3	26727545.3703308\\
4	60136977.0832444\\
5	106910181.481323\\
6	141388715.00905\\
7	167047158.564568\\
};
\addlegendentry{Euclidean ($\vtheta_2(\eTns_1, \eTns_2)$)}

\end{axis}
\end{tikzpicture}%
        \caption{Scaling}
        \label{fig:determinant_scal}
    \end{subfigure}
    \begin{subfigure}{0.32\textwidth}
        \centering
%
%
\definecolor{mycolor1}{rgb}{0.49000,0.18000,0.56000}%
\definecolor{mycolor2}{rgb}{0.30000,0.75000,0.93000}%
\begin{tikzpicture}
\footnotesize
\begin{axis}[%
width=4cm,
height=4cm,
at={(0,0)},
scale only axis,
xmin=1,
xmax=7,
tick label style={font=\scriptsize},
xtick={1,2,3,4,5,6,7},
xticklabels={{0},{0.1},{0.25},{0.5},{0.75},{0.9},{1}},
xlabel style={font=\color{white!15!black}},
xlabel={$x$},
ymin=6600000,
ymax=7200000,
ylabel style={font=\color{white!15!black}, yshift=-10pt},
ylabel={$\det(C)$},
axis background/.style={fill=white},
legend style={at={(0.0,1.4)}, anchor=north west, legend cell align=left, align=left, draw=none}
]
\addplot [color=mycolor1, line width=1.3pt, mark size=1.3pt, mark=*, mark options={solid, fill=mycolor1, mycolor1}]
  table[row sep=crcr]{%
1	6681886.34258271\\
2	6681886.79853228\\
3	6681886.86131581\\
4	6681885.99219314\\
5	6681886.16318406\\
6	6681885.82012458\\
7	6681885.99219315\\
};
\addlegendentry{Riemannian ($\vtheta_R(\vQ_1, \vQ_2)$)}

\addplot [color=mycolor2, dashed, line width=1.3pt, mark size=1.3pt, mark=*, mark options={solid, fill=mycolor2, mycolor2}]
  table[row sep=crcr]{%
1	6681886.34258271\\
2	6864801.26275338\\
3	7065077.45610887\\
4	7194618.34591418\\
5	7065077.27542058\\
6	6864800.97823262\\
7	6681885.99219315\\
};
\addlegendentry{Euclidean ($\vtheta_2(\eTns_1, \eTns_2)$)}

\end{axis}
\end{tikzpicture}%
        \caption{Rotation}
        \label{fig:determinant_rot}
    \end{subfigure}
    \begin{subfigure}{0.32\textwidth}
        \centering
%
%
\definecolor{mycolor1}{rgb}{0.49000,0.18000,0.56000}%
\definecolor{mycolor2}{rgb}{0.30000,0.75000,0.93000}%
\begin{tikzpicture}
\footnotesize
\begin{axis}[%
width=4cm,
height=4cm,
at={(0,0)},
scale only axis,
xmin=1,
xmax=7,
tick label style={font=\scriptsize},
xtick={1,2,3,4,5,6,7},
xticklabels={{0},{0.1},{0.25},{0.5},{0.75},{0.9},{1}},
xlabel style={font=\color{white!15!black}},
xlabel={$x$},
ymin=6500000,
ymax=9000000,
ylabel style={font=\color{white!15!black}, yshift=-12pt},
ylabel={$\det(C)$},
axis background/.style={fill=white},
legend style={at={(0.0,1.4)}, anchor=north west, legend cell align=left, align=left, draw=none}
]
\addplot [color=mycolor1, line width=1.3pt, mark size=1.3pt, mark=*, mark options={solid, fill=mycolor1, mycolor1}]
  table[row sep=crcr]{%
1	6681886.34258271\\
2	6681886.4565701\\
3	6681886.47226598\\
4	6681886.25498532\\
5	6681886.29773305\\
6	6681886.21196817\\
7	6681886.25498532\\
};
\addlegendentry{Riemannian ($\vtheta_R(\tV_1, \tV_2)$)}

\addplot [color=mycolor2, dashed, line width=1.3pt, mark size=1.3pt, mark=*, mark options={solid, fill=mycolor2, mycolor2}]
  table[row sep=crcr]{%
1	6681886.34258271\\
2	7402098.49071405\\
3	8235735.85226596\\
4	8872369.09213751\\
5	8413760.9353323\\
6	7538821.71806849\\
7	6681886.25498532\\
};
\addlegendentry{Euclidean ($\vtheta_2(\eTns_1, \eTns_2)$)}

\end{axis}
\end{tikzpicture}%
        \caption{Eigen-strain}
        \label{fig:determinant_eig}
    \end{subfigure}
    \caption{Comparison of the determinant of interpolated Kelvin matrices using Euclidean and Riemannian metrics for variations in scaling, rotation, and eigen-strain}
	\label{fig:determinant}
\end{figure}
\rvsn{In the case of scaling-based interpolation (\ffig{fig:determinant_scal}), the determinant of interpolants based on the Euclidean metric $\vtheta_2$ is consistently higher than that of the log-Euclidean distance $\vtheta_L$. On the other hand, in \ffigs{fig:determinant_rot}{fig:determinant_eig} for the respective rotation and eigen-strain variations, Euclidean interpolation suffers from the so-called \textit{swelling effect}, wherein the determinant of any interpolant for $0<x<1$ exceeds that of the endpoint matrices at $x=0$ and $x=1$. Notably, for fixed scaling parameters, the determinant of Riemannian interpolants $\vtheta_R$ remains unchanged.}

\rvsn{For more discussion on this, we refer the reader to \cite{JungSchwartzmanGroisser2015}, where the authors make detailed comparisons for second-order diffusion tensors.}

%
%
%
%
%
%
%
%
%
%
%
%
%
%


%
\section{Conclusion} \label{S:concl}

%

    \textcolor{black}{
    A novel framework for the probabilistic modelling of fourth-order random material tensor 
    fields has been introduced, with a focus on an important class of symmetric and positive 
    definite (SPD) tensors, such as elasticity tensors.}
    
    \textcolor{black}{
    A crucial aspect of material laws governed by such tensors is their invariance under 
    spatial symmetries. By employing the Kelvin notation and representing elasticity tensors 
    as Kelvin matrices their essential tensorial properties are preserved, so that the 
    spatial symmetries and invariances of elasticity tensors through rotations in physical space may
    reduce the Kelvin matrix to a block-diagonal form, the so-called group reduction. 
    Within each group-invariant subspace of this block-diagonal form, the Kelvin matrix 
    has been further simplified to a diagonal form through spectral decomposition.
    }
    
    \textcolor{black}{
    The group reduction, combined with spectral decomposition, forms the foundation of 
    our approach.  Based on this context, the stochastic representation has the ability to 
    separate the modelling of strength (Kelvin moduli), eigen-strain distribution, and spatial 
    orientation, allowing for independent control of each component.  This framework enables
    the consideration of random SPD matrices where specific invariance properties are known 
    for both the entire population and the mean, or where the ensemble exhibits a `lower' class 
    of invariance compared to the mean.  While the stochastic modelling of Kelvin elasticity 
    matrices adheres to the prior principles established for second-order tensors in 
    \citep{shivaEtal2021}, the inclusion of eigen-strain distributions represents a crucial 
    extension of these methods to higher-order tensors.
    }

    \textcolor{black}{
    Even-order SPD tensors form an open convex cone within the vector space of physically 
    symmetric tensors, meaning they constitute a differentiable manifold, but not a vector space. 
    In the modelling of random tensor fields, it is often preferable to find an unconstrained 
    representation in a linear vector space.  To achieve this, we consider the matrix exponential 
    and logarithm mappings.  Within the proposed probabilistic framework, Lie groups are associated 
    with diagonal SPD matrices and subgroups of orthogonal matrices, allowing for the independent 
    modelling of different aspects of the random tensor.  The product Lie group representation 
    is related to the associated Lie algebras for each individual component, the mapping between
    them being the component exponential map, which subsequently yields the full representation 
    of the Kelvin matrix.  Furthermore, the Lie representations as $n$-dimensional random fields 
    of Kelvin matrix parameters across all symmetry classes is outlined, including all eight 
    elastic symmetry classes in 3D and the four classes in 2D.  In summary, the dimension of 
    $n$ ranges from 2 to 6 in 2D, and from 2 to 21 in 3D, for isotropic to tri-clinic materials, respectively.
    }

    Given the nature of SPD tensors, calculating the average or mean requires special attention. 
    The conventional Euclidean mean is often not suitable, as it exhibits an undesirable property 
    known as the swelling effect.  Instead, we focus on the more general non-Euclidean Fr\'{e}chet mean,
    which is based on distance measurements other than the Euclidean metric.  Accordingly, 
    we establish criteria for defining an appropriate mean and metric specifically for elasticity. 
    The term `elasticity metric' or `elasticity distance' is proposed for this metric, which provides 
    fine control over measuring differences in spatial orientation, eigen-strain distribution, and Kelvin 
    moduli, allowing for the combination of these factors with adjustable weights.  Based on this 
    distance, one may then formulate the corresponding Fr\'{e}chet mean, referred to as the 
    `elasticity mean'.
    {To demonstrate this, as an example, we model a one-dimensional spatial field of orthotropic Kelvin matrices using interpolation based on the elasticity metric. The interpolated matrices are visualized by analysing their corresponding Young's modulus. The results reveal a notable difference in Riemannian and Euclidean metric-based interpolation.}

    \textcolor{black}{
    While this research focused on fourth-order SPD tensors, the stochastic modelling approach
    outlined here is sufficiently general to be applicable to any even-order physically symmetric 
    tensors, including those that are also positive definite.
    }

\appendix
\section{Appendix: Mean and Metric} \label{S:appx-metric}

%


The parametrisation of the open cone of fourth-order symmetric positive tensors $\mrm{Sym}^+(\C{T})$ 
is connected to the issue on what kind of metric to use on this manifold,
which in turn is connected with what kind of mean or average on should use.
Here we shall merely give a short overview,
for a fuller discussion of these topics, please refer to \citep{shivaEtal2021} and
the references therein, or also \citep{Oller_1995, PennecFillardAyache2004, Pennec2006, 
Moakher2006, Moakher2008, MoakherZerai2011, FeragenFuster2017, ThanwerdasPennec2019,
SommerFletcherPennec2020, Fletcher2020, GuiguiEtal2023, JungRooksEtAl2023, groissJungSchwman2023}.

\subsection{Fr\'echet Mean}  \label{SS:frechet}
Since antiquity different ways to compute a mean or average have been known, 
one just has to look at the classical Pythagorean means.  The one which is best known is of
course the arithmetic mean, which is based on the Euclidean metric $\vtheta_2$ as will be shown.
The other classical means may be obtained as Fr\'echet means --- to be introduced shortly --- 
with the variational characterisation below in \feq{eq:var-def-EuclMean}
via different metrics resp.\ distance functions $\vtheta$.

As noted before, the mean resp.\ average resp.\ expected value has a  variational
definition, where the metric of the underlying sample space appears naturally.
For the sake of simplicity, consider initially
a set of $n$ vectors $\vx_1,\dots,\vx_n \in \Rk$ --- obviously the
case $k = 1$ is the best known --- with $\Rk$ equipped with the standard Euclidean
norm, such that the squared Euclidean distance is 
$\vtheta_2(\vx,\vy)^2 = \nd{\vx - \vy}_2^2
= (\vx - \vy)^\trpos (\vx - \vy)$.  Then the arithmetic mean
$\vbar{x}^{(2)}$ is characterised by
\begin{align}  \label{eq:var-def-EuclMean}
  \D{E}_{\vtheta_2}(\{\vx_j \}) &:= \vbar{x}^{(2)} := \arg \min_{\vx \in\D{R}^k}
  \Psi_2(\vx), \quad \text{ where in this case } \\
  \label{eq:x-basedvar-E}
  \Psi_2(\vx) &:= \frac{1}{n} \sum_{j=1}^n \vtheta_2(\vx_j,\vx)^2,
\end{align}
and $\Psi_2(\vx)$ may be called the $\vx$-based variance.  The Euler-Lagrange
variational conditions of vanishing derivative of the functional $\Psi_2$ obviously yield the
usual definition of arithmetic mean.  This is easily extended from constant weights $w_j ≡ 1/n$
to not necessarily equal (probability) weights $w_j > 0$, $\sum_{j=1} w_j = 1$, 
and any vector space $\C{Y}$ equipped with an inner product and corresponding norm.
By taking just two vectors $\vx_1, \vx_2$ and weights $w_1 = \alpha, w_2 = 1 - \alpha$
with $0 \leq \alpha \leq 1$, it is easily gleaned that this furnishes the interpolation
between the two vectors on the shortest path connecting them --- a \emph{geodesic}, which in 
a Euclidean space is a straight line parallel to $\vx_1 - \vx_2$.

Going a step further, for a --- possibly continuous --- probability distribution $\D{P}$
of a $\C{Y}$-valued random variable $\ry(\omega)$ defined on a probability space
$(\Omega,\D{P})$ with $\C{Y}$ some normed vector space, the $\vx$-based variance is defined as
\begin{equation}  \label{eq:x-basedvar-P}
   \Psi_{\D{P}}(\vx) = \int_{\Omega} \vtheta_{\C{Y}}(\ry(\omega),\vx)^2\, \D{P}(\di \omega)
   \quad \text{ with } \quad \vtheta_{\C{Y}}(\ry(\omega),\vx) := \nd{\ry(\omega)-\vx}_{\C{Y}},
\end{equation}
and the variational characterisation \feq{eq:var-def-EuclMean} is still used to define
the arithmetic mean $\vbar{x}^{(\D{P})}$ or $\D{P}$-expected value $\D{E}_{{\D{P}}}(\ry) = 
\arg\min_{\vx} \Psi_{\D{P}}(\vx)$.
Observe that all that was needed is a metric  
space $\C{Y}$ with a metric or distance function $\vtheta_{\C{Y}}$.  Such a mean defined 
analogous to \feq{eq:var-def-EuclMean} on a metric space with a general $\vx$-based variance
$\Psi_{\vtheta}(\vx)$ as in \feq{eq:x-basedvar-P} with a general metric $\vtheta$
is called a \emph{Fr\'echet} mean \citep{Oller_1995, PennecFillardAyache2004,
Pennec2006, ThanwerdasPennec2019, SommerFletcherPennec2020, Fletcher2020}.

On the manifold of positive definite tensors $\Syp\subset \sy$, when using the Euclidean metric 
and the corresponding arithmetic average resp.\ mean derived from it, one observes
\citep{PennecFillardAyache2004, AndoLiMathias2004, Moakher2005, ArsignyFillardPennecEtAl2006, 
ArsignyFillardPennecEtAl2007, drydenEtal2009, DrydenEtal2010, WangSalehianEtAl2014, FujiiSeo2015,
FeragenFuster2017}
undesirable \emph{swelling, fattening, and shrinking} effects in interpolation
resp.\ averaging \citep{schwartzman_random_nodate, JungSchwartzmanGroisser2015, 
schwartzman_lognormal_2016, GroisserJungSchwartzman2017, groissJungSchwman2017}.
These undesirable effects will affect the corresponding mean in a similar manner.
This, as well as the wish to satisfy the desiderata for a mean formulated in
\fsec{SS:rep-requirem}, which are not satisfied by the arithmetic mean, but are 
easily seen to be inherited from the corresponding metric, is the reason to look at other 
distance functions on the cone of positive definite tensors.  In the following
an `elasticity distance' on $\Syp(\Strspc)$ will be defined via an `elasticity metric' on
$\sS = \ST \times \SO(\Strspc) \times \Dgp(\Strspc)$, which in turn can
be used to construct a corresponding `elasticity Fr\'echet mean' analogous to
\feqs{eq:var-def-EuclMean}{eq:x-basedvar-P}.

\subsection{Elasticity Distance}  \label{SS:metric}
The manifold of positive definite tensors $\Syp\subset \sy$ is geometrically an open cone in 
the vector space of symmetric tensors $\sy$, and may be equipped with a metric in different ways,
cf.\ \citep{Lang1995, CowinYang1997, PennecFillardAyache2004, AndoLiMathias2004, Moakher2005, 
Moakher2006, ArsignyFillardPennecEtAl2006, ArsignyFillardPennecEtAl2007, drydenEtal2009, 
DrydenEtal2010, FujiiSeo2015, FeragenFuster2017, MorinEtal2020}, as well as 
\citep{ArsignyFillardPennecEtAl2006,
ArsignyFillardPennecEtAl2007} and \citep{JungSchwartzmanGroisser2015, schwartzman_lognormal_2016, 
GroisserJungSchwartzman2017, groissJungSchwman2017, shivaEtal2021}.
In these non-Euclidean metrics mentioned, the convex cone $\Syp$ is metrically complete. 
These metrics also greatly reduce the swelling, fattening, and shrinking effects alluded to above.

As the approach in this work is to represent a positive definite tensor as
$\eTns = \tQ^\trpos \tV \tLbd \tV^\trpos \tQ$ with an
orthogonal $\tQ =\Trep(\vQ)\in \ST$ induced by a spatial rotation $\vQ \in \SO(\Physpc)$,
a strain resp.\ stress distribution $\tV \in \SO(\Strspc)$ and the Kelvin moduli resp.\
stiffnesses $\tLbd \in \Dgp(\Strspc)$,
the representation is characterised by the triple $(\vQ, \tV, \tLbd)$ from
the product manifold / product Lie group
$\sS = \SO(\Physpc) \times \SO(\Strspc) \times \Dgp(\Strspc)$.
Each of the factors of the product $\sS$ being a finite dimensional 
Lie group under matrix multiplication, 
the focus here is on how to endow such a locally compact Lie group with a Riemannian metric
and derive a metric structure from it.

Two instances when this can be done is when the Lie group is compact or when it is Abelian 
\citep{segalKunze78, Carmo1992, Lang1995, Spivak-1-1999, Jost2005, AlexandrinoBettiol2015},
as this allows the construction of a bi-invariant metric.  Observe that the whole
manifold resp.\ open cone $\Syp(\Strspc)$
is \emph{not} a Lie group under normal matrix multiplication, but it can be endowed with an
Abelian product \citep{ArsignyFillardPennecEtAl2006, ArsignyFillardPennecEtAl2007},
which turns it into a Lie group with Lie algebra $\sy(\Strspc)$, and
which for commuting elements of $\Syp(\Strspc)$ is the same as the usual matrix product.
But for the product Lie group $\sS$ note that $\SO(\Physpc)$ and
$\SO(\Strspc)$ are compact Lie groups, and $\Dgp(\Strspc)$ is a non-compact but Abelian Lie group.
As the focus in this work is to separate the effects of spatial orientation
as represented by $\vQ \in \SO(\Physpc)$, strain resp.\ stress distribution as 
represented by $\tV \in \SO(\Strspc)$, and stiffnesses as represented by 
$\tLbd \in \Syp(\Strspc)$, we shall not follow directly  
\citep{ArsignyFillardPennecEtAl2006, ArsignyFillardPennecEtAl2007} and use the above
mentioned Abelian Lie group structure on $\Syp(\Strspc)$, as their approach tends 
to ignore the possible non-commutativity of elements of $\Syp(\Strspc)$
and mixes orientation, stress distribution, and stiffness in the so-called \emph{log-Euclidean}
metric.  Here we rather follow \citep{WangSalehianEtAl2014, JungSchwartzmanGroisser2015, 
schwartzman_lognormal_2016, GroisserJungSchwartzman2017, groissJungSchwman2017, shivaEtal2021} and 
look for a metric on the product Lie group 
$\sS = \SO(\Physpc) \times \SO(\Strspc) \times \Dgp(\Strspc)$.

To give a bit of background for the construction of the elasticity metric to be introduced below, 
the general situation \citep{Carmo1992, Lang1995, Spivak-1-1999, Jost2005, AlexandrinoBettiol2015} 
is such that in a Lie group $G$,
viewed as a differentiable manifold, a neighbourhood $\C{U}(g)\subset G$ and tangent space 
$\mrm{T}_g G$ at each point $g\in G$ can easily be mapped to the identity element $e\in G$ via 
the group operation, e.g.\ $g^{-1}\C{U}(g) = \{ g^{-1}h \mid h\in \C{U}(g)\}$ is a neighbourhood
of $e\in G$, and the tangent spaces $\mrm{T}_g G$ and $\mrm{T}_e G$ are isomorphic.  
The tangent space $\mrm{T}_e G$ at the identity $e\in G$ is equivalent to the Lie algebra 
$\F{g}$ associated to $G$.  For the compact or Abelian Lie groups, any inner
product on the Lie algebra $\F{g}$ gives rise to a Riemannian structure on the whole
manifold resp.\ Lie group $G$, which allows one to measure distances on the group and 
Riemannian manifold $G$ at least locally.   
The mapping from the Lie algebra $\F{g}$ to the Lie group $G$
is given by the exponential map $\exp: \F{g} \to G$ of Lie group theory, 
which has a (generally only locally around $e \in G$) 
inverse $\log: G \to \F{g}$.   This in turn allows one to define geodesics as
curves of locally minimal length,
and hence the distance between points on the manifold as the minimal length 
of the geodesics joining them, cf.\ \citep{Carmo1992, Lang1995, Spivak-1-1999, 
Jost2005, AlexandrinoBettiol2015}, giving the manifold the structure of a complete
metric space.  In the Lie algebra $\F{g}$, which is a Euclidean vector space, geodesics 
through the origin are straight lines, and these are mapped by the exponential map to 
geodesics on the Lie group $G$ through the identity element $e \in G$, as 
$\F{g} \ni 0 \mapsto \exp(0) = e \in G$.  From the identity $e \in G$ the geodesics can 
be mapped to the neighbourhood of any other element $g\in G$ by the group operation.

On the compact Lie group of proper rotations of $\Rn$, the special orthogonal group 
$\SO(n)$ with the Lie algebra $\so(n)$ of skew symmetric matrices, this construction
gives the distance between two elements $\vQ_1, \vQ_2 
\in \SO(n)$ as  \citep{Lang1995, Moakher2002}:
\begin{equation}  \label{eq:SO-dist}
   \vtheta_R(\vQ_1, \vQ_2) := 
   \nd{\log (\vQ_1 \vQ_2^{-1})}_F  = \nd{\log (\vQ_1 \vQ_2^\trpos)}_F ,
\end{equation}
where $\nd{\vW}_F = \sqrt{\tr \vW^\trpos \vW}$ is the Frobenius norm.  Note that
by the group property $\vQ_1^\trpos \vQ_2 \in \SO(n)$, so that
$\log (\vQ_1^\trpos \vQ_2) = \vW \in \so(n)$ is a skew symmetric matrix.  This
yields a distance for the factors $\SO(\Physpc)$ resp.\ $\ST$ and $\SO(\Strspc)$
of the representation product Lie group $\sS$.

Diagonal positive definite matrices $\vLbd \in \Dgp(n)$ form an 
Abelian or commutative Lie group
under matrix multiplication, with Lie algebra $\dg(n) \cong \Rn$, the diagonal
matrices.  As  $\Dgp(n)$ is Abelian resp.\ commutative, one can now freely use the 
Log-Euclidean norm of \citep{ArsignyFillardPennecEtAl2006, ArsignyFillardPennecEtAl2007}
as a distance measure on $\Dgp(n)$. 
For two elements $\vLbd_1, \vLbd_2 \in \Dgp(n)$ this is given by
\begin{equation}  \label{eq:Diag-dist}
   \vtheta_L(\vLbd_1, \vLbd_2)    := \nd{\log (\vLbd_1 \vLbd_2^{-1})}_F 
   = \nd{\log \vLbd_1 - \log \vLbd_2}_F .
\end{equation}
By the group property $\vLbd_1 \vLbd_2^{-1} \in \Dgp(n)$, so that
$\log (\vLbd_1 \vLbd_2^{-1}) \in \dg(n)$ is a diagonal matrix.
Thus this yields a distance for the diagonal component $\Dgp(\Strspc)$ 
in the representation product Lie group $\sS$.

The product Lie group $\sS = \SO(\Physpc) \times \SO(\Strspc) \times \Dgp(\Strspc)$ has as 
identity element three copies of the appropriate identity matrices $\tns{e} := (\vI_{\Physpc},
\tI_{\Strspc},\tI_{\Strspc})$, and 
as Lie algebra the direct sum of the individual Lie algebras with a product exponential map: 
\begin{align}   \label{eq:prod-Lia-alg}
  \mrm{T}_{\tns{e}} \sS &=: \sSl := \so(\Physpc) \oplus \so(\Strspc) \oplus \dg(\Strspc),
  \quad \text{ with } \\  \label{eq:prod-Lia-exp}
  \exp_{\F{s}}:\, \F{s} &\ni (\vP,\tR,\tD) \mapsto 
      (\exp(\vP),\exp(\tR),\exp(\tD))=(\vQ,\tV,\tLbd) \in \sS .
\end{align}

By the construction given above, for each factor of the product Lie group $\sS$
one has a bi-invariant metric, and
we follow and extend \citep{WangSalehianEtAl2014, JungSchwartzmanGroisser2015, 
GroisserJungSchwartzman2017, groissJungSchwman2017, shivaEtal2021}, and define a 
Riemannian product metric on the product Lie group, here termed an ‘elasticity metric’.
For two elements $\eTns_1, \eTns_2 \in \Syp(\Strspc)$ with representations
$\eTns_j = \tQ_j^\trpos \tV_j \tLbd_j \tV_j^\trpos \tQ_j$, $j=1,2$,
the squared product distance $\tilde{\vtheta}_E^2$ is given by
\begin{equation}  \label{eq:elast-dist}
   \tilde{\vtheta}_E(\eTns_1, \eTns_2)^2 = \vtheta_L(\tLbd_1, \tLbd_2)^2 +
   c_{\Physpc} \vtheta_R(\vQ_1, \vQ_2)^2 + c_{\Strspc} \vtheta_R(\tV_1, \tV_2)^2 ,
\end{equation}
with two positive constants $c_{\Physpc}, c_{\Strspc}$, which allow one to tune the sensitivity
of the product distance $\tilde{\vtheta}_E$ to the different components in the product Lie group $\sS$.

As the representation $\eTns = \tQ^\trpos \tV \tLbd \tV^\trpos \tQ$ of a $\eTns \in \Syp(\Strspc)$ 
on $\sS$ is not unique \citep{FeragenFuster2017, GroisserJungSchwartzman2017, GroisserEtal2021}, 
one minimises \citep{JungRooksEtAl2023, groissJungSchwman2023}  
over all possible representations to give the `elasticity distance' $\vtheta_E$:
\begin{multline}  \label{eq:elast-metric}
   \hat{\vtheta}_E(\eTns_1, \eTns_2) =   \min  \{
    \tilde{\vtheta}_E(\tQ_1^\trpos \tV_1 \tLbd_1 \tV_1^\trpos \tQ_1 , 
    \tQ_2^\trpos \tV_2 \tLbd_2 \tV_2^\trpos \tQ_2 ) = \tilde{\vtheta}_E(\eTns_1, \eTns_2) \mid  \\
      (\vQ_j, \tV_j, \tLbd_j) \in \sS , 
      \tQ_j = \Trep(\vQ_j),\; \eTns_j = \tQ_j^\trpos \tV_j \tLbd_j \tV_j^\trpos \tQ_j;\;  j=1,2 \} .
\end{multline}
Unfortunately, this is not yet a metric --- it does not satisfy the triangle inequality,
see the discussion in \citep{FeragenFuster2017} --- 
but can be made into one by recognising that the triples $(\vQ, \tV, \tLbd) \in \sS$
which construct the same $\eTns = \tQ^\trpos \tV \tLbd \tV^\trpos \tQ$ form an
equivalence class in $\sS$, and we are really looking for a metric on the quotient space.
The proper construction for a metric on a quotient of a metric space is
\begin{multline}  \label{eq:elast-metric2}
   \vtheta_E(\eTns_a, \eTns_b) =   \inf  \{ \sum_{j=1}^{n-1}
     \hat{\vtheta}_E(\eTns_j, \eTns_{j+1}) \mid  n \ge 2, \; \eTns_1 = \eTns_a,\; \eTns_n = \eTns_b\; \\ 
       \eTns_j \in \Syp(\Strspc), j=1,\dots, n  \} ,
\end{multline}
i.e.\ by allowing `hops' $\eTns_j$ between $\eTns_a$ and $\eTns_b$.
Both expressions \feqs{eq:elast-metric}{eq:elast-metric2} are not easy to compute,
so that often one takes just $\tilde{\vtheta}_E$ from \feq{eq:elast-dist}, which
is a Riemannian metric, and should be good enough when the differences between
$\eTns_1$ and $\eTns_2$ are not large.

With the elasticity metric $\vtheta_E$ one may then define an elasticity Fr\'echet mean
or expected value $\D{E}_{\vtheta_E}$ for a random elasticity tensor $\eTns_r(\omega)$,
defined as a random variable $\eTns_r: \Omega \to \Syp(\Strspc)$:
\begin{align}  \label{eq:elast-mean}
   \D{E}_{\vtheta_E}(\eTns_r)  &:= \arg \min_{\Tbar{C}\in\Syp} \Psi_{\vtheta_E}(\Tbar{C}),
    \quad \text{ with}\\       \label{eq:elast-var-C}
    \Psi_{\vtheta_E}(\Tbar{C})  &:= \int_{\Omega} \vtheta_E(\eTns_r(\omega),\Tbar{C})^2\, 
       \D{P}(\di \omega) .
\end{align}
It is not difficult to see that the elasticity Fr\'echet mean $\D{E}_{\vtheta_E}$
in \feq{eq:elast-mean} will satisfy all the desiderata for a mean formulated in
\fsec{SS:rep-requirem} in case the $\Tbar{C}$-based variance $\Psi_{\vtheta_E}(\Tbar{C})$ 
in \feq{eq:elast-var-C} satisfies these.  But this follows from the fact that the elasticity metric 
$\vtheta_E$ in \feq{eq:elast-metric} satisfies the requirements for a metric in \fsec{SS:rep-requirem},
which in turn follows from the satisfaction of these requirements by the product distance
$\tilde{\vtheta}_E$ in \feq{eq:elast-dist}.

%
%
%
%
%
%
%
%
%
%
%
%


%
\section{Appendix: Rotations} \label{S:appx-Rotations}

%

A proper rotation is an orthogonal tensor $\vQ \in \SO(n)\subset\E{L}(\Rn)$ , 
i.e.\ $\vQ \vQ^\trpos = \vQ^\trpos \vQ = \vI$ (with the identity
$\vI = (\updelta_{ik})\in\SO(n)$ and the Kronecker-$\updelta$), and hence
$\vQ^{-1} = \vQ^\trpos$, with $\det\vQ = 1$.
These proper orthogonal matrices form the special orthogonal Lie group $\SO(n)$,
a differentiable manifold with a differentiable group operation, 
cf.\ \citep{Lang1995, Moakher2002, cardoso_2010}.  
This group is a connected subgroup of the group of all rotations $\SO(n)\subset \mrm{O}(n)$,
and contains the identity, $\vI \in\SO(n)$.


\subsection{Lie algebra and exponential map}  \label{SS:A-Lie-alg-exp}
The Lie algebra --- the tangent space at the identity --- of  
$\SO(n)$ is denoted by $\so(n)$ and is the vector space of \emph{skew-symmetric}
real matrices, i.e.\ $\vP = -\vP^\trpos$,  \citep{Lang1995, Moakher2002, cardoso_2010}.  
Like in any such Lie algebra, there is an exponential map $\exp: \so(n) \to \SO(n)$.  
Here, this is the common matrix exponential,
and in the case of $\SO(n)$ it is surjective, hence for any $\vQ \in \SO(n)$ there is
a $\vP \in \so(n)$ such that 
\begin{equation}   \label{eq:Lie-exp}
   \vQ = \exp(\vP) ,
\end{equation}
a representation which will be used in the sequel, as it allows one to work in the
vector space $\so(n)$.  The $\log: \SO(n) \to \so(n)$ can be defined as the
main branch of the inverse of \feq{eq:Lie-exp}, defined on a neighbourhood of
the identity $\vI \in \SO(n)$.  For the parametrisation of $\so(n)$, a linear bijective map
$\skwr: \D{R}^{\dim \so(n)} \to \so(n)$ will be defined, which by slight abuse of
notation will always be denoted by $\skwr$, regardless of the dimension $n$.

It may be of interest to recall that
the matrix Lie algebras (like $\so(n)$) have as Lie product the matrix commutator
\begin{equation}     \label{eq:Lie-alg-commut}
  [\vP_1, \vP_2] = \vP_1 \vP_2 - \vP_2 \vP_1.
\end{equation}
If one defines the group commutator on $\SO(n)$ as
\begin{equation}     \label{eq:Lie-grp-commut}
  [\vQ_1, \vQ_2]_{\SO} := \vQ_1 \vQ_2  \vQ_1^{-1} \vQ_2^{-1},
\end{equation}
then it holds with --- using \refeq{eq:Lie-exp} --- $\vQ_1 = \exp(\vP_1)$ and
$\vQ_2 = \exp(\vP_2)$, that
\begin{equation}     \label{eq:Lie-exp-commut}
   [\vQ_1, \vQ_2]_{\SO} = \exp([\vP_1, \vP_2]). 
\end{equation}

\subsection{Rotations in $\SO(\Physpc)$ in 3D and 2D}  \label{SS:A-rot-2-3}
In 3D with $n=d=3$, $\dim \so(3) = 3$, so $\so(3) \cong \D{R}^3$, which is a special
case.  For a vector 
$\vp = (p_i) \in \D{R}^3$, the corresponding skew matrix $\vP\in\so(3)$ is
\begin{equation}  \label{eq:rodrigues-P}
  \vP =  \skwr(\vp) := \begin{bmatrix}
    \phantom{-}0 & -p_3 & \phantom{-}p_2 \\
    \phantom{-}p_3 & \phantom{-}0 & -p_1 \\
    -p_2 & \phantom{-}p_1 & \phantom{-}0
  \end{bmatrix},  \quad \vP = -\vP^\trpos .
\end{equation}
It is sometimes convenient to write $\vp = \theta \vr$ with a unit vector ($\vr\cdot\vr =1$),
so that $| \theta |$ is the length of $\vp$.  Here
\begin{equation}  \label{eq:rodrigues-R}
  \vR :=  \skwr(\vr) \; \text{ is used, s.t. }\; \vP = \theta \vR.
\end{equation}

If $\D{R}^3$ is equipped with the usual vector cross 
product, it becomes a Lie algebra, one has for any $\vx\in \D{R}^3$ that 
$\vP \vx = \vp \times \vx$ with $\vP = \skwr(\vp)$,
and $\skwr$ becomes a Lie algebra isomorphism
\[
   [\vP_1, \vP_2] = \skwr(\vp_1 \times \vp_2), \; \text{ where } \vP_j = \skwr(\vp_j),\, j=1,2.
\]
According to Euler's theorem, a proper rotation $\vQ\in\SO(3)$ with rotation angle 
$\theta$ has a normalised eigenvector $\vr\in \D{R}^3$ with unit eigenvalue, the 
\emph{rotation axis}, such that $\vQ\vr = \vr$ --- and $\vP \vr = \vek{0}$, where
$\vP = \log \vQ$ --- and $\vQ$ may be represented via the Rodrigues formula
\begin{equation}  \label{eq:rodrigues}
  \vQ = \exp(\theta\vR) = \vI + (\sin\theta)\; \vR + (1-\cos\theta)\; \vR^2,\;
  \text{ where } \vR = \skwr(\vr).
\end{equation}

In 2D with $n=d=2$ one has $\dim \so(2) = 1$, hence the Lie algebra 
$\so(2)\cong \D{R}$ of 2D skew matrices and the 
Lie group $\SO(2)$ are one dimensional.  
This is equivalent to a 3D rotation around, say, the 3-axis,
and can easily be seen by only looking at a typical rotation with angle $\theta$
\begin{equation}  \label{eq:2Drotmat}
\vQ = \begin{bmatrix} \cos\theta & -\sin\theta  \\
                  \sin\theta & \phantom{-}\cos\theta  \end{bmatrix} .
\end{equation}
To write in 2D the equivalent of \refeq{eq:rodrigues}, a
convenient basis in $\so(2)$ is 
\begin{equation}  \label{eq:rodrigues-P2}
  \vR = \begin{bmatrix} 0 & -1 \\ 1 & \phantom{-}0 \end{bmatrix},
\end{equation}
and all matrices in $\so(2)$ are scalar multiples of it, so that
\begin{equation}  \label{eq:exp-Q2}
   \vQ= \exp(\theta\vR) = (\cos\theta)\, \vI + (\sin\theta)\, \vR 
   = \begin{bmatrix} \cos\theta & - \sin\theta \\ 
        \sin\theta & \phantom{-} \cos\theta \end{bmatrix}\in \SO(2).
\end{equation}

Observe that any $\vP \in \so(n)$ has purely imaginary or vanishing eigenvalues, and the 
exponential map has period $2 \uppi$ parallel to the imaginary axis, hence this is inherited
by the Rodrigues formulas \feqs{eq:rodrigues}{eq:exp-Q2}, as can also be seen by
the explicit expressions in terms of trigonometric function of the angle $\theta$.
Obviously a rotation of $\theta = \uppi$ about some axis is the same as a rotation
of $\theta = -\uppi$.  Hence those maps are only unique within a ball of radius $\uppi$
for the vector $\vp \in \D{R}^3$ in \feq{eq:rodrigues-P}, resp.\ the interval $]-\uppi,\uppi]$
for the angle $\theta$ in 2D in \feq{eq:exp-Q2}.

\subsection{Rotations in $\SO(\Strspc)$}  \label{SS:A-rot-Strspc}
What will be needed beyond the 2- and 3D cases just described is the modelling of
rotations in the strain-space $\Strspc \equiv \Rk.$  The maximum number of dimensions here
is $\dim \Strspc = k = 6$ for the 3D case, but occasionally only a subspace is needed.
The special orthogonal group $\SO(\Strspc)$ has dimension of its Lie algebra $\so(\Strspc)$,
the space of skew symmetric matrices.  As $\dim \so(\Rk) = k(k-1)/2 $, one obtains here
in the case $\Strspc \equiv \Rk \equiv \D{R}^6$ that $\dim \so(\D{R}^6) = 15$.  We shall describe
the mapping $\skwr: \D{R}^{\dim \so(\Strspc)} \to \so(\Strspc)$, 
already given for the 2- and 3D cases in the previous \fsec{SS:A-rot-2-3}.

Denote the canonical unit vectors in $\D{R}^{15}$ by $\ve_1,\dots,\ve_{15}$, and a general vector
$\vr = \sum_{j=1}^{15} p_j \ve_j \in \D{R}^{15}$.  Then we define the mapping $\skwr$, 
which maps $\vp$ to the skew-symmetric matrix $\tP \in \so(6)$, and which is completely defined
by its strictly upper triangle, via
\begin{equation}  \label{eq:so-6-R-15}
\tP = \skwr(\vp) :=
  \begin{bmatrix}
    0 &       p_{15} &       p_{10} & p_{6} & p_{3} & p_{1} \\
    \cdot & \phantom{0}0 &       p_{14} & p_{9} & p_{5} & p_{2} \\
    \cdot & \phantom{0}\cdot & \phantom{0}0 &       p_{13} & p_{8} & p_{4} \\
    \cdot & \phantom{0}\cdot & \phantom{0}\cdot & \phantom{0}0 &       p_{12} & p_{7} \\
    \cdot & \phantom{0}\cdot & \phantom{0}\cdot & \phantom{0}\cdot & \phantom{0}0 &      p_{11} \\
    \cdot & \phantom{0}\cdot & \phantom{0}\cdot & \phantom{0}\cdot & \phantom{0}\cdot & \phantom{0}0 
  \end{bmatrix} ,
\end{equation}
i.e.\ by the 5 upper diagonals,
and where the strictly lower triangle has to be filled-in anti-symmetrically.  Again,
by slight abuse of notation, we continue to designate this map by $\skwr$.  In case the
map is used for dimensions $3 < k < 6$, one may just use the strict upper triangle of
the $k$ upper diagonals, set the main diagonal to zero,
and complete the square matrix anti-symmetrically.  It may be noted that for dimensions
$k=2, 3$ this deviates from what was described in \feqs{eq:rodrigues-P}{eq:rodrigues-P2} by
a permutation of indices resp.\ sign.  This is due to the fact that in 3D one wants the
parameter vector $\vp$ to be the eigenvector resp.\ rotation axis.  For these cases $k=2, 3$
we shall stay with the convention as in \feqs{eq:rodrigues-P}{eq:rodrigues-P2}, and only
use \feq{eq:so-6-R-15} for $k > 3$.

\paragraph{$\SO(\Physpc)$-induced rotations in $\SO(\Strspc)$:}  
As already described in \fsec{SS:C-gen-notation}, a rotation $\vQ\in \SO(\Physpc)$ of the 
coordinate system in $\Physpc$ induces a transformation of
the stress and strain tensors as in \feq{eq:tens-Q-transform-q} 
\citep{mehrCow1990, MehrabadiCowinJaric1995}.  This is an element
$\tQ = \Trep(\vQ) \in \SO(\Strspc)$, given by an injective group homomorphism
$\Trep: \SO(\Physpc) \to \ST \subset \SO(\Strspc)$.
A similar correspondence exists on the corresponding Lie algebras:
$\tR = \trep(\vR) \in \so(\Strspc)$, given by a map 
$\trep: \so(\Physpc) \to \st \subset \so(\Strspc)$.

\paragraph{Rotations induced from 3D, i.e.\ $d=3$, $k=6$:}
In the 3D ($\Physpc \equiv \D{R}^3$ and $\Strspc \equiv \D{R}^6$) case of the Kelvin representation
in $\Strspc$, the rotation $\vQ$  is represented \citep{mehrCow1990, MehrabadiCowinJaric1995} 
by an orthogonal matrix 
$\tQ = \Trep(\vQ) \in \ST\subset\SO(\Strspc) \subset \Strspc^{\otimes 2} = \E{L}(\Strspc)$,
so that \feq{eq:tens-Q-transform-q} becomes $\Ttil{s} = \tnb{Q}\strsv$ and 
$\Ttil{e} = \tQ\strnv$, where in terms of the components $(Q_{ik}) = \vQ\in\SO(3)$ in 3D
the map $\Trep: \SO(3) \to \ST \subset \SO(\Strspc)$ is
\begin{multline}  \label{eq:nVoigt-orth}
  \tQ    =  \Trep(\vQ) :=\\
{ \scriptsize
 \begin{bmatrix} 
   Q_{11}^2 & Q_{12}^2 & Q_{13}^2 & \sqz Q_{12}Q_{13} & \sqz Q_{11}Q_{13} & \sqz Q_{11}Q_{12} \\  
   Q_{21}^2 & Q_{22}^2 & Q_{23}^2 & \sqz Q_{22}Q_{23} & \sqz Q_{21}Q_{23} & \sqz Q_{22}Q_{21} \\  
   Q_{31}^2 & Q_{32}^2 & Q_{33}^2 & \sqz Q_{33}Q_{32} & \sqz Q_{33}Q_{31} & \sqz Q_{31}Q_{32} \\  
   \sqz Q_{21}Q_{31} & \sqz Q_{22}Q_{32} & \sqz Q_{23}Q_{33} & Q_{22}Q_{33}+ Q_{23}Q_{32} & Q_{21}Q_{33}+ Q_{23}Q_{31} & Q_{21}Q_{32}+ Q_{22}Q_{31} \\  
   \sqz Q_{11}Q_{31} & \sqz Q_{12}Q_{32} & \sqz Q_{13}Q_{33} & Q_{12}Q_{33}+ Q_{13}Q_{32} & Q_{11}Q_{33}+ Q_{13}Q_{31} & Q_{11}Q_{32}+ Q_{12}Q_{31} \\  
   \sqz Q_{11}Q_{21} & \sqz Q_{12}Q_{22} & \sqz Q_{13}Q_{23} & Q_{12}Q_{23}+ Q_{13}Q_{22} & Q_{11}Q_{23}+ Q_{13}Q_{21} & Q_{11}Q_{22}+ Q_{12}Q_{21}  
   \end{bmatrix}.
   } 
\end{multline}

To translate \feq{eq:rodrigues}, where $\vQ = \exp(\theta\vR)$ and $\vR=\skwr(\vr)$ is
formed from the normalised rotation axis vector $\vr = (r_j) \in \D{R}^3$ as in \feq{eq:rodrigues-P},
one may define a map $\trep: \so(3) \to \st\subset\so(6)$ (and combine it with 
$\skwr: \D{R}^3 \to \so(3)$ in \feq{eq:rodrigues-P})
\begin{equation}  \label{eq:rodrigues-P6}
  \tR = \trep(\vR) := { \scriptsize \begin{bmatrix}
  0 & 0 & 0 & 0 & \phantom{-}\sqz\,r_2 & -\sqz\,r_3 \\
  0 & 0 & 0 & -\sqz\,r_1 & 0 & \phantom{-}\sqz\,r_3 \\
  0 & 0 & 0 & \phantom{-}\sqz\,r_1 & -\sqz\,r_2 & 0 \\
  0 & \phantom{-}\sqz\,r_1 & -\sqz\,r_1 &  0 & \phantom{-}r_3 & -r_2 \\
   -\sqz\,r_2 & 0 & \phantom{-}\sqz\,r_2 & -r_3 & 0 & \phantom{-}r_1 \\
   \phantom{-}\sqz\,r_3 & -\sqz\,r_3 & 0 & \phantom{-}r_2 & -r_1 & 0
  \end{bmatrix} }  = \trep(\skwr(\vr)).
\end{equation}
As the Lie group $\SO(3)$ and its Lie algebra $\so(3)$ are three-dimensional, 
the image $\ST := \im \Trep \subset \SO(\Strspc)$ of $\Trep$ is a three-dimensional 
Lie subgroup $\ST$ of the full 15-dimensional Lie group $\SO(\Strspc)$
of proper rotations on $\Strspc$, and a similar relation holds for the Lie algebras:
\begin{equation}  \label{eq:def-ST-st}
  \ST := \Trep(\SO(\Physpc)) \subset \SO(\Strspc), \quad
  \st := \trep(\so(\Physpc)) \subset \so(\Strspc).
\end{equation}

It may be gleaned from \feq{eq:rodrigues-P6}, using \feq{eq:so-6-R-15},
 that the canonical unit vectors $\vg_1,
\vg_2, \vg_3 \in \D{R}^3$ are thus represented in $\D{R}^{15}$ by 
\begin{align}  \label{eq:span-Q6}
  \vh_1 &= \trep(\vg_1) = -\sqz \ve_9 + \sqz \ve_{13} + \ve_{11}; \\ \label{eq:span-Q6-2}
  \vh_2 &= \trep(\vg_2) = \phantom{-}\sqz \ve_3 - \sqz \ve_{8\phantom{3}} - \ve_{7}, \text{ and } 
     \\ \label{eq:span-Q6-3}
  \vh_3 &= \trep(\vg_3) = -\sqz \ve_1 + \sqz \ve_{2\phantom{3}}  + \ve_{12},
\end{align}
by which one extends the map $\trep: \so(3) \to \st\subset\so(6)$ to the
parameter spaces, and by slight abuse of notation we keep the same notation $\trep$.
Thus, the three vectors $\{\vh_j\}$ in \feqss{eq:span-Q6}{eq:span-Q6-2}{eq:span-Q6-3} 
span a subspace of parameters for spatial rotations, 
$\C{Q} := \spn \{\vh_1, \vh_2, \vh_3\} \subset \D{R}^{15}$, which
according to the schema \feq{eq:so-6-R-15} is mapped by $\skwr$ onto the
subspace of spatially induced rotations $\skwr(\spn \{\vh_j\}) = \st\subset\so(6)$.
A spatial rotation with rotation vector (cf.\ \feeqs{eq:rodrigues-P}{eq:rodrigues})
$ \theta \,\vr = \theta \,(r_1,r_2,r_3) \in \D{R}^3$ is thus mapped onto 
$\vq = \theta \, \sum_{j=1}^3 r_j \vh_j \in \C{Q}$, and then onto 
$\theta \, \tR$, see \feq{eq:rodrigues-P6}.

The Rodrigues formula \feq{eq:rodrigues} then becomes \citep{cowMehr1995a, MehrabadiCowinJaric1995}:
\begin{multline}  \label{eq:rodrigues6}
  \tQ =  \Trep(\vQ) = \exp(\theta\tR) =
   \tI + (\sin\theta)\; \tR + (1-\cos\theta) \tR^2 \\ +  
    \frac{1}{3}(\sin\theta)\;(1-\cos\theta) (\tR + \tR^3) + 
    \frac{1}{6}(1-\cos\theta)^2 (\tR^2+\tR^4),
\end{multline}
such that 
$\tQ=\Trep(\vQ) = \exp(\theta\vR) = \exp(\theta\;\trep(\vR)) = \exp(\theta\;\trep(\skwr(\vr)))$.

It is easy to see that the orthogonal complement $\C{Q}^\perp \subset \D{R}^{15}$
--- orthogonal w.r.t.\ the canonical inner product in $\D{R}^{15}$ --- of the
subspace $\C{Q} \subset \D{R}^{15}$ of parameters for spatial rotations 
is spanned by the vectors
\begin{align}  \nonumber
   \vk_1 &= \ve_{4}; \quad \vk_2 = \ve_{5}; \quad \vk_3 = \ve_6; \quad
   \vk_4 = \ve_{10}; \quad \vk_5 = \ve_{14}; \quad \vk_{6} = \ve_{15}; \\
   \vk_{7} &= \ve_{9} + \ve_{13}; \quad
   \vk_{8} = \ve_{3} + \ve_{8}; \quad
   \vk_{9} = \ve_{1} + \ve_{2}; \label{eq:span-V6}  \quad 
   \vk_{10} = \ve_{9} - \ve_{13} + 2\sqz\ve_{11};\\ 
   \vk_{11} &= \ve_{3} - \ve_{8} + 2\sqz\ve_{7}; \quad 
   \vk_{12} = \ve_{1} - \ve_{2} + 2\sqz\ve_{12}. \nonumber  
\end{align}
A rotation in $\SO(\Strspc)$ not originating from a spatial rotation is thus
described by parameters $(p_1,\dots,p_{12})$ and through the vector
$\vp = \sum_{j=1}^{12} p_j \vk_j \in \C{Q}^\perp$, which is then mapped onto $\tP$ in
\feq{eq:so-6-R-15}.

\paragraph{Rotations induced from 2D, i.e.\ $d=2$, $k=3$:}
As in the 3D case, in the 2D instance (with $\Physpc \equiv \D{R}^2$ and $\Strspc \equiv \D{R}^3$) 
of the Kelvin representation in $\Strspc$, the rotation $\vQ\in\SO(2)$  is 
represented \citep{mehrCow1990, MehrabadiCowinJaric1995} by an orthogonal matrix 
$\tQ = \Trep(\vQ) \in \ST\subset\SO(\Strspc) \subset \Strspc^{\otimes 2} = \E{L}(\Strspc)$.
Again, by a slight abuse of notation the transform is again denoted by $\Trep:\SO(2)\to\ST$,
and similarly for $\trep: \so(2) \to \st\subset\so(\Strspc)$ later.
In this instance again \feq{eq:tens-Q-transform-q} becomes $\Ttil{s} = \tnb{Q}\strsv$ and 
$\Ttil{e} = \tQ\strnv$, where in terms of the components $(Q_{ik}) = \vQ\in\SO(2)$ in 2D
given by \feq{eq:2Drotmat}, the mapping $\Trep: \SO(2) \to \ST \subset \SO(\Strspc)$ is
\begin{multline}  \label{eq:nVoigt-orth2D}
  \tQ    =  \Trep(\vQ) := { \small
   \begin{bmatrix} 
   \cos^2\theta & \sin^2\theta &  -\sqz \;\sin\theta \cos\theta \\  
   \sin^2\theta & \cos^2\theta &  \phantom{-}\sqz\; \sin\theta \cos\theta \\  
   \phantom{-}\sqz\; \sin\theta \cos\theta & -\sqz\; \sin\theta \cos\theta & \cos^2\theta - \sin^2\theta  
   \end{bmatrix} } \\ =
  { \small \frac{1}{2}
 \begin{bmatrix} 
   1+\cos 2\theta & 1-\cos 2\theta &  -\sqz\; \sin 2\theta \\  
   1-\cos 2\theta & 1+\cos 2\theta &  \phantom{-}\sqz\; \sin 2\theta \\  
   \phantom{-}\sqz\; \sin 2\theta & -\sqz\; \sin 2\theta & \quad\;\;\, 2\;\cos 2\theta  
   \end{bmatrix} } \in \ST.
\end{multline}
Just as in \feq{eq:rodrigues6}, this also can be written in the form of an exponential
\begin{equation}  \label{eq:rodrigues3}
  \tQ = \exp(\theta\tR) =
   \tI + \frac{\sin(2\theta)}{2}\; \tR - \frac{1+\cos(2\theta)}{4} \tR^2 \in \ST,
\end{equation}
where from \feq{eq:rodrigues-P6} one sees that in \feq{eq:rodrigues3}, with $\vR\in\so(2)$ given
by \feq{eq:rodrigues-P2}, one has that
\begin{equation}  \label{eq:rodrigues-P23}
  \tR = \trep(\vR) := \begin{bmatrix}
  0 & 0  & -\sqz \\
  0 & 0  & \phantom{-}\sqz \\
   \phantom{-}\sqz & -\sqz & 0 
  \end{bmatrix} \in \st
\end{equation}
is a constant matrix.  This is a reflection of the fact that
the matrices $\tQ\in \ST \subset \SO(\Strspc)$ of the form \refeq{eq:nVoigt-orth2D},
resulting from an underlying spatial $\vQ\in \SO(2)$ in $\Physpc=\D{R}^2$ are only
a one-dimensional Lie subgroup $\ST\subset \SO(\Strspc)$ of $\SO(2)=\SO(\Physpc)$-induced rotations of 
$\SO(\Strspc)$.  And as $\SO(2)$ and $\so(2)$ are one-dimensional, hence so are their images $\ST
\subset \SO(\Strspc)$ and $\st\subset \so(\Strspc)$ under $\Trep$ resp.\ $\trep$ in \feq{eq:def-ST-st}. 

As $\dim \SO(\Strspc) = 3$, the parameter space is $\D{R}^3$ with the canonical basis 
$\{ \vg_1, \vg_2, \vg_3 \}$.  One may observe with the help of \feq{eq:rodrigues-P} that
$\vh_1 := - \sqz (\vg_1 + \vg_2)$ is mapped onto $\tR$ in \feq{eq:rodrigues-P23}, and thus 
$\spn \{ \vh_1 \} = \C{Q} \subset \D{R}^3$ is the subspace that describes spatial rotations,
such that with a rotation angle $\theta$ one has with \feq{eq:rodrigues}
\begin{equation}  \label{eq:q-to-Q-in-23}
  \C{Q} \to \st \to \ST \subset \SO(\Strspc); \quad
  q := \theta  \mapsto q\, \vh_1 \mapsto \theta \tR \mapsto \tQ = \exp(\theta \tR). 
\end{equation}

Obviously, not every $\tU\in\SO(\Strspc)$ is of that spatially induced form, and in the parameter
space one may take the orthogonal complement $\C{Q}^\perp$ of $\C{Q}$ with basis
\begin{equation}  \label{eq:Q-perp}
 \C{Q}^\perp = \spn \{ \vk_1 := \sqz(\vg_1 - \vg_2), \vk_2 := \vg_3 \}.
\end{equation}
A not spatially induced rotation in $\SO(\Strspc)$ is thus described by two parameters
$(s_1,s_2)$, which lead with the use of \feqs{eq:rodrigues-P}{eq:rodrigues} to a mapping
\begin{equation}   \label{eq:s-to-P-in-23}
  \C{Q}^\perp \to \st^\perp \to \SO(\Strspc); \quad
  (s_1,s_2) \mapsto \vp := s_1 \vk_1 + s_2 \vk_2 \mapsto \tP \mapsto \tV = \exp(\tP) .
\end{equation}



%


%
%
%
%
%
%
%
%
%
%
%


%
\section{Appendix: Notation} \label{S:appx-Notation}

%

%
%

With physical space given by $\Physpc \equiv \Rd$ (the physically interesting instances
are $d=1,2,3$), we are interested in elements of various tensor powers $\Physpc^{\otimes q}$. 

Denote vectors as $\vx = (x_i) \in \Physpc$ with lower case boldface, 
second order tensors (identified with linear maps)
as $\vQ = (Q_{ik})\in\Physpc\otimes\Physpc = \Physpc^{\otimes 2} \cong \E{L}(\Physpc)$
 with upper case boldface,
the identification given via the matrix-vector product  $\vy = (y_i)=  \vQ\vx = (Q_{ik} x_k)$,
where Einstein's summation convention over repeated indices is used.
  Linear maps are identified with the associated matrices.  
Matrix-matrix multiplication resp.\ the concatenation of maps in $\E{L}(\Physpc)$ is denoted
similarly as $\vP=\vQ\vR = (Q_{ik}R_{kj}) = (P_{ij})$.

Tensors of order $q>2$ will be denoted as $\mat{w} = (\mrm{w}_{i_1 i_2\dots i_q}) = 
(\mrm{w}_{\F{i}}) \in  \Physpc^{\otimes q}$ in lower case bold face roman, and their components
in lower case normal face roman, with a multi-index $\F{i} = (i_1, i_2,\dots, i_q) \in \D{N}^q$
denoted in lower case fraktur.



\subsection{Kelvin notation}  \label{SS:C-Kelvin}
%
Applying the foregoing to the case of linear elasticity, the state of the material
is described by the symmetric infinitesimal strain and Cauchy stress tensors 
$\strnt,\strst \in \SymPh$, where
$\SymPh := \E{L}_s(\Physpc) = \sy(\Physpc)\subset \Physpc^{\otimes 2}$ 
is the space
of symmetric 2nd order tensors / matrices resp.\ symmetric linear maps  of $\Physpc$: 
$\strnt = (\veps_{ij}) = (\veps_{ji}) = \strnt^\trpos$, and similarly for $\strst$.  
In $\SymPh$ the inner product is induced by the one in $\Physpc$, and denoted by 
$\strst:\strnt = \ip{\strst}{\strnt} = (\sigma_{ik}\vepsilon_{ik})$,
a double contraction.

The strain and stress tensor are connected through Hooke's generalised law 
\feqs{eq:Hooke-1-p}{eq:Hooke-tns} by the elasticity tensor 
$\etns = (\mrm{c}_{i_1 i_2 i_3 i_4}) \in \SymTen := \E{L}_s(\SymPh) \subset \Physpc^{\otimes 4}$,
which according to \feq{eq:tns-as-linmap} can be seen as a symmetric linear map on $\SymPh$.
Hooke's elasticity tensor $\etns$ is a 4th order tensor in the space  
$\etns \in \SymTen \subset \Physpc^{\otimes 4}$ with major (physical) symmetry 
$\mrm{c}_{ik\ell m} = \mrm{c}_{\ell mik}$, as it is a symmetric linear map 
$\etns \in \E{L}_s(\SymPh)$ on $\SymPh$ (cf.\ \feq{eq:map-sym}), and minor symmetries 
$\mrm{c}_{ik\ell m} = \mrm{c}_{ki\ell m} = \mrm{c}_{ikm\ell}$,
as it operates on symmetric second-order tensors in $\SymPh$, cf.\ \citep{Moakher2008}.
\begin{equation}  \label{eq:Hooke-tns}
  \text{Hooke's generalised law: }\quad
     \strst = (\sigma_{i_1 i_2}) = \etns : \strnt = (\mrm{c}_{i_1 i_2 i_3 i_4} \veps_{i_3 i_4}).
\end{equation}  
For $\dim\,\Physpc =d=3$ one has $\dim\, \SymPh = 6$, and for $\dim\,\Physpc =d=2$ 
it is $\dim\, \SymPh = 3$.
For $\dim\,\Physpc =d=3$ one has $\dim \SymTen = 21$, and for $\dim\,\Physpc =d=2$ 
it is $\dim \SymTen = 6$.  The elasticity tensor is usually also SPD, see \feq{eq:map-SPD}.


\paragraph{Hooke's law in Kelvin's matrix notation:}  
As already alluded to following \feq{eq:tns-as-linmap} as well as \feqs{eq:Hooke-1-p}{eq:Hooke-tns}, 
one may associate a linear map to the elasticity tensor $\etns \in \SymTen$.
It is often simplest done by choosing a basis in the strain-stress space
$\SymPh$.  One such way to assign a matrix to the elasticity
tensor $\etns \in \SymTen$ is the well-known 
\emph{Voigt notation} by choosing different bases in $\SymPh$
for strains and for stresses.  Unfortunately, the resulting matrix does not transform as a 
tensor \citep{mehrCow1990, Moakher2008}, and also would make the formulation of eigenvectors 
awkward \citep{helbig2013}. 
Therefore it is advantageous to choose the same basis for stresses and strains, and 
thereby defining a new space $\Strspc = \Rk$, where $k=\dim\,\SymPh$, 
i.e.\ $k=6$ for $d=3$ and $k=3$ in case $d=2$.  This gives a one-to-one 
correspondence from $\SymPh$ to $\Strspc$,
a map which will be designated by $\vrep: \SymPh \to \Strspc$.
This is classically known as the \emph{Kelvin notation}, cf.\ \citep{mehrCow1990, Moakher2008, 
helbig2013}.

\paragraph{Kelvin notation in 3D:}
Following Kelvin \citep{kelvin1856, kelvin1878}, see also \citep{Helbig1996, helbig2013} for
an account in modern notation, one chooses in 3D for $d=3$ and $k=6=\dim\SymPh$
\emph{one} basis for strains and stresses in $\SymPh$ \citep{mehrCow1990, Moakher2008}, 
such that the stress and strain $\strst, \strnt \in \SymPh$ is represented by
\begin{align}  \label{eq:Kv-vrep3-strs}
  \strsv &:= (\tns{s}_i) = \vrep(\strst) = \vrep((\sigma_{ij}))
  := [\sigma_{11}, \sigma_{22}, \sigma_{33}, 
     \sqz\,\sigma_{23}, \sqz\,\sigma_{13}, \sqz\,\sigma_{12}]^\trpos \in  \Strspc \\
   \label{eq:Kv-vrep3-strn}
  \strnv &:= (\tns{e}_i) = \vrep(\strnt) = \vrep((\veps_{ij}))
  := [\veps_{11}, \veps_{22}, \veps_{33}, 
     \sqz\,\veps_{23}, \sqz\,\veps_{13}, \sqz\,\veps_{12}]^\trpos \in \Strspc .
\end{align}
This way the elastic energy is $\frk{1}{2}(\strst:\strnt) = \frk{1}{2}(\strsv\cdot\strnv)
=\frk{1}{2}(\strsv^\trpos\strnv)$.  While this would also be true in the Voigt notation,
in the Kelvin notation one has additionally $\strst:\strst = \strsv^\trpos\strsv$ and
$\strnt:\strnt = \strnv^\trpos\strnv$, as the same inner product is used for stresses and strains,
facilitating the consideration of eigenvalue problems for the elasticity tensor.

Note that elements of $\Strspc$ such as $\strsv$ are denoted by lower case bold sans-serif
letters to distinguish them from vectors in $\Physpc$ such as $\vx$.  Linear
maps $\E{L}(\Strspc)$ on $\Strspc$ like the Kelvin matrix $\eTns \in \E{L}(\Strspc)$ 
later in \feq{eq:Hooke-vec} are denoted by upper case bold sans-serif letters
to distinguish them from linear maps on $\Physpc$ like $\vQ\in\E{L}(\Physpc)$,
but the matrix notation from $\Physpc \equiv \Rd$ is also used on $\Strspc \equiv \Rk$.

Sometimes it is useful to consider other, slightly different, basis vectors in $\Strspc$ 
than the canonical unit vectors $\te_1,\dots,\te_6 \in \Strspc$ 
implied in \feqs{eq:Kv-vrep3-strs}{eq:Kv-vrep3-strn}, which come from the Cartesian
form of the tensors $\strst, \strnt \in \SymPh$.  Recall that in 3D the trace of the strain
tensor, $\tr \strnt$, is the volume
change, and $- \frk{1}{3}\tr \strst$ is the pressure, and one would like to separate this
from shear effects.  To achieve this distinction one may introduce the orthonormal vectors
\begin{alignat}{2}  
  \label{eq:n-rep3D}
  \tn &= \frac{1}{\sqrt{3}} (\te_1 + \te_2 + \te_3)&{}=&{}\frac{1}{\sqrt{3}}\;
       [\phantom{-}1,\phantom{-}1,\phantom{-}1,\phantom{-}0,\phantom{-}0,\phantom{-}0]^\trpos , \\ 
  \label{eq:y-rep3D}
  \ty &= \frac{1}{\sqrt{2}} (\te_2 - \te_1 )& =& \frac{1}{\sqrt{2}}\;
       [-1,\phantom{-}1,\phantom{-}0,\phantom{-}0,\phantom{-}0,\phantom{-}0]^\trpos , \text{ and}\\
  \label{eq:z-rep3D}
  \tz &= \frac{1}{\sqrt{6}} (\te_1 + \te_2 - 2\te_3)& =& \frac{1}{\sqrt{6}}\;
       [\phantom{-}1,\phantom{-}1,-2,\phantom{-}0,\phantom{-}0,\phantom{-}0]^\trpos .
\end{alignat}
Thus a projection on $\tn \in \Strspc$ yields a quantity proportional to volume 
change resp.\ pressure, and the projections onto the orthonormal vectors 
$\ty, \tz, \te_4, \te_5, \te_6 \in \Strspc$ are pure shears.

The linear vector representation map $\vrep: \SymPh \to \Strspc$ in 
\feqs{eq:Kv-vrep3-strs}{eq:Kv-vrep3-strn} induces following \feq{eq:tns-as-linmap} 
a linear representation map 
$\Vrep: \SymTen \to \E{L}_s(\Strspc)$, which
maps $\etns\in\SymTen$ onto a symmetric matrix $\eTns \in \E{L}_s(\Strspc)  = \sy(\Strspc)$, 
$\etns \mapsto \eTns = \Vrep(\etns)$, such that Hooke's generalised law in \feq{eq:Hooke-tns} becomes
\begin{equation}  \label{eq:Hooke-vec}
  \strsv = \eTns \strnv .
\end{equation}
This map  $\eTns = \Vrep(\etns)$ can be
given in terms of the tensor components of $\etns = (\mrm{c}_{ik\ell m})$, $(i,k,\ell,m,=1,\dots,3)$,
yielding the 3D \emph{Kelvin matrix} \feq{eq:Kelvin-C-3D-1} with 
$\dim \SymTen = 21$ independent components.

\paragraph{Kelvin notation in 2D:}
In 2D for $d=2$, $\Physpc=\D{R}^2$, and $k=3=\dim\SymPh$, one again chooses
the same basis \citep{mehrCow1990, Moakher2008} for strains and stresses
$\strst, \strnt \in \SymPh$, the representation now being --- compare 
\feqs{eq:Kv-vrep3-strs}{eq:Kv-vrep3-strn}:
\begin{align}  \label{eq:Kv-vrep2-strs}
  \strsv &:= \vrep(\strst) := [\sigma_{11}, \sigma_{22}, \sqz\,\sigma_{12}]^\trpos \in  \Strspc \\
   \label{eq:Kv-vrep2-strn}
  \strnv &:= \vrep(\strnt) := [\vepsilon_{11}, \vepsilon_{22}, \sqz\,\vepsilon_{12}]^\trpos \in\Strspc .
\end{align}

As in the 3D case above, sometimes one needs slightly different basis 
vectors in $\Strspc$ than the canonical unit vectors $\te_1,\te_2,\te_3 \in \Strspc$ 
implied in \feqs{eq:Kv-vrep2-strs}{eq:Kv-vrep2-strn}.  In 2D $\tr \strnt$ is the area
change,  and one would like to separate this from shear effects.  To this end 
introduce the orthonormal vectors
\begin{alignat}{2}    \label{eq:n-rep2D}
  \tn &= \frac{1}{\sqrt{2}} (\te_1 + \te_2)&{}=&{}\frac{1}{\sqrt{2}}\;
       [\phantom{-}1,\phantom{-}1,\phantom{-}0]^\trpos , \\ 
  \label{eq:y-rep2D}
  \ty &= \frac{1}{\sqrt{2}} (\te_2 - \te_1 )& =& \frac{1}{\sqrt{2}}\;
       [-1,\phantom{-}1,\phantom{-}0]^\trpos .
\end{alignat}
Then a projection on $\tn \in \Strspc$ yields a quantity proportional to volume 
change resp.\ pressure, and the projections onto the orthonormal vectors 
$\ty, \te_3 \in \Strspc$ are pure shears.

Again, the linear vector representation $\vrep: \SymPh \to \Strspc$ in 
\feqs{eq:Kv-vrep2-strs}{eq:Kv-vrep2-strn} induces a linear representation 
$\Vrep: \SymTen \to \E{L}_s(\Strspc) = \sy(\Strspc)$, mapping $\etns\in\SymTen$ onto a 
matrix $\eTns \in  \E{L}_s(\Strspc)$, and is given in terms of the 
tensor components of $\etns = (\mrm{c}_{ik\ell m})$, $(i,k,\ell,m,=1,\dots,2)$, by
the 2D Kelvin matrix \feq{eq:Kelvin-C-2D-1} with $\dim \SymTen = 6$ independent components.


%
%
%
%
%
%
%
%
%
%


\section*{Acknowledgments}
We are grateful to acknowledge that
the research reported in this publication was partly supported by funding 
from the Deutsche Forschungsgemeinschaft (DFG) (sks),
and a Gay-Lussac Humboldt Research Award (hgm).


\bibliographystyle{hgmplain-1}

\bibliography{\thebib/jabbrevlong,\thebib/references%
}

\providecommand{\bysame}{\leavevmode\hbox to3em{\hrulefill}\thinspace}
\providecommand{\MR}{\relax\ifhmode\unskip\space\fi MR }
\providecommand{\MRhref}[2]{%
  \href{http://www.ams.org/mathscinet-getitem?mr=#1}{#2}
}
\providecommand{\href}[2]{#2}
\begin{thebibliography}{100}

\bibitem{AlexandrinoBettiol2015}
M.~M. Alexandrino and R.~G. Bettiol, \emph{{Lie Groups and Geometric Aspects of
  Isometric Actions}}, Springer, 2015.

\bibitem{AndoLiMathias2004}
T.~Ando, C.-K. Li, and R.~Mathias, \emph{Geometric means}, Linear Algebra and
  Its Applications \textbf{385} (2004), 305--334, \href
  {http://dx.doi.org/10.1016/j.laa.2003.11.019}
  {\path{doi:10.1016/j.laa.2003.11.019}}.

\bibitem{annin2008}
B.~D. Annin and N.~I. Ostrosablin, \emph{Anisotropy of elastic properties of
  materials}, J Appl Mech and Tech Phys \textbf{49} (2008), 998--1014.

\bibitem{ArsignyFillardPennecEtAl2006}
V.~Arsigny, P.~Fillard, X.~Pennec, and N.~Ayache, \emph{{Log-Euclidean} metrics
  for fast and simple calculus on diffusion tensors}, Magn Reson Med.
  \textbf{56} (2006), no.~2, 411--421, \href
  {http://dx.doi.org/10.1002/mrm.20965} {\path{doi:10.1002/mrm.20965}}.

\bibitem{ArsignyFillardPennecEtAl2007}
\bysame, \emph{Geometric means in a novel space structure of symmetric positive
  definite matrices}, SIAM J. Matrix Anal. Appl. \textbf{29} (2007), no.~1,
  328--347, \href {http://dx.doi.org/10.1137/05063799}
  {\path{doi:10.1137/05063799}}.

\bibitem{ashman_continuous_1984}
R.~Ashman, S.~Cowin, W.~Van~Buskirk, and J.~Rice, \emph{A continuous wave
  technique for the measurement of the elastic properties of cortical bone},
  Journal of Biomechanics \textbf{17} (1984), no.~5, 349--361.

\bibitem{auffrayEtal2014}
N.~Auffray, B.~Kolev, and M.~Petitot, \emph{On anisotropic polynomial relations
  for the elasticity tensor}, J Elast \textbf{115} (2014), 77--103, \href
  {http://dx.doi.org/10.1007/s10659-013-9448-z}
  {\path{doi:10.1007/s10659-013-9448-z}}.

\bibitem{BabuskaNobileTemp07}
I.~Babu\v{s}ka, F.~Nobile, and R.~Tempone, \emph{On obtaining effective
  elasticity tensors with entries zeroing method}, SIAM Journal on Numerical
  Analysis \textbf{45} (2007), no.~3, 1005--1034, \href
  {http://dx.doi.org/10.1137/050645142} {\path{doi:10.1137/050645142}}.

\bibitem{BachmayrCohenDeVoreEtAl2017}
M.~Bachmayr, A.~Cohen, R.~{DeVore}, and G.~Migliorati, \emph{Sparse polynomial
  approximation of parametric elliptic {PDEs}. {Part II}: lognormal
  coefficients}, ESAIM: M2AN \textbf{51} (2017), 341--363, \href
  {http://dx.doi.org/10.1051/m2an/2016051} {\path{doi:10.1051/m2an/2016051}}.

\bibitem{BachmayrCohenMigliorati2018}
M.~Bachmayr, A.~Cohen, and G.~Migliorati, \emph{Representations of {Gaussian}
  random fields and approximation of elliptic {PDEs} with lognormal
  coefficients}, J. Fourier. Anal. Appl. \textbf{24} (2018), 621--649, \href
  {http://dx.doi.org/10.1007/s00041-017-9539-5}
  {\path{doi:10.1007/s00041-017-9539-5}}.

\bibitem{Backus1970}
G.~Backus, \emph{A geometric picture of anisotropic elasticity tensors}, Review
  of Geophysics and Space Physics \textbf{8} (1970), no.~3, 633--671.

\bibitem{BaerheimHelbig1993}
R.~Baerheim and K.~Helbig, \emph{Decomposition of the anisotropic elastic
  tensor in base tensors,}, Canadian Journal of Exploration Geophysics
  \textbf{29} (1993), no.~1, 41--50.

\bibitem{Baerheim1993}
R.~Baerheim, \emph{Harmonic decomposition of the anisotropic elasticity
  tensor}, Q. J. Mech. appl. Math. \textbf{46} (1993), no.~3, 391--418, \href
  {http://dx.doi.org/10.1093/qjmam/46.3.391}
  {\path{doi:10.1093/qjmam/46.3.391}}.

\bibitem{Baerheim1998a}
\bysame, \emph{Classification of symmetry by means of {Maxwell} multipoles},
  Quarterly Journal of Mechanics and Applied Mathematics \textbf{51} (1998),
  no.~1, 73--104, \href {http://dx.doi.org/10.1093/qjmam/51.1.73}
  {\path{doi:10.1093/qjmam/51.1.73}}.

\bibitem{blin1996}
A.~Blinowski, J.~{Ostrowska-Maciejewska}, and J.~Rychlewski,
  \emph{Two-dimensional {H}ooke's tensors --- isotropic decomposition,
  effective symmetry criteria}, Arch Mech \textbf{48} (1996), 325--345.

\bibitem{bonaEtal2004}
A.~B\'ona, I.~Bucataru, and M.~A. Slawinski, \emph{Material symmetries of
  elasticity tensors}, Quart J Mech Appl Math \textbf{57} (2004), 584--598.

\bibitem{BrowaeysChevrot2004}
J.~Browaeys and S.~Chevrot, \emph{Decomposition of the elastic tensor and
  geophysical applications}, Geophys. J. Int. \textbf{159} (2004), 667--678,
  \href {http://dx.doi.org/10.1111/j.1365-246X.2004.02415.x}
  {\path{doi:10.1111/j.1365-246X.2004.02415.x}}.

\bibitem{BucataruSlawinski2008}
I.~Bucataru and M.~A. Slawinski, \emph{Invariant properties for finding
  distance in space of elasticity tensors} [online], arXiv: 0712.1082
  [cond-mat.mtrl-sci], 2008, \href {http://arxiv.org/abs/0712.1082}
  {\path{arXiv:0712.1082}}, \href {http://dx.doi.org/10.48550/arXiv.0712.1082}
  {\path{doi:10.48550/arXiv.0712.1082}}.

\bibitem{bona_coordinate-free_2007}
A.~Bóna, I.~Bucataru, and M.~A. Slawinski, \emph{Coordinate-free
  {Characterization} of the {Symmetry} {Classes} of {Elasticity} {Tensors}},
  Journal of Elasticity \textbf{87} (2007), no.~2--3, 109--132, \href
  {http://dx.doi.org/10.1007/s10659-007-9099-z}
  {\path{doi:10.1007/s10659-007-9099-z}}.

\bibitem{cardoso_2010}
J.~Cardoso and F.~Leite, \emph{Exponentials of skew-symmetric matrices and
  logarithms of orthogonal matrices}, Journal of Computational and Applied
  Mathematics \textbf{233} (2010), no.~11, 2867--2875, \href
  {http://dx.doi.org/10.1016/j.cam.2009.11.032}
  {\path{doi:10.1016/j.cam.2009.11.032}}.

\bibitem{cigdem2010}
\c{C}. Din\c{c}kal, \emph{Decomposition of elastic constant tensor into
  orthogonal parts}, Ph.D. thesis, Graduate School of Natural and Applied
  Sciences, Middle East Technical University, Ankara, 2010.

\bibitem{Dinckal2012}
\bysame, \emph{Irreducible decomposition of elastic constant tensor for
  anisotropic engineering materials}, Materials, Methods \& Technologies
  \textbf{6} (2012), no.~1, 71--92, Available from:
  \url{http://www.scientific-publications.net/download/materials-methods-and-technologies-2012-1.pdf}.

\bibitem{DinckalAkgoez2010}
\c{C}. Din\c{c}kal and Y.~Akg\"oz, \emph{Decomposition of elastic constant
  tensor into orthogonal parts}, International Journal of Engineering, Science
  and Technology \textbf{2} (2010), no.~6, 22--46.

\bibitem{chenNORTA2001}
H.~Chen, \emph{Initialization for {NORTA}: {G}eneration of random vectors with
  specified marginals and correlations}, INFORMS Journal on Computing
  \textbf{13} (2001), no.~4, 312--331, \href
  {http://dx.doi.org/10.1287/ijoc.13.4.312.9736}
  {\path{doi:10.1287/ijoc.13.4.312.9736}}.

\bibitem{Ciarlet1978}
P.~G. Ciarlet, \emph{{The Finite Element Method for Elliptic Problems}},
  North-Holland, 1978.

\bibitem{Ciarlet2013}
\bysame, \emph{{Linear and Nonlinear Functional Analysis with Applications}},
  SIAM, 2013.

\bibitem{Cowin2013}
S.~Cowin, \emph{{Continuum Mechanics of Anisotropic Materials}}, Springer,
  2013, \href {http://dx.doi.org/10.1007/978-1-4614-5025-2}
  {\path{doi:10.1007/978-1-4614-5025-2}}.

\bibitem{cowin_identification_1987}
S.~Cowin and M.~Mehrabadi, \emph{On the identification of material symmetry for
  anisotropic elastic materials}, Quart J Mech Appl Math. \textbf{40} (1987),
  451--476.

\bibitem{CowinMehr1989}
S.~C. Cowin and M.~M. Mehrabadi, \emph{Identification of the elastic symmetry
  of bone and other materials}, J Biomechanics \textbf{22} (1989), no.~6/7,
  503--515.

\bibitem{cowMehr1992}
S.~Cowin and M.~Mehrabadi, \emph{The structure of the linear anisotropic
  elastic symmetries}, J Mech Phys Solids \textbf{40} (1992), 1459--1471.

\bibitem{cowMehr1995}
\bysame, \emph{Anisotropic symmetries of linear elasticity}, ASME Appl Mech Rev
  \textbf{48} (1995), 247--285.

\bibitem{cowMehr1995a}
\bysame, \emph{The mirror symmetries of anisotropic elasticity}, IUTAM
  Symposium on Anisotropy, Inhomogeneity and Nonlinearity in Solid Mechanics
  (D.~Parker and A.~England, eds.), Solid Mechanics and its Applications,
  vol.~39, Springer, 1995, pp.~31--36, \href
  {http://dx.doi.org/10.1007/978-94-015-8494-4}
  {\path{doi:10.1007/978-94-015-8494-4}}.

\bibitem{CowinYang1997}
S.~Cowin and G.~Yang, \emph{Averaging anisotropic elastic constant data}, J
  Elast \textbf{46} (1997), 151--180.

\bibitem{Cressie1993}
N.~A.~C. Cressie, \emph{{Statistics for Spatial Data}}, Wiley Series in
  Probability and Statistics, John Wiley \& Sons, 1993, \href
  {http://dx.doi.org/10.1002/9781119115151} {\path{doi:10.1002/9781119115151}}.

\bibitem{DungSchwabX2022}
D.~{D\~ung}, V.~K. Nguyen, C.~Schwab, and J.~Zech, \emph{Analyticity and
  sparsity in uncertainty quantification for {PDEs} with {G}aussian random
  field inputs} [online], arXiv: 2201.01912 [math.NA], 2022, \href
  {http://arxiv.org/abs/2201.01912} {\path{arXiv:2201.01912}}, \href
  {http://dx.doi.org/10.48550/arXiv.2201.01912}
  {\path{doi:10.48550/arXiv.2201.01912}}.

\bibitem{DanekEtAl2015}
T.~Danek, M.~Kochetov, and M.~A. Slawinski, \emph{Effective elasticity tensors
  in context of random errors}, J Elast \textbf{121} (2015), 55--67, \href
  {http://dx.doi.org/10.1007/s10659-015-9519-4}
  {\path{doi:10.1007/s10659-015-9519-4}}.

\bibitem{DanekSlawinski2015}
T.~Danek and M.~A. Slawinski, \emph{On choosing effective elasticity tensors
  using a {Monte} {Carlo} method}, Acta Geophysica \textbf{63} (2015), no.~1,
  45--61, \href {http://dx.doi.org/10.2478/s11600-013-0197-y}
  {\path{doi:10.2478/s11600-013-0197-y}}.

\bibitem{deGrootMazur1984}
S.~R. {de Groot} and P.~Mazur, \emph{{Non-Equlibrium Thermodynamics}}, Dover,
  1984.

\bibitem{DesmoratAuffraytEtAl2022}
R.~Desmorat, N.~Auffray, B.~Desmorat, M.~Olive, and B.~Kolev, \emph{Minimal
  functional bases for elasticity tensor symmetry classes}, J Elast
  \textbf{147} (2022), 201--228, \href
  {http://dx.doi.org/10.1007/s10659-021-09872-2}
  {\path{doi:10.1007/s10659-021-09872-2}}.

\bibitem{Desmorat2009}
R.~Desmorat, \emph{Décomposition de {Kelvin} et concept de contraintes
  effectives multiples pour les matériaux anisotropes}, C. R. Mecanique
  \textbf{337} (2009), 733--738.

\bibitem{Carmo1992}
M.~P. {do Carmo}, \emph{{Riemannian Geometry}}, Birkh\"auser, 1992.

\bibitem{DobrillaEtAl2023}
S.~Dobrilla, H.~G. Matthies, and A.~Ibrahimbegovic, \emph{Considerations on the
  identifiability of fracture and bond properties of reinforced concrete}, Int
  J Numer Methods Eng \textbf{124} (2023), 3662--3686, \href
  {http://dx.doi.org/10.1002/nme.7289} {\path{doi:10.1002/nme.7289}}.

\bibitem{drydenEtal2009}
I.~L. Dryden, A.~Koloydenko, and D.~Zhou, \emph{Non-{E}uclidean statistics for
  covariance matrices, with applications to diffusion tensor imaging}, Ann.
  Appl. Stat. \textbf{3} (2009), 1102--1123, \href
  {http://dx.doi.org/10.1214/09-AOAS249} {\path{doi:10.1214/09-AOAS249}}.

\bibitem{DrydenEtal2010}
I.~L. Dryden, X.~Pennec, and J.-M. Peyrat, \emph{Power {E}uclidean metrics for
  covariance matrices with application to diffusion tensor imaging} [online],
  arXiv: 1009.3045 [stat.ME], 2010, \href {http://arxiv.org/abs/1009.3045}
  {\path{arXiv:1009.3045}}, \href {http://dx.doi.org/10.48550/arXiv.1009.3045}
  {\path{doi:10.48550/arXiv.1009.3045}}.

\bibitem{EspigHackbuschEtAl2014}
M.~Espig, W.~Hackbusch, A.~Litvinenko, H.~G. Matthies, and P.~Wähnert,
  \emph{Efficient low-rank approximation of the stochastic {Galerkin} matrix in
  tensor formats}, Computers \& Mathematics With Applications \textbf{67}
  (2014), no.~4, 818--829.

\bibitem{faessStiefel1992}
A.~F\"assler and E.~Stiefel, \emph{{Group Theoretical Methods and Their
  Applications}}, Birkh\"auser, 1992.

\bibitem{Fed1968}
F.~I. Fedorov, \emph{{Theory of Elastic Waves in Crystals}}, Plenum Press,
  1968.

\bibitem{FeragenFuster2017}
A.~Feragen and A.~Fuster, \emph{Geometries and interpolations for symmetric
  positive definite matrices}, Modeling, Analysis, and Visualization of
  Anisotropy (T.~Schultz, E.~Özarslan, and I.~Hotz, eds.), Mathematics and
  Visualization, Springer, 2017, pp.~85--113, \href
  {http://dx.doi.org/10.1007/978-3-319-61358-1_5}
  {\path{doi:10.1007/978-3-319-61358-1_5}}.

\bibitem{FittEtAl2019}
D.~Fitt, H.~Wyatt, T.~E. Woolley, and L.~A. Mihai, \emph{Uncertainty
  quantification of elastic material responses: testing, stochastic calibration
  and {B}ayesian model selection}, Mechanics of Soft Materials \textbf{1}
  (2019), 13, \href {http://dx.doi.org/10.1007/s42558-019-0013-1}
  {\path{doi:10.1007/s42558-019-0013-1}}.

\bibitem{Fletcher2020}
T.~Fletcher, \emph{Statistics on manifolds}, {Riemannian} Geometric Statistics
  in Medical Image Analysis (X.~Pennec, S.~Sommer, and T.~Fletcher, eds.),
  Elsevier, 2020, pp.~39--74, \href {http://dx.doi.org/10.1016/b978-
  0-12-814725-2.00008-x} {\path{doi:10.1016/b978- 0-12-814725-2.00008-x}}.

\bibitem{forteVianel1996}
S.~Forte and M.~Vianello, \emph{Symmetry classes for elasticity tensors}, J
  Elast \textbf{43} (1996), 81--108, \href
  {http://dx.doi.org/10.1109/TMI.2003.815059}
  {\path{doi:10.1109/TMI.2003.815059}}.

\bibitem{ForteVianello1998}
\bysame, \emph{Functional bases for transversely isotropic and transversely
  hemitropic invariants of elasticity tensors}, Q. J. Mech. Appl. Math.
  \textbf{51} (1998), 543--552, \href
  {http://dx.doi.org/10.1093/qjmam/51.4.543}
  {\path{doi:10.1093/qjmam/51.4.543}}.

\bibitem{FrancoisEtAl1998}
M.~L.~M. François, G.~Geymonat, and Y.~Berthaud, \emph{On choosing effective
  elasticity tensors using a {Monte} {Carlo} method}, Acta Geophysica
  \textbf{35} (1998), no.~31--32, 4091--4106, \href
  {http://dx.doi.org/10.1016/S0020-7683(97)00303-X}
  {\path{doi:10.1016/S0020-7683(97)00303-X}}.

\bibitem{FujiiSeo2015}
J.~I. Fujii and Y.~Seo, \emph{On the {A}ndo–{L}i–{M}athias mean and the
  {K}archer mean of positive definite matrices}, Linear and Multilinear Algebra
  \textbf{63} (2015), no.~3, 639--649, \href
  {http://dx.doi.org/10.1080/03081087.2014.896359}
  {\path{doi:10.1080/03081087.2014.896359}}.

\bibitem{GhoshHendNORTA03}
S.~Ghosh and S.~G. Henderson, \emph{Behavior of the {NORTA} method for
  correlated random vector generation as the dimension increases}, ACM
  Transactions on Modeling and Computer Simulation (TOMACS) \textbf{13} (2003),
  no.~3, 276--294, \href {http://dx.doi.org/10.1145/937332.937336}
  {\path{doi:10.1145/937332.937336}}.

\bibitem{GierlachDanek2018}
B.~Gierlach and T.~Danek, \emph{On obtaining effective elasticity tensors with
  entries zeroing method}, Geology, Geophysics \& Environment \textbf{44}
  (2018), no.~2, 259--274, \href {http://dx.doi.org/10.7494/geol.2018.44.2.259}
  {\path{doi:10.7494/geol.2018.44.2.259}}.

\bibitem{Gneiting2002}
T.~Gneiting, \emph{Compactly supported correlation functions}, Journal of
  Multivariate Analysis \textbf{83} (2002), 493--508, \href
  {http://dx.doi.org/10.1006/jmva.2001.2056}
  {\path{doi:10.1006/jmva.2001.2056}}.

\bibitem{GneitingEtAl2010}
T.~Gneiting, W.~Kleiber, and M.~Schlather, \emph{{M}at\'ern cross-covariance
  functions for multivariate random fields}, Journal of The American
  Statistical Association \textbf{105} (2010), no.~491, 1167--1177, \href
  {http://dx.doi.org/10.1198/jasa.2010.tm09420}
  {\path{doi:10.1198/jasa.2010.tm09420}}.

\bibitem{grigoriu2012}
M.~Grigoriu, \emph{Stochastic systems: Uncertainty quantification and
  propagation}, Springer, 2012, \href
  {http://dx.doi.org/10.1007/978-1-4471-2327-9}
  {\path{doi:10.1007/978-1-4471-2327-9}}.

\bibitem{Grigoriu2016}
\bysame, \emph{Microstructure models and material response by extreme value
  theory}, SIAM/ASA J Uncertainty Quantification \textbf{4} (2016), 190--217,
  \href {http://dx.doi.org/10.1137/15M1006453} {\path{doi:10.1137/15M1006453}}.

\bibitem{groissJungSchwman2017}
D.~Groisser, S.~Jung, and A.~Schwartzman, \emph{Foundations for a
  scaling-rotation geometry in the space of symmetric positive-definite
  matrices} [online], arXiv: 1702.03237 [math.MG], 2017, \href
  {http://arxiv.org/abs/1702.03237} {\path{arXiv:1702.03237}}, \href
  {http://dx.doi.org/10.48550/arXiv.1702.03237}
  {\path{doi:10.48550/arXiv.1702.03237}}.

\bibitem{GroisserJungSchwartzman2017}
\bysame, \emph{Geometric foundations for scaling-rotation statistics on
  symmetric positive definite matrices: Minimal smooth scaling-rotation curves
  in low dimensions}, Electronic Journal of Statistics \textbf{17} (2017),
  1092--1159, \href {http://dx.doi.org/10.1214/17-EJS1250}
  {\path{doi:10.1214/17-EJS1250}}.

\bibitem{GroisserEtal2021}
\bysame, \emph{Uniqueness questions in a scaling-rotation geometry on the space
  of symmetric positive-definite matrices}, Differential Geometry and its
  Applications \textbf{79} (2021), 101798, \href
  {http://dx.doi.org/10.1016/j.difgeo.2021.101798}
  {\path{doi:10.1016/j.difgeo.2021.101798}}.

\bibitem{groissJungSchwman2023}
\bysame, \emph{A genericity property of {Fréchet} sample means on {Riemannian}
  manifolds} [online], arXiv: 2309.13823 [math.PR], 2023, \href
  {http://arxiv.org/abs/2309.13823} {\path{arXiv:2309.13823}}, \href
  {http://dx.doi.org/10.48550/arXiv.2309.13823}
  {\path{doi:10.48550/arXiv.2309.13823}}.

\bibitem{GuiguiEtal2023}
N.~Guigui, N.~Miolane, and X.~Pennec, \emph{Introduction to {Riemannian}
  geometry and geometric statistics: From basic theory to implementation with
  {Geomstats}}, Foundations and Trends in Machine Learning \textbf{16} (2023),
  no.~3, 329--493, \href {http://dx.doi.org/10.1561/2200000098}
  {\path{doi:10.1561/2200000098}}.

\bibitem{GuilleminotSoize2011}
J.~Guilleminot and C.~Soize, \emph{Non-{G}aussian positive-definite
  matrix-valued random fields with constrained eigenvalues: application to
  random elasticity tensors with uncertain material symmetries}, Int. J. Numer.
  Methods Eng. \textbf{88} (2011), no.~11, 1128--1151.

\bibitem{GuilleminotSoize2012}
\bysame, \emph{Generalized stochastic approach for constitutive equation in
  linear elasticity: a random matrix model}, Int. J. Numer. Methods Eng.
  \textbf{90} (2012), no.~5, 613--635, \href
  {http://dx.doi.org/10.1002/nme.3338} {\path{doi:10.1002/nme.3338}}.

\bibitem{GuilleminotSoize2012a}
\bysame, \emph{Stochastic modeling of anisotropy in multiscale analysis of
  heterogeneous materials: A comprehensive overview on random matrix
  approaches}, Mechanics of Materials \textbf{44} (2012), 35--46, \href
  {http://dx.doi.org/10.1016/j.mechmat.2011.06.003}
  {\path{doi:10.1016/j.mechmat.2011.06.003}}.

\bibitem{GuilleminotSoize2013}
\bysame, \emph{On the statistical dependence for the components of random
  elasticity tensors exhibiting material symmetry properties}, J. Elasticity
  \textbf{111} (2013), no.~2, 109--130.

\bibitem{GuilleminotSoize2013a}
\bysame, \emph{Stochastic model and generator for random fields with symmetry
  properties: application to the mesoscopic modeling of elastic random media},
  Multiscale Model. Simul. \textbf{11} (2013), no.~3, 840--870.

\bibitem{GuilleminotSoize2014}
\bysame, \emph{Itô {SDE}-based generator for a class of non-{G}aussian
  vector-valued random fields in uncertainty quantification}, SIAM J. Sci.
  Comput. \textbf{36} (2014), no.~6, A2763--A2786.

\bibitem{GuilleminotSoizeGhanemR.2011}
J.~Guilleminot, C.~Soize, and R.~G. Ghanem, \emph{Stochastic representation for
  anisotropic permeability tensor random fields}, Int. J. Numer. Anal. Meth.
  Geomech. \textbf{36} (2011), no.~13, 1592--1608, \href
  {http://dx.doi.org/10.1002/nag.1081} {\path{doi:10.1002/nag.1081}}.

\bibitem{gupta_matrix_1999}
A.~Gupta and D.~Nagar, \emph{Matrix {Variate} {Distributions}}, Monographs and
  {Surveys} in {Pure} and {Applied} {Mathematics}, Taylor \& Francis, 1999.

\bibitem{Hauberg2015}
S.~Hauberg, \emph{Principal curves on {Riemannian} manifolds}, {IEEE}
  Transactions on Pattern Analysis and Machine Intelligence \textbf{38} (2015),
  no.~9, 1915--1921, \href {http://dx.doi.org/10.1109/tpami.2015.2496166}
  {\path{doi:10.1109/tpami.2015.2496166}}.

\bibitem{Helbig1996}
K.~Helbig, \emph{Kelvin and the early history of seismic anisotropy}, Seismic
  Anisotropy (E.~Fjær, R.~M. Holt, and J.~S. Rathore, eds.), General Series,
  Society of Exploration Geophysicists, 1996, pp.~15--36, \href
  {http://dx.doi.org/10.1190/1.9781560802693.ch2}
  {\path{doi:10.1190/1.9781560802693.ch2}}.

\bibitem{helbig2013}
\bysame, \emph{Review paper: {W}hat {K}elvin might have written about
  {E}lasticity}, Geophysical Prospecting \textbf{61} (2013), 1--20, \href
  {http://dx.doi.org/10.1111/j.1365-2478.2011.01049.x}
  {\path{doi:10.1111/j.1365-2478.2011.01049.x}}.

\bibitem{HerrmannSchwab2019}
L.~Herrmann and C.~Schwab, \emph{Multilevel quasi-{M}onte {C}arlo integration
  with product weights for elliptic {PDEs} with lognormal coefficients}, ESAIM:
  Mathematical Modelling and Numerical Analysis \textbf{53} (2019), 1507--1552,
  \href {http://dx.doi.org/10.1051/m2an/2019016}
  {\path{doi:10.1051/m2an/2019016}}.

\bibitem{HoangSchwab2014}
V.~H. Hoang and C.~Schwab, \emph{{$N$-Term} {W}iener chaos approximation rates
  for elliptic {PDEs} with lognormal {G}aussian random inputs}, Mathematical
  Models and Methods in Applied Sciences \textbf{24} (2014), no.~4, 797--826,
  \href {http://dx.doi.org/10.1142/S0218202513500681}
  {\path{doi:10.1142/S0218202513500681}}.

\bibitem{AIbrahimbEtAl_SN-2022}
A.~Ibrahimbegovic, H.~G. Matthies, S.~Dobrilla, E.~Karavelić, R.~A. {Mejia
  Nava}, C.~U. Nguyen, M.~S. Sarfaraz, A.~Stanić, and J.~Vondřejc,
  \emph{Synergy of stochastics and inelasticity at multiple scales: novel
  {B}ayesian applications in stochastic upscaling and fracture size and scale
  effects}, Springer Nature Applied Sciences \textbf{4} (2022), no.~7, 191,
  \href {http://dx.doi.org/10.1007/s42452-022-04935-y}
  {\path{doi:10.1007/s42452-022-04935-y}}.

\bibitem{AI-HGM-EK-2020}
A.~Ibrahimbegovic, H.~G. Matthies, and E.~Karaveli\'c, \emph{Reduced model of
  macro-scale stochastic plasticity identification by {B}ayesian inference:
  Application to quasi-brittle failure of concrete}, Comput. Methods Appl.
  Mech. Engrg. \textbf{372} (2020), 113428, \href
  {http://dx.doi.org/10.1016/j.cma.2020.113428}
  {\path{doi:10.1016/j.cma.2020.113428}}.

\bibitem{ItinHehl2013}
Y.~Itin and F.~Hehl, \emph{The constitutive tensor of linear elasticity: Its
  decompositions, {Cauchy} relations, null {Lagrangians}, and wave
  propagation}, J. Math. Phys. \textbf{54} (2013), 042903, \href
  {http://dx.doi.org/10.1063/1.4801859} {\path{doi:10.1063/1.4801859}}.

\bibitem{ItinHehl2015}
\bysame, \emph{Irreducible decompositions of the elasticity tensor under the
  linear and orthogonal groups and their physical consequences}, Journal of
  Physics: Conference Series \textbf{597} (2015), 012046, \href
  {http://dx.doi.org/10.1088/1742-6596/597/1/012046}
  {\path{doi:10.1088/1742-6596/597/1/012046}}.

\bibitem{JammalamadakaKozubowski2017}
S.~Jammalamadaka and T.~Kozubowski, \emph{A general approach for obtaining
  wrapped circular distributions via mixtures}, Sankhy\=a : The Indian Journal
  of Statistics \textbf{79} (2017), 133--157, \href
  {http://dx.doi.org/10.1007/s13171-017-0096-4}
  {\path{doi:10.1007/s13171-017-0096-4}}.

\bibitem{jammalamadaka_topics_2001}
S.~R. Jammalamadaka and A.~Sengupta, \emph{Topics in circular statistics},
  Series on multivariate analysis, World Scientific, River Edge, N.J, 2001.

\bibitem{Jaynes-2003}
E.~T. Jaynes, \emph{Probability theory: The logic of science}, Cambridge
  University Press, 2003.

\bibitem{Jones1998}
H.~F. Jones, \emph{{Groups, Representations and Physics}}, Taylor \& Francis,
  1998.

\bibitem{Jost2005}
J.~Jost, \emph{{Riemannian} geometry and geometric analysis}, 4th ed.,
  Springer, 2005.

\bibitem{JungRooksEtAl2023}
S.~Jung, B.~Rooks, D.~Groisser, and A.~Schwartzman, \emph{Averaging symmetric
  positive-definite matrices on the space of eigen-decompositions} [online],
  arXiv: 2306.12025 [stat.ME], 2023, \href {http://arxiv.org/abs/2306.12025}
  {\path{arXiv:2306.12025}}, \href
  {http://dx.doi.org/10.48550/arXiv.2306.12025}
  {\path{doi:10.48550/arXiv.2306.12025}}.

\bibitem{JungSchwartzmanGroisser2015}
S.~Jung, A.~Schwartzman, and D.~Groisser, \emph{Scaling-rotation distance and
  interpolation of symmetric positive-definite matrices}, SIAM J. Matrix Anal.
  Appl. \textbf{36} (2015), no.~3, 1180--1201, \href
  {http://dx.doi.org/10.1137/140967040} {\path{doi:10.1137/140967040}}.

\bibitem{KlapperHahn2006}
H.~Klapper and T.~Hahn, \emph{Point-group symmetry and physical properties of
  crystals}, International Tables for Crystallography (T.~Hahn, ed.), vol.~A,
  Springer, 2006, pp.~804--808, \href
  {http://dx.doi.org/10.1107/97809553602060000521}
  {\path{doi:10.1107/97809553602060000521}}.

\bibitem{KowalczykOstrowska2009}
K.~Kowalczyk–Gajewska and J.~Ostrowska–Maciejewska, \emph{Review on
  spectral decomposition of {Hooke's} tensor for all symmetry groups of linear
  elastic material}, Engng. Trans. \textbf{57} (2009), no.~3--4, 145--183.

\bibitem{Lang1995}
S.~Lang, \emph{{Differential and Riemannian Manifolds}}, Graduate Texts in
  Mathematics, vol. 160, Springer, 1995.

\bibitem{Litvinenko_Keyes2019}
A.~Litvinenko, D.~Keyes, V.~Khoromskaia, B.~N. Khoromskij, and H.~G. Matthies,
  \emph{{Tucker} tensor analysis of {Mat{\'{e}}rn} functions in spatial
  statistics}, Computational Methods in Applied Mathematics \textbf{19} (2019),
  no.~1, 101--122, \href {http://dx.doi.org/10.1515/cmam-2018-0022}
  {\path{doi:10.1515/cmam-2018-0022}}.

\bibitem{LordPowell2014}
G.~J. Lord, C.~E. Powell, and T.~Shardlow, \emph{An introduction to
  computational stochastic {PDEs}}, Cambridge University Press, 2014, \href
  {http://dx.doi.org/10.1017/CBO9781139017329}
  {\path{doi:10.1017/CBO9781139017329}}.

\bibitem{Love1944}
A.~E.~H. Love, \emph{{A Treatise on the Mathematical Theory of Elasticity}},
  4th ed., Dover, 1944.

\bibitem{malgrange_symmetry_2014}
C.~Malgrange, C.~Ricolleau, and M.~Schlenker, \emph{Symmetry and {Physical}
  {Properties} of {Crystals}}, Springer, 2014, \href
  {http://dx.doi.org/10.1007/978-94-017-8993-6}
  {\path{doi:10.1007/978-94-017-8993-6}}.

\bibitem{malyarenko_statistically_2014}
A.~Malyarenko and M.~Ostoja-Starzewski, \emph{Statistically isotropic tensor
  random fields: {Correlation} structures}, Mathematics and Mechanics of
  Complex Systems \textbf{2} (2014), no.~2, 209--231.

\bibitem{silvestrov_spectral_2016}
\bysame, \emph{Spectral {Expansion} of {Three}-{Dimensional} {Elasticity}
  {Tensor} {Random} {Fields}}, Engineering {Mathematics} {I}, vol. 178,
  Springer, Cham, 2016, pp.~281--300.

\bibitem{malyarenko_random_2017}
\bysame, \emph{A {Random} {Field} {Formulation} of {Hooke}’s {Law} in {All}
  {Elasticity} {Classes}}, J Elast \textbf{127} (2017), no.~2, 269--302, \href
  {http://dx.doi.org/10.1007/s10659-016-9613-2}
  {\path{doi:10.1007/s10659-016-9613-2}}.

\bibitem{altenbach_tensor_2018}
\bysame, \emph{Tensor {Random} {Fields} in {Continuum} {Mechanics}},
  Encyclopedia of {Continuum} {Mechanics}, Springer, Berlin, Heidelberg, 2018,
  pp.~1--9.

\bibitem{malyarenko_tensor-valued_2019}
\bysame, \emph{Tensor-{Valued} {Random} {Fields} for {Continuum} {Physics}},
  Cambridge University Press, 2019.

\bibitem{marzouk:2007:JCP}
Y.~Marzouk, H.~Najm, and L.~Rahn, \emph{Stochastic spectral methods for
  efficient {B}ayesian solution of inverse problems}, Journal of Computational
  Physics \textbf{224} (2007), no.~2, 560--586.

\bibitem{marzouk:2009:CompPhys}
Y.~Marzouk and D.~Xiu, \emph{A stochastic collocation approach to {B}ayesian
  inference in inverse problems}, Communications in Computational Physics
  \textbf{6} (2009), no.~4, 826--847.

\bibitem{Matern_1986}
B.~Mat\'ern, \emph{{Spatial Variation}}, Springer, 1986.

\bibitem{Matthies2008}
H.~G. Matthies, \emph{Stochastic finite elements: {C}omputational approaches to
  stochastic partial differential equations}, Z. Angew. Math. Mech. (ZAMM)
  \textbf{88} (2008), no.~11, 849--873, \href
  {http://dx.doi.org/10.1002/zamm.200800095}
  {\path{doi:10.1002/zamm.200800095}}.

\bibitem{Matthies2005}
H.~G. Matthies and A.~Keese, \emph{{Galerkin} methods for linear and nonlinear
  elliptic stochastic partial differential equations}, Computer Methods in
  Applied Mechanics and Engineering \textbf{194} (2005), no.~12--16,
  1295--1331, \href {http://dx.doi.org/10.1016/j.cma.2004.05.027}
  {\path{doi:10.1016/j.cma.2004.05.027}}.

\bibitem{MatthiesZanderEtAl2016}
H.~G. Matthies, E.~Zander, B.~Rosi\'c, A.~Litvinenko, and O.~Pajonk,
  \emph{Inverse problems in a {B}ayesian setting}, Computational Methods for
  Solids and Fluids (A.~Ibrahimbegovic, ed.), Computational Methods in Applied
  Sciences, vol.~41, Springer, 2016, pp.~245--286, \href
  {http://dx.doi.org/10.1007/978-3-319-27996-1_10}
  {\path{doi:10.1007/978-3-319-27996-1_10}}.

\bibitem{hgm07}
H.~G. Matthies, \emph{Uncertainty quantification with stochastic finite
  elements}, Encyclopedia of Computational Mechanics, vol. Part 1:
  Fundamentals, John Wiley \& Sons, 2007, \href
  {http://dx.doi.org/10.1002/0470091355.ecm071}
  {\path{doi:10.1002/0470091355.ecm071}}.

\bibitem{hgm18}
\bysame, \emph{Uncertainty quantification and {Bayesian} inversion},
  Encyclopedia of Computational Mechanics, vol. Part 1: Fundamentals, John
  Wiley \& Sons, 2nd ed., 2018, pp.~1--51, \href
  {http://dx.doi.org/10.1002/9781119176817.ecm2071}
  {\path{doi:10.1002/9781119176817.ecm2071}}.

\bibitem{hgmRO-1-2018}
H.~G. Matthies and R.~Ohayon, \emph{Analysis of parametric models: Linear
  methods and approximations}, Advances in Computational Mathematics
  \textbf{45} (2019), 2555--2586, \href
  {http://dx.doi.org/10.1007/s10444-019-09735-4}
  {\path{doi:10.1007/s10444-019-09735-4}}.

\bibitem{hgmRO-3-2019}
\bysame, \emph{Parametric models analysed with linear maps}, Adv. Model. and
  Simul. in Eng. Sci. \textbf{7} (2020), 41, \href
  {http://dx.doi.org/10.1186/s40323-020-00172-3}
  {\path{doi:10.1186/s40323-020-00172-3}}.

\bibitem{MatthiesZanderRosicEtAl2016}
H.~G. Matthies, E.~Zander, B.~V. Rosi\'c, and A.~Litvinenko, \emph{Parameter
  estimation via conditional expectation: a {Bayesian} inversion}, Adv. Model.
  and Simul. in Eng. Sci. \textbf{3} (2016), 24, \href
  {http://dx.doi.org/10.1186/s40323-016-0075-7}
  {\path{doi:10.1186/s40323-016-0075-7}}.

\bibitem{mehrCow1990}
M.~Mehrabadi and S.~Cowin, \emph{Eigentensors of linear anisotropic elastic
  materials}, Quart J Mech Appl Math. \textbf{43} (1990), 15--41.

\bibitem{MehrabadiCowinJaric1995}
M.~Mehrabadi, S.~Cowin, and J.~Jaric, \emph{Six-dimensional orthogonal tensor
  representation of the rotation about an axis in three dimensions},
  International Journal of Solids and Structures \textbf{32} (1995), no.~3/4,
  439--449.

\bibitem{Moakher2002}
M.~Moakher, \emph{Means and averaging in the group of rotations}, SIAM J.
  Matrix Anal. Appl. \textbf{24} (2002), no.~1, 1--16, \href
  {http://dx.doi.org/10.1137/S089547980138387}
  {\path{doi:10.1137/S089547980138387}}.

\bibitem{Moakher2005}
\bysame, \emph{A differential geometric approach to the geometric mean of
  symmetric positive-definite matrices}, SIAM J. Matrix Anal. Appl. \textbf{26}
  (2005), no.~3, 735--747, \href {http://dx.doi.org/10.1137/S08954798034369}
  {\path{doi:10.1137/S08954798034369}}.

\bibitem{Moakher2006}
\bysame, \emph{On the averaging of symmetric positive-definite tensors}, J
  Elast \textbf{82} (2006), 273--296, \href
  {http://dx.doi.org/10.1007/s10659-005-9035-z}
  {\path{doi:10.1007/s10659-005-9035-z}}.

\bibitem{Moakher2008}
\bysame, \emph{Fourth-order {Cartesian} tensors: old and new facts, notions and
  applications}, Quarterly Journal of Mechanics and Applied Mathematics
  \textbf{61} (2008), no.~2, 181--203, \href
  {http://dx.doi.org/10.1093/qjmam/hbm027} {\path{doi:10.1093/qjmam/hbm027}}.

\bibitem{Moakher2009}
\bysame, \emph{The algebra of fourth-order tensors with application to
  diffusion {MRI}}, Visualization and Processing of Tensor Fields. Mathematics
  and Visualization (D.~Laidlaw and J.~Weickert, eds.), Springer, 2009,
  pp.~57--80, \href {http://dx.doi.org/10.1007/978-3-540-88378-4_4}
  {\path{doi:10.1007/978-3-540-88378-4_4}}.

\bibitem{MoakherNorris2006}
M.~Moakher and A.~N. Norris, \emph{The closest elastic tensor of arbitrary
  symmetry to an elasticity tensor of lower symmetry}, J Elast \textbf{85}
  (2006), 215--263, \href {http://dx.doi.org/10.1007/s10659-006-9082-0}
  {\path{doi:10.1007/s10659-006-9082-0}}.

\bibitem{MoakherZerai2011}
M.~Moakher and M.~Zéraï, \emph{The {Riemannian} geometry of the space of
  positive-definite matrices and its application to the regularizationof
  positive-definite matrix-valued data}, J Math Imaging Vis \textbf{40} (2011),
  171--187, \href {http://dx.doi.org/10.1007/s10851-010-0255-x}
  {\path{doi:10.1007/s10851-010-0255-x}}.

\bibitem{MorinEtal2020}
L.~Morin, P.~Gilormini, and K.~Derrien, \emph{Generalized {Euclidean} distances
  for elasticity tensors}, J Elast \textbf{138} (2020), 221--232, \href
  {http://dx.doi.org/10.1007/s10659-019-09741-z}
  {\path{doi:10.1007/s10659-019-09741-z}}.

\bibitem{mosegaard:2002:IP}
K.~Mosegaard and M.~Sambridge, \emph{{Monte} {C}arlo analysis of inverse
  problems}, Inverse Problems \textbf{18} (2002), no.~3, 29--54.

\bibitem{newn2005}
R.~E. Newnham, \emph{Properties of materials: Anisotropy, symmetry, structure},
  Oxford University Press, 2005.

\bibitem{NielsenBhatia2013}
F.~Nielsen and R.~Bhatia (eds.), \emph{Matrix information geometry}, Berlin,
  Springer, 2013.

\bibitem{Nikabadze2009}
M.~U. Nikabadze, \emph{On some problems of tensor calculus. {I}}, J Math Sci
  \textbf{161} (2009), 668--697, \href
  {http://dx.doi.org/10.1007/s10958-009-9595-8}
  {\path{doi:10.1007/s10958-009-9595-8}}.

\bibitem{Nikabadze2016}
\bysame, \emph{Eigenvalue problems of a tensor and a tensor-block matrix
  ({TMB}) of any even rank with some applications in mechanics}, Generalized
  Continua as Models for Classical and Advanced Materials (H.~Altenbach and
  S.~Forest, eds.), Advanced Structured Materials, vol.~42, Springer, 2016,
  pp.~279--317, \href {http://dx.doi.org/10.1007/978-3-319-31721-2_14}
  {\path{doi:10.1007/978-3-319-31721-2_14}}.

\bibitem{Nikabadze2017a}
\bysame, \emph{Eigenvalue problem for tensors of even rank and its applications
  in mechanics}, J Math Sci \textbf{221} (2017), 174--204, \href
  {http://dx.doi.org/10.1007/s10958-017-3226-6}
  {\path{doi:10.1007/s10958-017-3226-6}}.

\bibitem{Nikabadze2017}
\bysame, \emph{Topics on tensor calculus with applications to mechanics}, J
  Math Sci \textbf{225} (2017), 1--194, \href
  {http://dx.doi.org/10.1007/s10958-017-3467-4}
  {\path{doi:10.1007/s10958-017-3467-4}}.

\bibitem{NordmannEtAl2018}
J.~Nordmann, M.~Aßmus, and H.~Altenbach, \emph{Visualising elastic anisotropy:
  theoretical background and computational implementation}, Continuum Mech
  Thermodyn \textbf{30} (2018), 689--708, \href
  {http://dx.doi.org/10.1007/s00161-018-0635-9}
  {\path{doi:10.1007/s00161-018-0635-9}}.

\bibitem{nordmann_visualising_2018}
J.~Nordmann, M.~Aßmus, and H.~Altenbach, \emph{Visualising elastic anisotropy:
  theoretical background and computational implementation}, Continuum Mechanics
  and Thermodynamics \textbf{30} (2018), no.~4, 689--708 (en), \href
  {http://dx.doi.org/10.1007/s00161-018-0635-9}
  {\path{doi:10.1007/s00161-018-0635-9}}.

\bibitem{NouySoize2014}
A.~Nouy and C.~Soize, \emph{Random fields representations for stochastic
  elliptic boundary value problems and statistical inverse problems}, European
  Journal of Applied Mathematics \textbf{25} (2014), no.~3, 339--373, \href
  {http://dx.doi.org/10.1017/S0956792514000072}
  {\path{doi:10.1017/S0956792514000072}}.

\bibitem{nye_physical_1984}
J.~F. Nye, \emph{Physical properties of crystals: their representation by
  tensors and matrices}, Oxford University Clarendon Press, 1984.

\bibitem{OdenDemko2010}
J.~T. Oden and L.~Demkowicz, \emph{{Applied Functional Analysis}}, Chapman and
  Hall, 2010.

\bibitem{OdenReddy2012}
J.~T. Oden and J.~N. Reddy, \emph{{An Introduction to the Mathematical Theory
  of Finite Elements}}, Dover, 2012.

\bibitem{oliveAuffray2013}
M.~Olive and N.~Auffray, \emph{Symmetry classes for even order tensors}, Math
  and Mech of Complex Systems \textbf{1} (2013), 177--210, \href
  {http://dx.doi.org/10.2140/memocs.2013.1.177}
  {\path{doi:10.2140/memocs.2013.1.177}}.

\bibitem{OliveEtal2017}
M.~Olive, B.~Kolev, and N.~Auffrey, \emph{A minimal integrity basis for the
  elasticity tensor}, Arch. Rational Mech. Anal. \textbf{226} (2017), 1--31,
  \href {http://dx.doi.org/10.1007/s00205-017-1127-y}
  {\path{doi:10.1007/s00205-017-1127-y}}.

\bibitem{OliveKolevEtAl2018}
M.~Olive, B.~Kolev, B.~Desmorat, and R.~Desmorat, \emph{Harmonic factorization
  and reconstruction of the elasticity tensor}, J Elast \textbf{132} (2018),
  67--101, \href {http://dx.doi.org/10.1007/s10659-017-9657-y}
  {\path{doi:10.1007/s10659-017-9657-y}}.

\bibitem{Oller_1995}
J.~M. Oller and J.~M. Corcuera, \emph{Intrinsic analysis of statistical
  estimation}, The Annals of Statistics \textbf{23} (1995), no.~5, 1562--1581,
  \href {http://dx.doi.org/10.1214/aos/1176324312}
  {\path{doi:10.1214/aos/1176324312}}.

\bibitem{ostoja-starzewski_microstructural_2007}
M.~Ostoja-Starzewski, \emph{Microstructural {Randomness} and {Scaling} in
  {Mechanics} of {Materials}}, Chapman and Hall/CRC, August 2007.

\bibitem{Ostrosablin1984}
N.~Ostrosablin, \emph{On the structure of the tensor of elasticity moduli.
  {E}lastic eigenstates}, Continuum Dynamics \textbf{66} (1984), 11--125.

\bibitem{Ostrosablin1986}
\bysame, \emph{Elasticity moduli and eigenstates for materials of
  crystallographic syngonies}, Continuum Dynamics \textbf{75} (1986), 113--125.

\bibitem{Ostrosablin1986a}
\bysame, \emph{On the structure of the tensor of elasticity moduli and
  classification ofanisotropic materials}, Zh. Prikl. Mekh. Tekh. Fiz.
  \textbf{4} (1986), 127--135.

\bibitem{PajonkRosicEtAl2012}
O.~Pajonk, B.~V. Rosi\'c, A.~Litvinenko, and H.~G. Matthies, \emph{A
  deterministic filter for non-{Gaussian} {Bayesian} estimation}, Physica D:
  Nonlinear Phenomena \textbf{241} (2012), no.~7, 775--788, \href
  {http://dx.doi.org/10.1016/j.physd.2012.01.001}
  {\path{doi:10.1016/j.physd.2012.01.001}}.

\bibitem{Pennec2006}
X.~Pennec, \emph{Intrinsic statistics on {Riemannian} manifolds: Basic tools
  for geometric measurements}, Journal of Mathematical Imaging and Vision
  \textbf{25} (2006), no.~1, 127--154, \href
  {http://dx.doi.org/10.1007/s10851-006-6228-4}
  {\path{doi:10.1007/s10851-006-6228-4}}.

\bibitem{PennecFillardAyache2004}
X.~Pennec, P.~Fillard, and N.~Ayache, \emph{A {R}iemannian framework for tensor
  computing}, Research Report RR-5255, INRIA, 2004, Available from:
  \url{https://hal.inria.fr/inria-00070743}.

\bibitem{PennecSommerFletcher2020}
X.~Pennec, S.~Sommer, and T.~Fletcher (eds.), \emph{{Riemannian} geometric
  statistics in medical image analysis}, The Elsevier and MICCAI Society Book
  Series, Academic Press, 2020.

\bibitem{raoJammaSengup2001}
S.~{Rao Jammalamadaka} and A.~Sengupta, \emph{Topics in circular statistics},
  World Scientific, 2001.

\bibitem{Richards1963}
P.~Richards, \emph{Symmetry in numerical matrix calculations}, SIAM Review
  \textbf{5} (1963), no.~3, 234--248.

\bibitem{RosicSykoraPajonkEtAl2016}
B.~Rosi\'c, J.~S\'ykora, O.~Pajonk, K.~A., and H.~G. Matthies, \emph{Comparison
  of numerical approaches to {Bayesian} updating}, Computational Methods for
  Solids and Fluids (A.~Ibrahimbegovic, ed.), Computational Methods in Applied
  Sciences, vol.~41, Springer, 2016, pp.~427--461, \href
  {http://dx.doi.org/10.1007/978-3-319-27996-1_16}
  {\path{doi:10.1007/978-3-319-27996-1_16}}.

\bibitem{RosicKucerovaSykoraEtAl2013}
B.~V. Rosi\'c, A.~Ku\v{c}erov\'a, J.~S\'ykora, O.~Pajonk, A.~Litvinenko, and
  H.~G. Matthies, \emph{Parameter identification in a probabilistic setting},
  Engineering Structures \textbf{50} (2013), 179--196, \href
  {http://dx.doi.org/10.1016/j.engstruct.2012.12.029}
  {\path{doi:10.1016/j.engstruct.2012.12.029}}.

\bibitem{RosicLitvinenkoEtAl2012}
B.~V. Rosi\'c, A.~Litvinenko, O.~Pajonk, and H.~G. Matthies,
  \emph{Sampling-free linear {Bayesian} update of polynomial chaos
  representations}, Journal of Computational Physics \textbf{231} (2012),
  no.~17, 5761--5787, \href {http://dx.doi.org/10.1016/j.jcp.2012.04.044}
  {\path{doi:10.1016/j.jcp.2012.04.044}}.

\bibitem{rych1984}
J.~Rychlewski, \emph{On {H}ooke’s law}, Prikl. Matem. Mekhan. \textbf{48}
  (1984), 420--435, Translated into English as Journal of Mathematics and
  Mechanics (Oxford) (1984) 303--314.

\bibitem{rych2000}
\bysame, \emph{A qualitative approach to {H}ooke’s tensor. {Part I}}, Arch
  Mech \textbf{52} (2000), 737--759.

\bibitem{rych2001}
\bysame, \emph{A qualitative approach to {H}ooke’s tensor. {Part II}}, Arch
  Mech \textbf{53} (2001), 45--63.

\bibitem{SavvasPapai2020}
D.~Savvas, I.~Papaioannou, and G.~Stefanou, \emph{{B}ayesian identification and
  model comparison for random property fields derived from material
  microstructure}, Comput. Methods Appl. Mech. Engrg. \textbf{365} (2020),
  113026, \href {http://dx.doi.org/10.1016/j.cma.2020.113026}
  {\path{doi:10.1016/j.cma.2020.113026}}.

\bibitem{schwartzman_random_nodate}
A.~Schwartzman, \emph{Random ellipsoids and false discovery rates: Statistics
  for diffusion tensor imaging data}, Ph.D. thesis, Stanford University, 2006.

\bibitem{schwartzman_lognormal_2016}
\bysame, \emph{Lognormal {Distributions} and {Geometric} {Averages} of
  {Symmetric} {Positive} {Definite} {Matrices}: {Lognormal} {Positive}
  {Definite} {Matrices}}, International Statistical Review \textbf{84} (2016),
  no.~3, 456--486.

\bibitem{segalKunze78}
I.~E. Segal and R.~A. Kunze, \emph{Integrals and operators}, $2^{nd}$ ed.,
  Springer, Berlin, 1978.

\bibitem{shivaEtal2021}
S.~K. Shivanand, B.~Rosi\'c, and H.~G. Matthies, \emph{Stochastic modelling of
  symmetric positive-definite material tensors}, Journal of Computational
  Physics \textbf{505} (2024), 112883, \href
  {http://dx.doi.org/10.1016/j.jcp.2024.112883}
  {\path{doi:10.1016/j.jcp.2024.112883}}.

\bibitem{slaw2007}
M.~A. Slawinski, \emph{Waves and rays in elastic continua}, 2nd ed., World
  Scientific, 2007.

\bibitem{soize_nonparametric_2000}
C.~Soize, \emph{A nonparametric model of random uncertainties for reduced
  matrix models in structural dynamics}, Probabilistic Engineering Mechanics
  \textbf{15} (2000), no.~3, 277--294, \href
  {http://dx.doi.org/10.1016/S0266-8920(99)00028-4}
  {\path{doi:10.1016/S0266-8920(99)00028-4}}.

\bibitem{soize_2006}
\bysame, \emph{Non-{G}aussian positive-definite matrix-valued random fields for
  elliptic stochastic partial differential operators}, Computer Methods in
  Applied Mechanics and Engineering \textbf{195} (2006), no.~1--3, 26--64,
  \href {http://dx.doi.org/10.1016/j.cma.2004.12.014}
  {\path{doi:10.1016/j.cma.2004.12.014}}.

\bibitem{Soize2017}
C.~Soize, \emph{{Uncertainty Quantification}}, Interdisciplinary Applied
  Mathematics, vol.~47, Springer, 2017, \href
  {http://dx.doi.org/10.1007/978-3-319-54339-0}
  {\path{doi:10.1007/978-3-319-54339-0}}.

\bibitem{Soize2021a}
\bysame, \emph{Stochastic elliptic operators defined by non-{G}aussian random
  fields with uncertain spectrum} [online], arXiv: 2106.07706 [cs.NA], 2021,
  \href {http://arxiv.org/abs/2106.07706} {\path{arXiv:2106.07706}}, \href
  {http://dx.doi.org/10.48550/arXiv.2106.07706}
  {\path{doi:10.48550/arXiv.2106.07706}}.

\bibitem{SommerFletcherPennec2020}
S.~Sommer, T.~Fletcher, and X.~Pennec, \emph{Introduction to differential and
  {Riemannian} geometry}, {Riemannian} Geometric Statistics in Medical Image
  Analysis (X.~Pennec, S.~Sommer, and T.~Fletcher, eds.), Elsevier, 2020,
  pp.~3--37, \href {http://dx.doi.org/10.1016/b978- 0-12-814725-2.00008-x}
  {\path{doi:10.1016/b978- 0-12-814725-2.00008-x}}.

\bibitem{Spivak-1-1999}
M.~Spivak, \emph{{A Comprehensive Introduction to Differential Geometry}}, 3rd
  ed., vol.~1, Publish or Perish, 1999.

\bibitem{StaberGuilleminot2017}
B.~Staber and J.~Guilleminot, \emph{Stochastic modeling and generation of
  random fields of elasticity tensors: A unified information-theoretic
  approach}, C. R. M\'ecanique \textbf{345} (2017), 399--416, \href
  {http://dx.doi.org/10.1016/j.crme.2017.05.001}
  {\path{doi:10.1016/j.crme.2017.05.001}}.

\bibitem{StahnMuellerBertram2020}
O.~Stahn, W.~Müller, and A.~Bertram, \emph{Distances of stiffnesses to
  symmetry classes}, J Elast \textbf{141} (2020), 349--361, \href
  {http://dx.doi.org/10.1007/s10659-020-09787-4}
  {\path{doi:10.1007/s10659-020-09787-4}}.

\bibitem{AMStuart2010}
A.~M. Stuart, \emph{Inverse problems: a {Bayesian} perspective}, Acta Numerica
  \textbf{19} (2010), 451--559, \href
  {http://dx.doi.org/10.1017/S0962492910000061}
  {\path{doi:10.1017/S0962492910000061}}.

\bibitem{sutcl1992}
S.~Sutcliffe, \emph{Spectral decomposition of the elasticity tensor}, J Appl
  Mech \textbf{59} (1992), 762--773.

\bibitem{tarantola:2005}
A.~Tarantola, \emph{{Inverse Problem Theory and Methods for Model Parameter
  Estimation}}, SIAM, 2005.

\bibitem{ThanwerdasPennec2019}
Y.~Thanwerdas and X.~Pennec, \emph{Exploration of balanced metrics on symmetric
  positive definite matrices} [online], arXiv: 1909.03852 [math.DG], 2019,
  \href {http://arxiv.org/abs/1909.03852} {\path{arXiv:1909.03852}}, \href
  {http://dx.doi.org/10.48550/arXiv.1909.03852}
  {\path{doi:10.48550/arXiv.1909.03852}}.

\bibitem{ThanwerdasPennec2019-2}
\bysame, \emph{Is affine-invariance well defined on {SPD} matrices? {A}
  principled continuum of metrics} [online], arXiv: 1906.01349 [math.DG], 2019,
  \href {http://arxiv.org/abs/1906.01349} {\path{arXiv:1906.01349}}, \href
  {http://dx.doi.org/10.48550/arXiv.1906.01349}
  {\path{doi:10.48550/arXiv.1906.01349}}.

\bibitem{ThanwerdasPennec2021}
\bysame, \emph{{$\mathrm{O}(n)$-invariant} {R}iemannian metrics on {SPD}
  matrices} [online], arXiv: 2109.05768 [math.DG], 2021, \href
  {http://arxiv.org/abs/2109.05768} {\path{arXiv:2109.05768}}, \href
  {http://dx.doi.org/10.48550/arXiv.2109.05768}
  {\path{doi:10.48550/arXiv.2109.05768}}.

\bibitem{ThanwerdasPennec2023}
\bysame, \emph{{$\mathrm{O}(n)$-invariant} {R}iemannian metrics on {SPD}
  matrices}, Linear Algebra and its Applications \textbf{661} (2023), 163--201,
  \href {http://dx.doi.org/10.1016/j.laa.2022.12.009}
  {\path{doi:10.1016/j.laa.2022.12.009}}.

\bibitem{kelvin1856}
W.~K. {Thomson (Lord~Kelvin)}, \emph{{XXI Elements} of a mathematical theory of
  elasticity. {Part 1}, {O}n stresses and strains}, Philosophical Transactions
  of the Royal Society \textbf{166} (1856), 481--498.

\bibitem{kelvin1878}
\bysame, \emph{Elasticity}, Encyclopedia Britannica \textbf{7} (1878),
  796--825.

\bibitem{tinder2008}
R.~F. Tinder, \emph{Tensor properties of solids: Phenomenological development
  of the tensor properties of crystals}, Morgan \& Claypool, 2008, \href
  {http://dx.doi.org/10.2200/S00057ED1V01Y200712ENG04}
  {\path{doi:10.2200/S00057ED1V01Y200712ENG04}}.

\bibitem{ting1996}
T.~C.~T. Ting, \emph{Positive definiteness of anisotropic elastic constants},
  Math and Mech of Solids \textbf{1} (1996), 301--314.

\bibitem{TodhunterPearson1960}
I.~Todhunter and K.~Pearson, \emph{A history of the theory of elasticity and of
  the strength of materials: From {Galilei} to {Lord Kelvin}}, vol.~2, Dover,
  1960.

\bibitem{torquato:2001}
S.~Torquato, \emph{{Random Heterogeneous Materials}}, Springer, 2001.

\bibitem{Walpole1984}
L.~J. Walpole, \emph{Fourth-rank tensors of the thirty-two crystal classes:
  Multiplication tables}, Proc. R. Soc. London \textbf{391} (1984), 149--179,
  \href {http://dx.doi.org/10.1098/rspa.1984.0008}
  {\path{doi:10.1098/rspa.1984.0008}}.

\bibitem{WangSalehianEtAl2014}
Y.~Wang, H.~Salehian, G.~Cheng, and B.~Vemuri, \emph{Tracking on the product
  manifold of shape and orientation for tractography from diffusion {MRI}},
  Proceedings of the IEEE Conference on Computer Vision and Pattern Recognition
  {(CVPR)}, 2014, pp.~3051--3056, \href
  {http://dx.doi.org/10.1109/cvpr.2014.390} {\path{doi:10.1109/cvpr.2014.390}}.

\bibitem{WeberGluegeBertram2019}
M.~Weber, R.~Glüge, and A.~Bertram, \emph{Distance of a stiffness tetrad to
  the symmetry classes of linear elasticity}, International Journal of Solids
  and Structures \textbf{156--157} (2019), 281--293, \href
  {http://dx.doi.org/10.1016/j.ijsolstr.2018.08.021}
  {\path{doi:10.1016/j.ijsolstr.2018.08.021}}.

\bibitem{yosida-fa-1980}
K.~Yosida, \emph{Functional analysis}, $6^{th}$ ed., Springer, Berlin, 1980.

\bibitem{zhang-malyarenko_fractal_2022}
X.~Zhang, A.~Malyarenko, E.~Porcu, and M.~Ostoja-Starzewski,
  \emph{Elastodynamic problem on tensor random fields with fractal and {Hurst}
  effects}, Meccanica \textbf{57} (2022), 957--970, \href
  {http://dx.doi.org/10.1007/s11012-021-01424-1}
  {\path{doi:10.1007/s11012-021-01424-1}}.

\bibitem{ZhengZou2000}
Q.-S. Zheng and W.-N. Zou, \emph{Irreducible decompositions of physical tensors
  of high orders}, Journal of Engineering Mathematics \textbf{37} (2000),
  273--288.

\end{thebibliography}

\end{document}